\documentclass{aa}
\usepackage{natbib}
\usepackage{graphicx}
\usepackage{epsfig}
\usepackage{amsmath}
\usepackage{dirtytalk}
\usepackage{txfonts}
\usepackage{hyperref}
\usepackage{url}
\usepackage{xcolor}
\hypersetup{
colorlinks,
linkcolor={red!80!black},
citecolor={blue!80!black},
urlcolor={blue!80!black}
}

\def\cm3{cm$^{-3}$}
\def\kms{km~s$^{-1}$}

\def\msun{M$_{\odot}$}

\def\one{\ts {\,\sc i}}
\def\two{\ts {\,\sc ii}}

\def\beq{\begin{equation}}
\def\eeq{\end{equation}}

\def\lesssim{\mathrel{\hbox{\rlap{\hbox{\lower4pt\hbox{$\sim$}}}\hbox{$<$}}}}
\def\gtrsim{\mathrel{\hbox{\rlap{\hbox{\lower4pt\hbox{$\sim$}}}\hbox{$>$}}}}

\def\one{{\,\sc i}}
\def\two{{\,\sc ii}}

\def\vblob{$V_{\rm blob}$}
\def\dvblob{$\delta V_{\rm blob}$}

\newcommand{\code}[1]{\texttt{#1}}

\def\v1d{{\code{V1D}}}

\def\cmfgen{{\code{CMFGEN}}}

\newcommand{\iso}[2]{\ensuremath{^{#1}\rm{#2}}}

\def\apj{ApJ}

\def\apjl{ApJL}
\def\aap{A\&A}
\def\araa{ARA\&A}

\def\mnras{MNRAS}
\def\nat{Nature}

\def\physrep{Phys.~Rep.}

\def\nifs{\iso{56}Ni}
\def\cofs{\iso{56}Co}

\begin{document}

   \title{Polarization signatures of a high-velocity scatterer in nebular-phase spectra of Type II supernovae}
   \titlerunning{Polarization signatures of a high-velocity scatterer in SNe II}

\author{
   Luc Dessart\inst{\ref{inst1}}
  \and
   D. John Hillier\inst{\ref{inst2}}
  \and
   Douglas C. Leonard\inst{\ref{inst3}}
}

\institute{
Institut d'Astrophysique de Paris, CNRS-Sorbonne Universit\'e,  98 bis boulevard Arago, F-75014 Paris, France.\label{inst1}
  \and
    Department of Physics and Astronomy \& Pittsburgh Particle Physics, Astrophysics, and Cosmology Center (PITT PACC),  University of Pittsburgh, 3941 O'Hara Street, Pittsburgh, PA 15260, USA.\label{inst2}
  \and
    Department of Astronomy, San Diego State University, San Diego, CA 92182-1221, USA.\label{inst3}
}

   \date{Received; accepted}
  \abstract{
Type II supernovae (SNe) often exhibit a linear polarization, arising from free-electron scattering, with complicated optical signatures, both in the continuum and in lines. Focusing on the early nebular phase, at a SN age of 200\,d, we conduct a systematic study of the polarization signatures associated with a \nifs\  ``blob'' that breaks spherical symmetry. Our ansatz, supported by non-local thermodynamic equilibrium radiative transfer calculations, is that the primary role of such a \nifs\ blob is to boost the local density of free electrons, which is otherwise reduced following recombination in Type II SN ejecta. Using 2D polarized radiation transfer modeling, we explore the influence of such an electron-density enhancement, varying its magnitude $N_{\rm e, fac}$, its velocity location \vblob, and its spatial extent. For plausible $N_{\rm e, fac}$ values of a few tens, a high-velocity blob can deliver a continuum polarization $P_{\rm cont}$ of $0.5-1.0$\,\% at 200\,d. Our simulations reproduce the analytic scalings for $P_{\rm cont}$, and in particular the linear growth with the blob radial optical depth. The most constraining information is, however, carried by polarized line photons. For a high \vblob, the polarized spectrum appears as a replica of the full spectrum, scaled down by a factor 100 to 1000 (i.e., $1/P_{\rm cont}$), and redshifted by an amount \vblob $(1-\cos \alpha_{\rm los})$, where $\alpha_{\rm los}$ is the line of sight angle. As \vblob\ is reduced, the redshift decreases and the replication deteriorates. Lines whose formation region overlap with the blob appear weaker and narrower in the polarized flux. Because of its dependence on inclination ($\propto \sin^2 \alpha_{\rm los}$), the polarization preferentially reveals asymmetries in the plane perpendicular to the line of sight ($\alpha_{\rm los}=$\,90\,deg). This property also weakens the broadening of lines in the polarized flux. With the adequate choice of electron-density enhancement, some of these results may apply to asymmetric explosions in general, or to the polarization signatures from newly-formed dust in the outer ejecta.
}

   \maketitle

\section{Introduction}
\label{sect_intro}

One possible source of polarization in core-collapse supernovae (SNe) is the presence of \nifs\ at high velocity in the ejecta but confined within a restricted solid angle \citep{chugai_pol_87a_92,chugai_04dj_06,D20_12aw_pol}.\footnote{A detailed introduction to SN spectropolarimetry is provided in \citet{D20_12aw_pol} and is therefore not repeated here.} The \nifs\ and \cofs\ nuclei will influence the surrounding gas through the persistent decay power produced, thereby influencing the temperature and the  ionization of the gas. The resulting free-electron density distribution may no longer be symmetrically distributed above the emitting inner layers and may thus yield a residual polarization. This scenario was explored in detail by \citet{D20_12aw_pol} for the analysis of the spectropolarimetric observations of the Type II-Plateau SN\,2012aw.

 These observations, however, were all taken during the photospheric phase, stopping right at the transition to the nebular phase (at the fall off from the plateau). Late-time spectropolarimetric observations of nearby Type II SNe are, however, possible. Such observations have been obtained for example for SNe 1999em \citep{leonard_99em_specpol_01}, 2004dj \citep{leonard_04dj_06}, 2004et
\citep{leonard_pol_09,leonard_08bk_12}, 2006ov and 2006my \citep{chornock_pol_10}, 2017gmr \citep{nagao_17gmr_pol_19}, and  as part of our own observation programme for SN 2013ej.  Intriguing spectropolarimetric properties are seen in some of these data and will be presented in a forthcoming publication (Leonard et al., in prep).  Here we provide a theoretical framework within which one may  interpret such nebular-phase spectropolarimetric observations.

In the next section we discuss evidence for asymmetries, including those associated with the \nifs\ distribution, in core-collapse SN ejecta. To explore the potential influence of \nifs\ asymmetries on the polarization of nebular phase spectra we design a toy model which utilizes a localized enhancement in free electrons to produce an asymmetric model.
We discuss the validity of this heuristic approach, its advantages and its limitations. Section~\ref{sect_mod} describes the numerical setup for the grid of simulations performed in this study. Section~\ref{sect_ref} contains a description of the results for a representative 2D configuration, while subsequent sections present the results obtained when the blob properties are varied. We discuss the impact of a change in the velocity of the blob (section~\ref{sect_vblob}), in the associated enhancement in electron density (section~\ref{sect_nefac}), in the blob opening angle (section~\ref{sect_opening_angle}), in the blob velocity width (Section~\ref{sect_dvblob}), and in the assumption of mirror symmetry (aka one blob versus two blobs; section~\ref{sect_mirror}). We present our conclusions in section~\ref{sect_conc}. An appendix gives useful information for the interpretation of the results discussed in the main text.

\begin{figure}
\epsfig{file=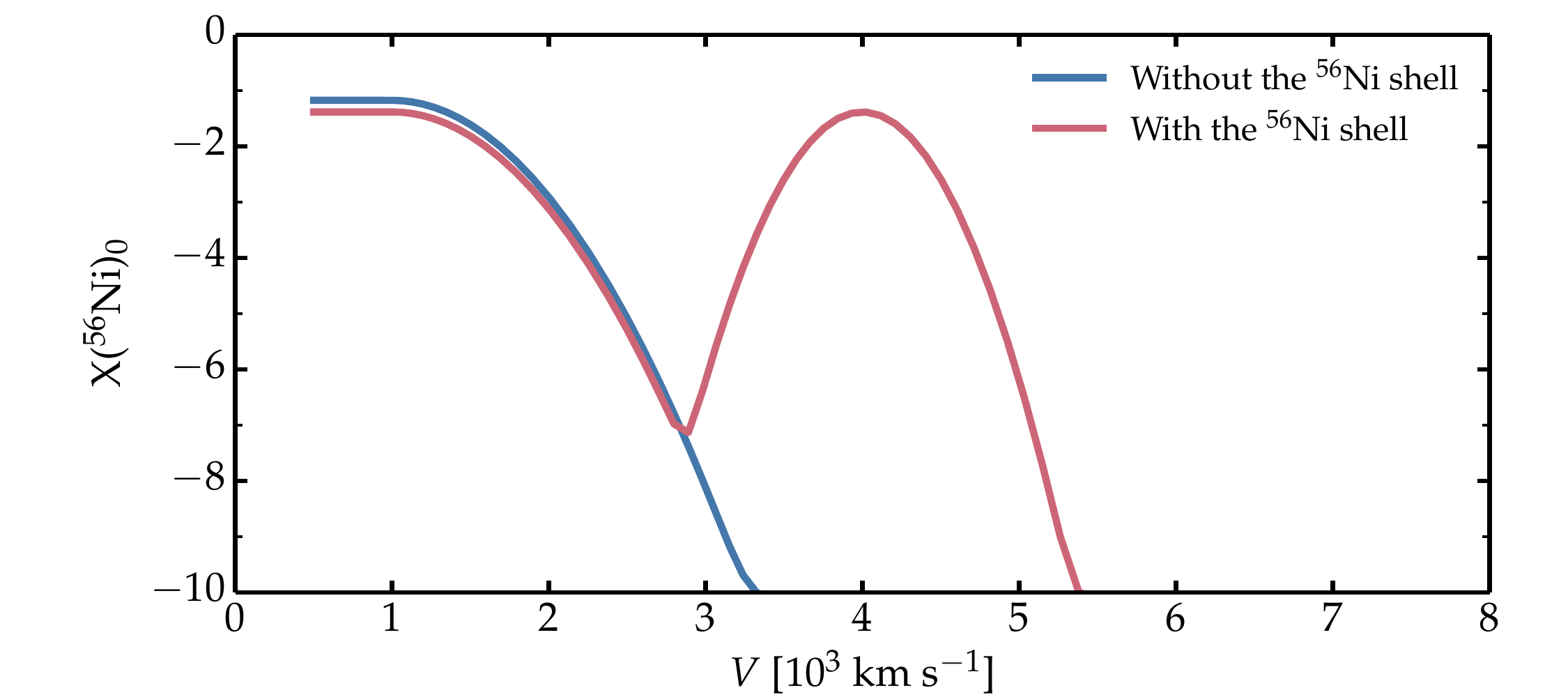, width=9.2cm}
\epsfig{file=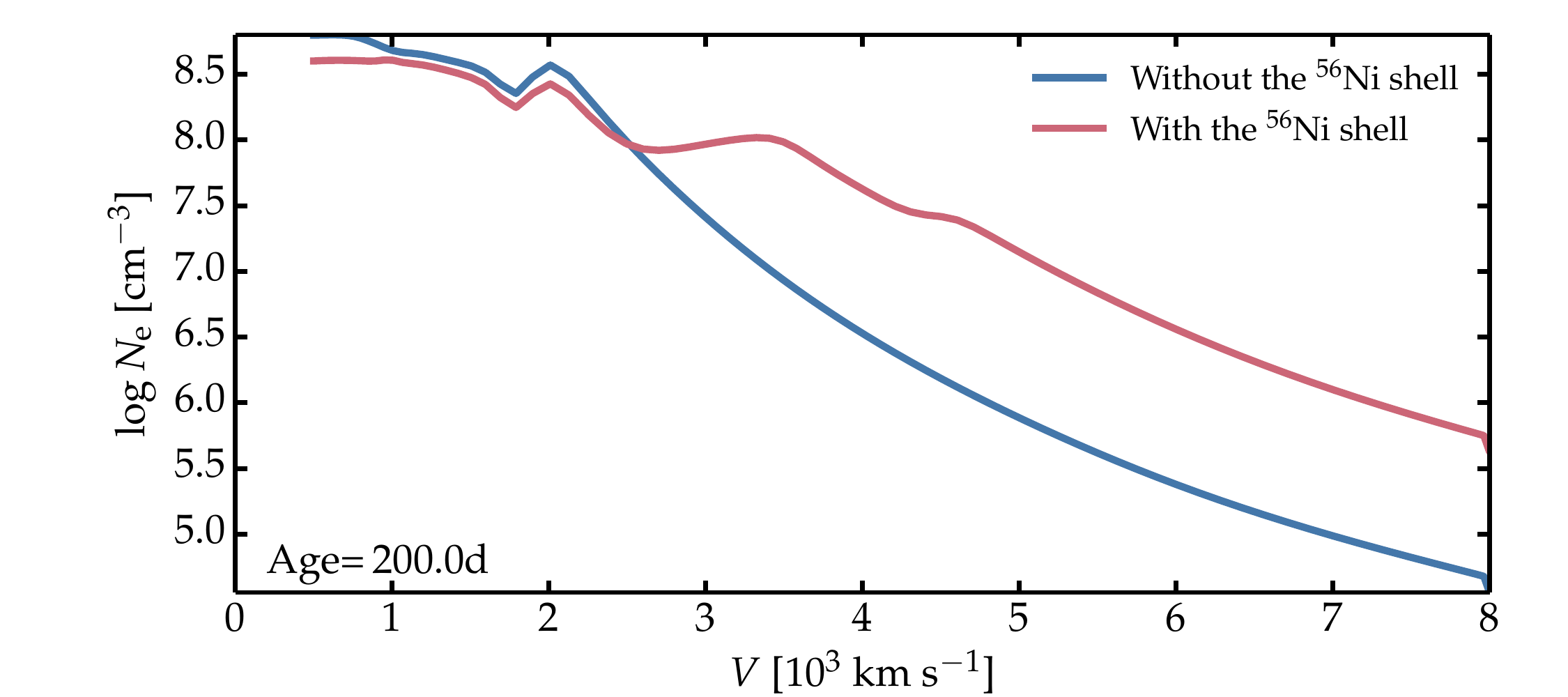, width=9.2cm}
\epsfig{file=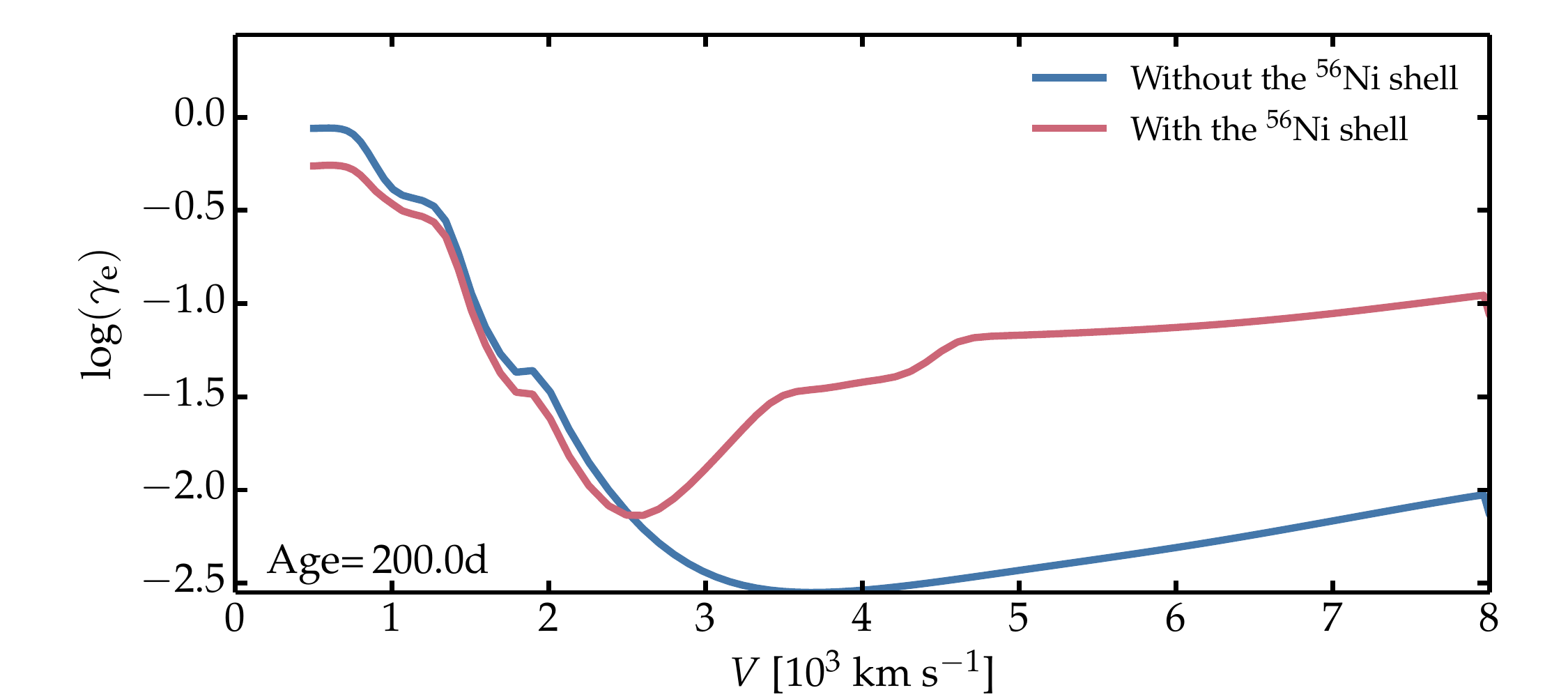,width=9.2cm}
\caption{Top: Profile of the \nifs\ mass fraction versus velocity for the model with a \nifs-rich shell centered at 4000\,\kms\ and the original model without it. Both models have the same total \nifs\ mass of 0.08\,\msun.  Middle: Electron-density profile for the models at top and computed with \cmfgen\ at 200\,d after explosion. The radial electron-scattering optical depth associated with the \nifs-rich shell is 0.15. Bottom: Same as middle, but now showing the mean number of electrons per nucleon $\gamma_{\rm e}$, which is equal to \hbox{$N_{\rm e} \bar{A} m_{\rm H} / \rho$}, where $\rho$ is the mass density and $\bar{A}$ is the mean atomic weight of the material. Conditions are far from complete ionization (especially for hydrogen) so the ionization and the electron density could be enhanced even further. [See section~\ref{sect_intro} for discussion]
\label{fig_cmfgen}
}
\end{figure}

\section{Physical motivation}
\label{sect_motivation}

\subsection{Context}

The notion that core-collapse SN ejecta exhibit extensive chemical mixing started with the observations of SN\,1987A. Indeed,  the smooth rising optical brightness and the high-energy radiation observed after about 200\,d in SN\,1987A suggest the mixing of \nifs\ out to $3000-4000$\,\kms\ (e.g. \citealt{sn1987A_rev_90}). In parallel, numerical simulations of core-collapse SN explosions suggest a strong breaking of spherical
symmetry on small and large scales. This arises both from the intrinsic multi-dimensional nature of the explosion mechanism and the shock wave propagation in a stratified massive star progenitor \citep{muller_87A_91,kifonidis_00,sasi_03,wongwathanarat_15_3d}. Recently, \citet{gabler_3dsn_21} have extended a set of 3D neutrino-driven explosion simulations until late times, thereby providing useful information on the 3D chemical and density structure expected to hold in standard Type II SNe once they have evolved to hundreds of days after explosion.

The simulations of \citet{gabler_3dsn_21} show that the \nifs\ distribution can take all sorts of shapes, including elongated narrow fingers or individual blobs detached from an overall quasi-spherical distribution confined to the inner ejecta. In some cases, the entire asymmetric distribution of \nifs\ is limited to one or two protrusions (their model W15),  or exhibits many fingers spread uniformly over all directions, including isolated blobs (their model B15). Both the \nifs\ and the mass density distribution are asymmetric.

  What is clear is that \nifs\ enhancements at high velocity, reaching inside the outer  H-rich ejecta are routinely predicted in such simulations. In the  B15 and B15x simulations of \citet{gabler_3dsn_21}, the high velocity blobs or narrow fingers subtend an angle of order 10\,deg with a \nifs\ mass fraction (prior to decay) of a few 0.01 (see their Fig.~10). The associated decay heating (which is assumed to be local and not subject to radiative losses) causes these \nifs-rich regions to expand and create lower-density regions, surrounded by a dense shell of compressed, swept-up H-rich material, with no compression beyond.

In reality, $\gamma$-rays emitted in the decay of \nifs\ and \cofs\ travel some distance and are absorbed by the surrounding H-rich material (see, for example \citealt{d12_snibc} for a discussion of this process, and their Figs.~1 and 2 for illustration). Hence, high-velocity \nifs\ blobs are expected to raise the free-electron density of the H-rich material that surrounds them. Because the 3D distribution of \nifs\ is generally asymmetric, the associated non-local influence of decay heating is expected to produce a cocoon of enhanced ionization around them, which could produce in turn a polarization signature. Determining the boost to the electron density around these \nifs\ blobs requires a full 3D non-local thermodynamic equilibrium (non-LTE) radiative transfer simulation. Such simulations, accounting for the asymmetric geometry and inhomogeneities, are impracticable.  Hence, in the next section, we present the results from two 1D \cmfgen\ simulations that quantify the potential effect of a high-velocity \nifs\ abundance enhancement in Type II SN ejecta.

\subsection{Influence of high-velocity $^{56}$Ni in 1D radiative transfer calculations of a Type II SN ejecta}
\label{sect_cmfgen}

To investigate the influence of an asymmetric \nifs\ distribution on the SN radiation during the nebular phase, we used a similar ejecta structure to model x1p5 (which arises from a 15\msun\ star initially) described by \citet{HD19}. We computed two simulations at 200\,d with the 1D non-LTE radiative transfer code \cmfgen\ \citep{HD12}. In the first simulation, we adopted a smooth \nifs\ distribution, while in the second one, the \nifs\ distribution exhibits a bump at 4000\,\kms -- the \nifs\ profile was renormalized so that the total \nifs\ is the same and equal to 0.08\,\msun\ in both models (top panel of Fig.~\ref{fig_cmfgen}).  In 1D, this bump corresponds to a shell of enhanced \nifs\ mass fraction spherically distributed around the ejecta center (the \nifs\ mass associated with that spherical shell, at velocities $\geq$\,2500\,\kms, is about 0.03\,\msun). Despite this enhancement, the \nifs\ mass fraction remains around or below 0.01, so that effectively \nifs\ is always mixed with other species, in particular with H and He for velocities of 2000\,\kms\ or more. We then solved the 1D steady-state non-LTE radiative transfer problem at 200\,d for that ejecta using the same approach as used in \citet{HD19}. The details of the computation (e.g., model atoms, etc) are here unimportant since we are mostly interested in the influence of the \nifs\ enhancement on the gas properties. As can be seen in the middle panel of Fig.~\ref{fig_cmfgen}, the electron density $N_{\rm e}$ is a factor of ten greater beyond 2500\,\kms\ in the model with the \nifs-rich shell at 4000\,\kms (the same holds for the mean number of electrons per nucleon; bottom panel of Fig.~\ref{fig_cmfgen}). Although this shell has a characteristic width of 1000\,\kms, its influence on the electron density is widespread. This is caused in part by the non-local deposition of decay power by $\gamma$-rays, since a larger fraction of the \nifs\ is located at lower density (longer $\gamma$-ray mean free path) in the model with the high velocity \nifs. This is also associated with some $\gamma$-ray leakage so that only 83\,\% of the decay power is absorbed in the model  with high velocity \nifs. In comparison, the trapping of $\gamma$-rays is nearly complete in the other model (99.8\,\% of the decay power is absorbed). In a Type II SN, the influence of a high velocity \nifs\ shell  would first increase and then decrease when the $\gamma$-ray leakage is so large that the decay power absorbed becomes dominated by the contribution from positrons, which would occur after several years.

The polarization arising from a spherical high velocity \nifs-rich shell is zero. However, by limiting the opening angle of the \nifs-rich material, such as in a \nifs\ blob or finger, a residual polarization will result. Due to transfer effects, the blob will also influence its surroundings. The extent of the blob influence will depend on the radial and angular sizes of the blob -- a small mass of \nifs\ confined to a blob can potentially influence a solar mass or more of ejecta material (see Appendix~\ref{sect_blob_mass}). In this paper, we tend to use interchangeably the \nifs\ blob and its associated electron-density enhancement because in our approach the \nifs\ is microscopically mixed with other species at any ejecta location. In nature, this microscopic mixing does not take place during the dynamical phase of the explosion, so that a \nifs\ blob would be essentially pure \nifs\ and be surrounded by material with a distinct composition (a high velocity \nifs\ blob would be surrounded by material rich in H and He in a Type II SN -- i.e., a ``cocoon'', as discussed in the previous section). In that case, the \nifs\ blob would boost the ionization of that surrounding material, thereby producing a similar configuration to what we assume in this work.\footnote{In the work of \citet{gabler_3dsn_21}, radiative losses are neglected and the decay heating is treated as local. These assumptions overestimate the heating of the \nifs-rich regions and their expansion. Hence, relaxing these assumptions would yield \nifs-rich blobs and fingers subtending a smaller angular extent than they obtain. Detailed  calculations are required to determine how large this reduction is.}

In the 1D model with a \nifs-rich shell placed around 4000\,\kms, the total mass of \nifs\ in that shell is about 0.03\,\msun, while the total \nifs\ mass contained in the ejecta is 0.08\,\msun\ (the latter applies to both models). When we restrict the angular size of that \nifs\ enhancement to a small solid angle (as in a blob), the corresponding reduction in mass goes as $0.5 (1-\cos\beta)$, where $\beta$ is the (blob) half opening angle. For an opening angle of 10\,deg, the \nifs\ blob mass would be only 0.0076\,\msun. This value is comparable to those obtained in the simulations of \citet{gabler_3dsn_21} ---  their simulation B15x, for example, has about 10\,\% of the \nifs\ mass beyond 2500\,\kms.

The composition of the \nifs-rich blob is determined by explosive nucleosynthesis and may be a mixture primarily of iron-group elements and He (see, for example \citealt{WH07}). The mixing observed in  3D explosion simulations is primarily macroscopic (microscopic mixing, if present, is probably caused by numerical diffusion), and consists in the shuffling of material in space with little alteration of the original mixture. Hence, the \nifs\ blob should not be microscopically mixed with H-rich material. Instead, a high velocity \nifs-rich blob should retain its original composition (the one it had at the time of explosion) but be surrounded by the H-rich material present in the outer ejecta. This effect is not captured in the \cmfgen\ simulation above since we performed a macroscopic and a microscopic mixing of elements. But in reality, the \nifs-rich material should eventually be surrounded by a cocoon of partially ionized H-rich material, while material more distant from this heating source would be cooler and more recombined. This configuration motivated the heuristic exploration we present in the following sections. Being an exploration, some adopted parameters may not be fully pertinent for the \nifs\ blob case. However, they may offer some useful insights into other sources of asymmetries, such as that caused by an asymmetric explosion energy.

\begin{figure*}
\epsfig{file=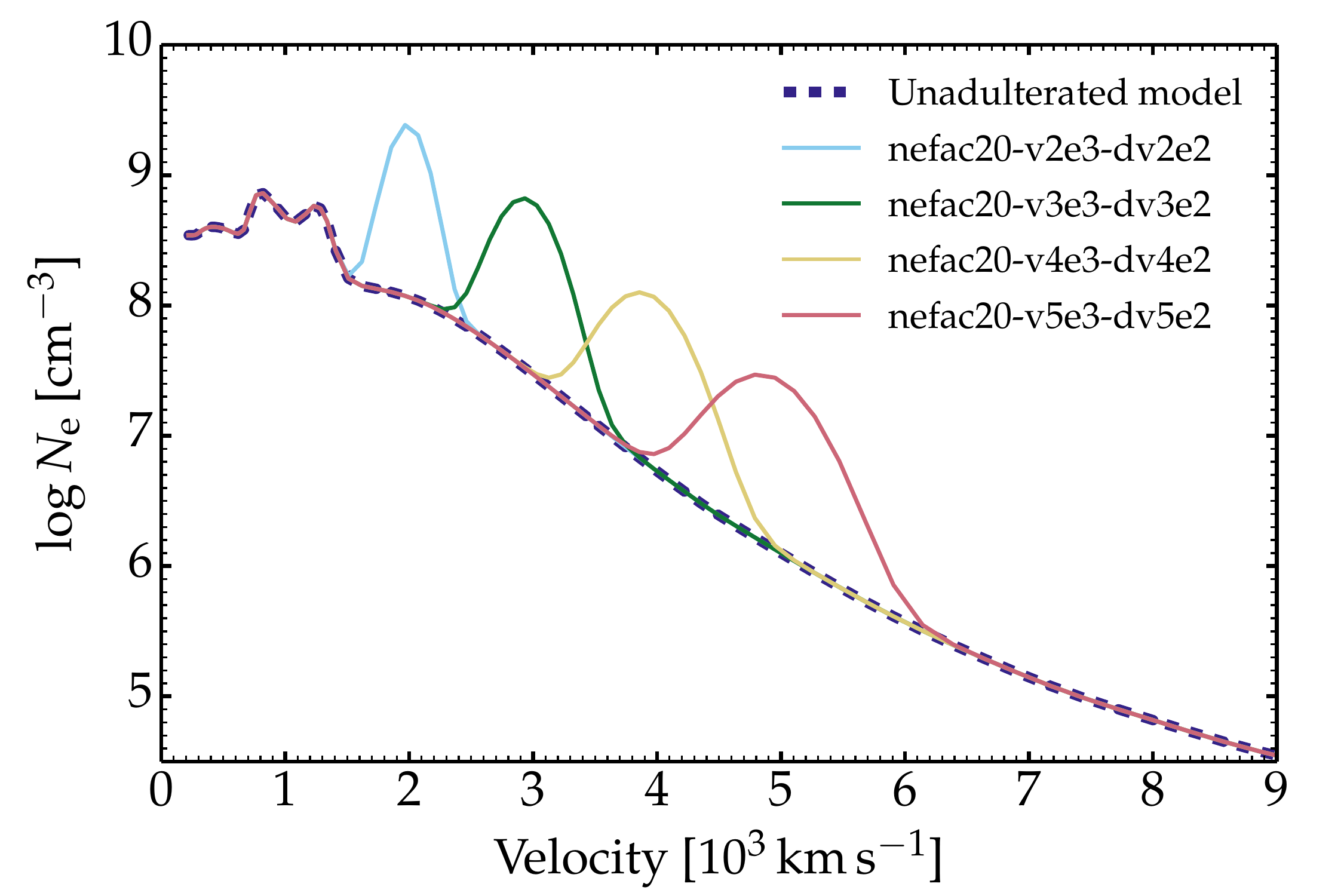, width=6.2cm}
\epsfig{file=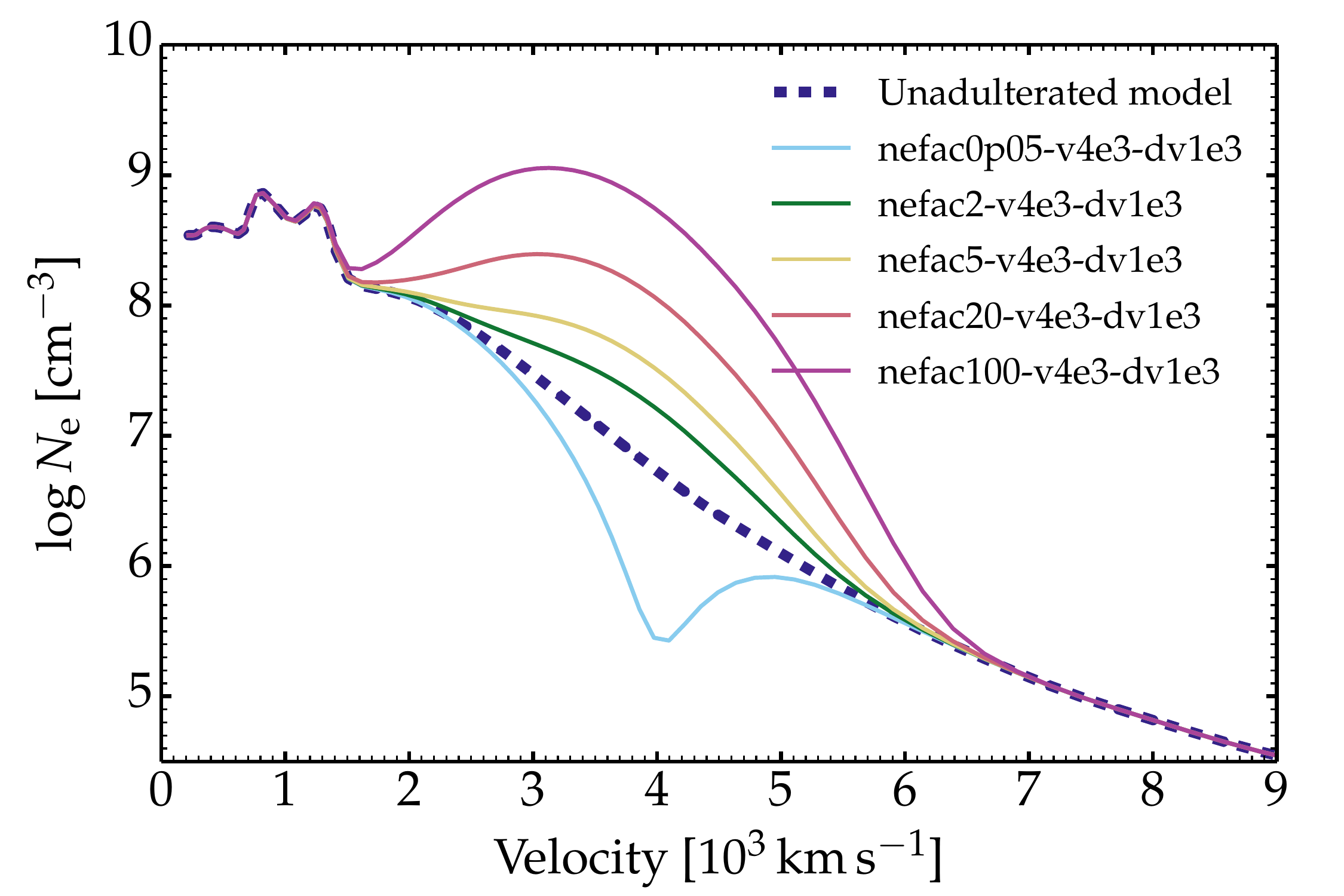, width=6.2cm}
\epsfig{file=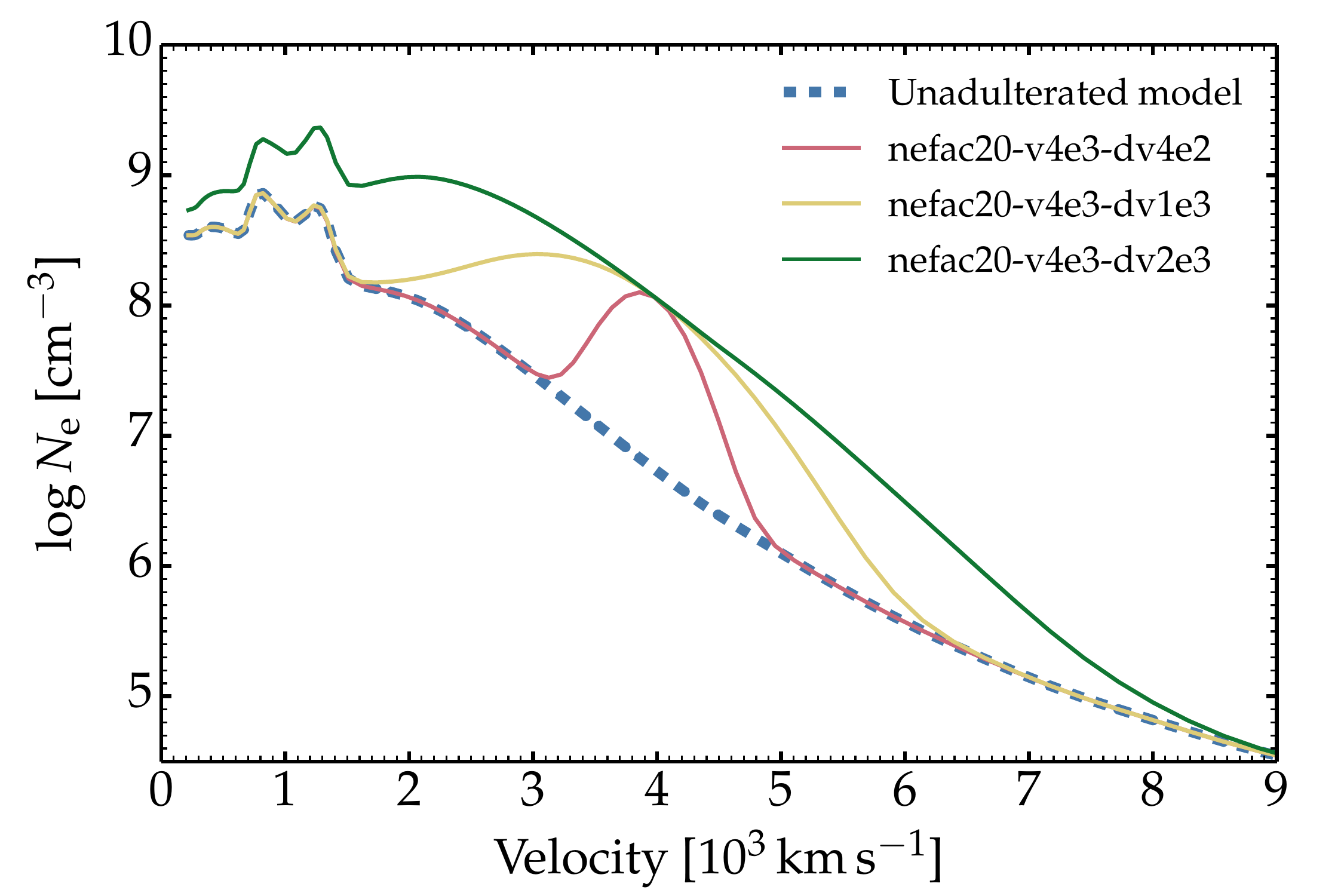, width=6.2cm}
\epsfig{file=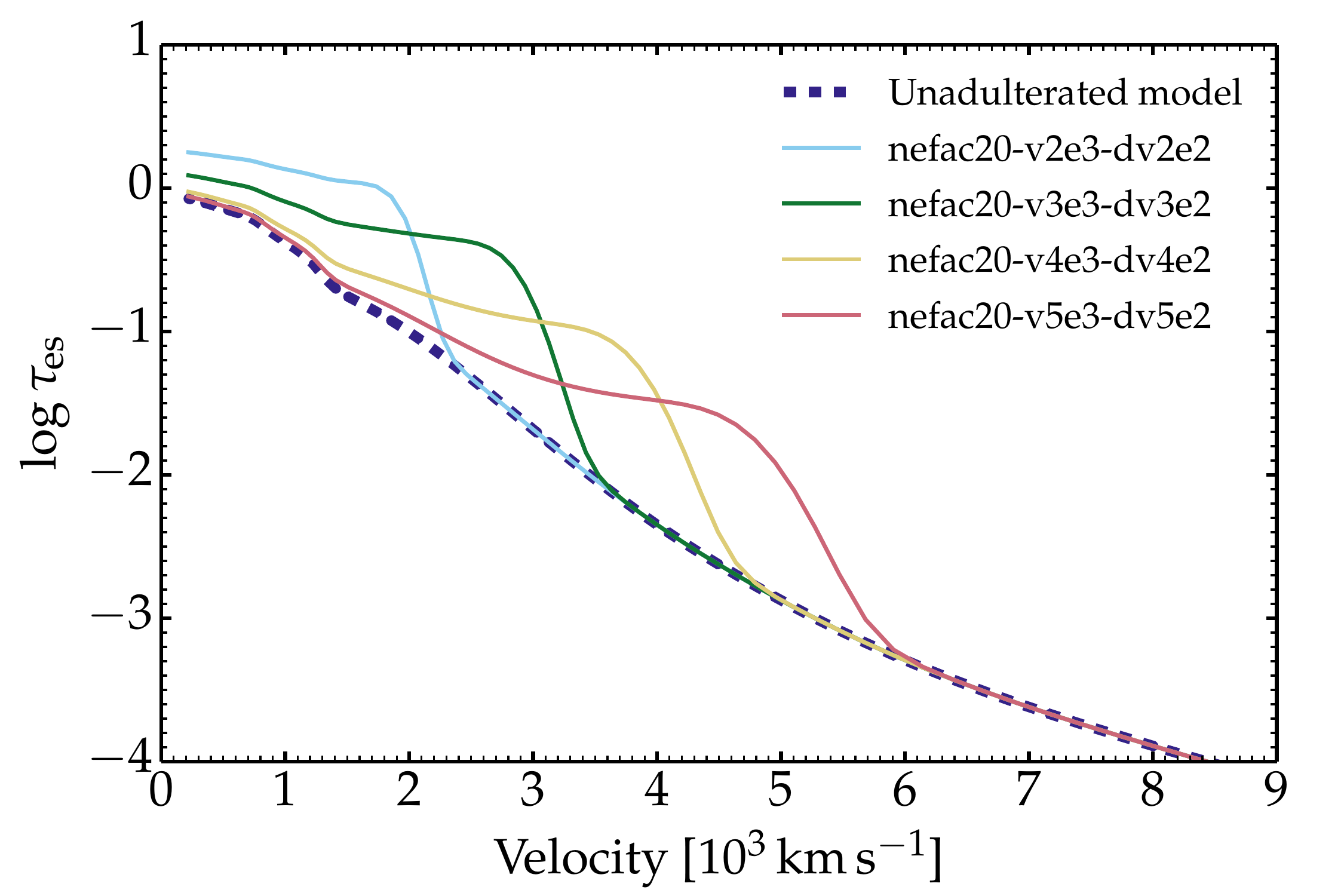, width=6.2cm}
\epsfig{file=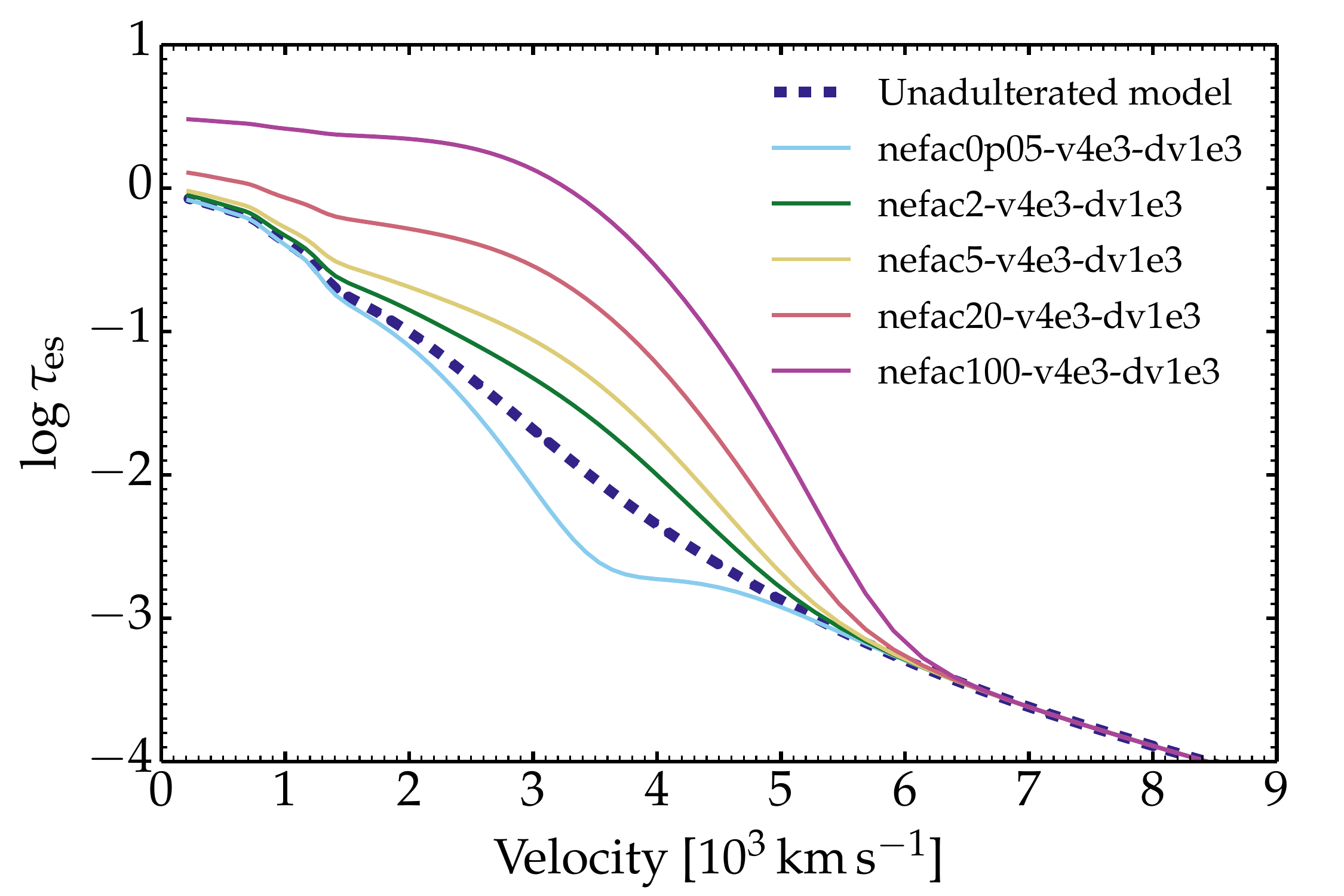, width=6.2cm}
\epsfig{file=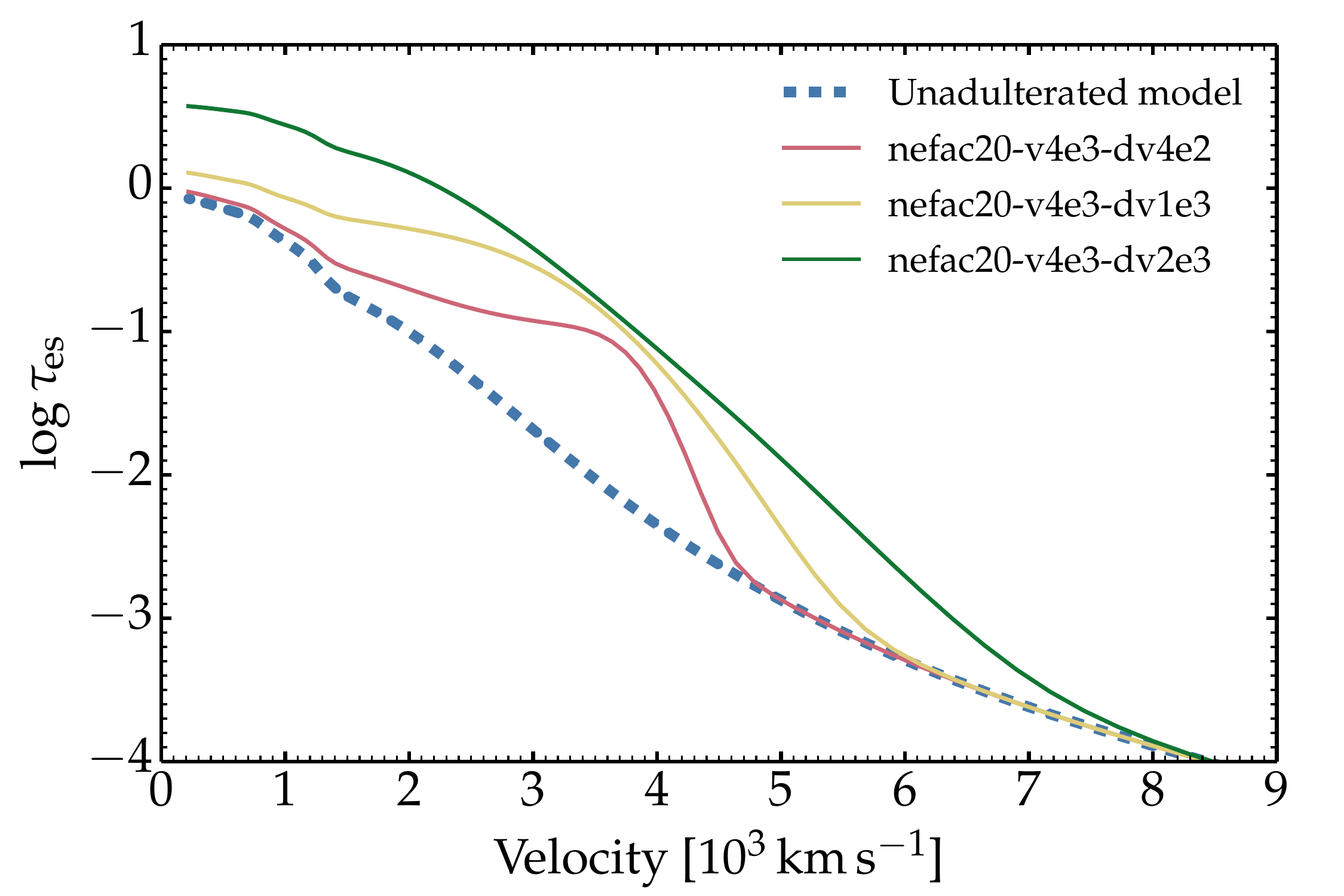, width=6.2cm}
\caption{Illustration of the adjustments to the electron density in the adulterated model in order to mimic the influence of a \nifs\ blob -- these configurations would be shells in a 1D context but correspond to a blob in our 2D hybrid model setup. The blob radial (electron-scattering) optical depth is equal to the difference in radial optical depth (bottom row panels) along a direction intersecting the blob and one that does not (see values in rightmost column of Table~\ref{tab_set}). The total radial electron scattering optical depth of the unadulterated model is 0.84. Each column represents one set of models that differ in the value of $V_{\rm blob}$ (left column; fixed $\delta V_{\rm blob}/V_{\rm blob}$ and $N_{\rm e,fac}$), of $N_{\rm e,fac}$ (middle column; fixed $V_{\rm blob}$  and $\delta V_{\rm blob}$), and of $\delta V_{\rm blob}$ (right column; fixed $V_{\rm blob}$  and $N_{\rm e,fac}$). Note that, in the middle panel, model nefac0p05-v4e3-dv1e3 represents a \nifs\ ``hole", where  $N_{\rm e}$ is reduced relative to the unadulterated model (this model is discussed in the appendix); see text for details.
\label{fig_blob_properties}
}
\end{figure*}

\begin{figure}
\epsfig{file=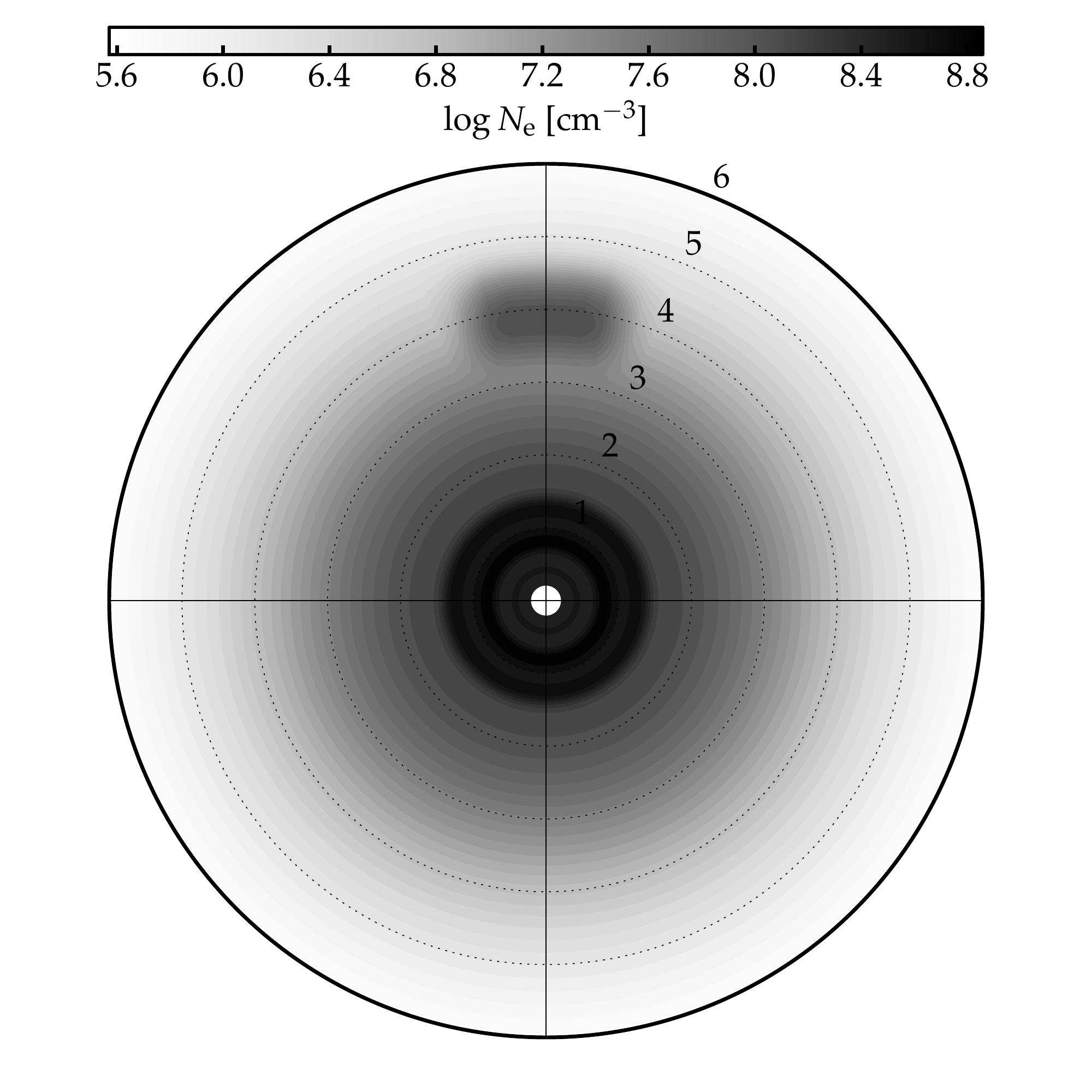, width=9.2cm}
\caption{Two-dimensional distribution of the free-electron density $N_{\rm e}$ for a Type II SN model at 200\,d characterized by a blob along the symmetry axis (here lying along the vertical direction; the inclination is 90\,deg) with characteristics corresponding to model nefac20-v4e3-dv4e2 (see Table.~\ref{tab_set}) and an opening angle of 20\,deg. The number given along the one o'clock direction corresponds to the velocity in units of 1000\,\kms.
\label{fig_2d_ed}
}
\end{figure}

\begin{table}
\caption{List of blob configurations adopted together with the corresponding radial electron-scattering optical depth of the ``blob". The main blob characteristics are the associated free-electron density enhancement $N_{\rm e,fac}$, the blob central velocity \vblob, and  the characteristic velocity extent of the blob \dvblob\ (see section~\ref{sect_mod} and Eq.~\ref{eq_nefac} for details). These are combined with the unadulterated model x1p5 to form a 2D hybrid model characterized by a chosen ``blob" half-opening angle $\beta$ (see section~\ref{sect_mod} for details).
\label{tab_set}
}
\begin{tabular}{lcccc}
\hline
Model                 &  $N_{\rm e,fac}$ & $V_{\rm blob}$ & $\delta V_{\rm blob}$ & $\tau_{\rm blob}$  \\
\hline
   nefac20-v2e3-dv2e2 & 20.0   & 2e3 & 2e2 & 0.933       \\
   nefac20-v3e3-dv3e2 & 20.0   & 3e3 & 3e2 & 0.383       \\
   nefac20-v4e3-dv4e2 & 20.0   & 4e3 & 4e2 & 0.098       \\
   nefac20-v5e3-dv5e2 & 20.0   & 5e3 & 5e2 & 0.029       \\
\hline
   nefac20-v4e3-dv4e2 & 20.0   & 4e3 & 4e2 & 0.098       \\
   nefac20-v4e3-dv1e3 & 20.0   & 4e3 & 1e3 & 0.436       \\
   nefac20-v4e3-dv2e3 & 20.0   & 4e3 & 2e3 & 2.934       \\
\hline
    nefac2-v4e3-dv1e3 & 2      & 4e3 & 1e3 & 0.044       \\
    nefac5-v4e3-dv1e3 & 5      & 4e3 & 1e3 & 0.109       \\
   nefac20-v4e3-dv1e3 & 20     & 4e3 & 1e3 & 0.436       \\
  nefac100-v4e3-dv1e3 & 100    & 4e3 & 1e3 & 2.178       \\
\hline
\end{tabular}
\end{table}

\section{Numerical setup of blob simulations}
\label{sect_mod}

The ansatz of our present study, motivated by the results of 1D non-LTE radiative transfer calculations and 3D explosion simulations of core-collapse SNe (see Section~\ref{sect_motivation} and Fig.~\ref{fig_cmfgen}, as well as \citealt{D20_12aw_pol}), is that the main effect of such a \nifs\ blob is to boost the electron density. This boost occurs over a spatial extent that depends on the blob properties (e.g., size and location). We assume, for simplicity, that other sources of opacity and emissivity (lines, bound-free processes) are not affected (for example, the change in composition due to a \nifs-rich blob would be a modest rise in iron mass fraction at late times). In practice, the line emission scales as the density squared, but the electron scattering optical depth is linear in density. In this sense we are not confining the emission to the inner region -- this confinement will occur naturally. A small blob, even with an enhanced density, won't contribute much to the emission. One can thus investigate the influence of a \nifs\ blob by directly varying the electron density.

We thus proceed by taking a \cmfgen\ model for a Type II SN (model x1p5 at 200\,d from the study of \citealt{HD19}) and apply a variety of adjustments to the electron density profile in order to mimic the presence of a \nifs\ blob.\footnote{Some chemical mixing is applied as in \citet{HD19} using the standard technique. A superior technique is to shuffle shells of distinct composition in the unmixed ejecta, in order to generate a macroscopic mixing with no microscopic mixing, as discussed in \citet{DH20_shuffle}.} With this flexible approach, a wide range of blob properties can be investigated and their polarized signatures analyzed. It is important to realize that the blob represents the asymmetric part in the whole \nifs\ distribution and the SN ejecta. It can be thought of as a protrusion of \nifs\ at larger velocity or a local enhancement in \nifs\ fraction within an already \nifs-rich region. It can be combined with a spherical distribution of \nifs\ up to large velocity, which by itself would produce no residual polarization (but may impact the light curve or the line profiles by modulating the local heating rate and the magnitude of non-thermal effects).

\begin{figure*}
\epsfig{file=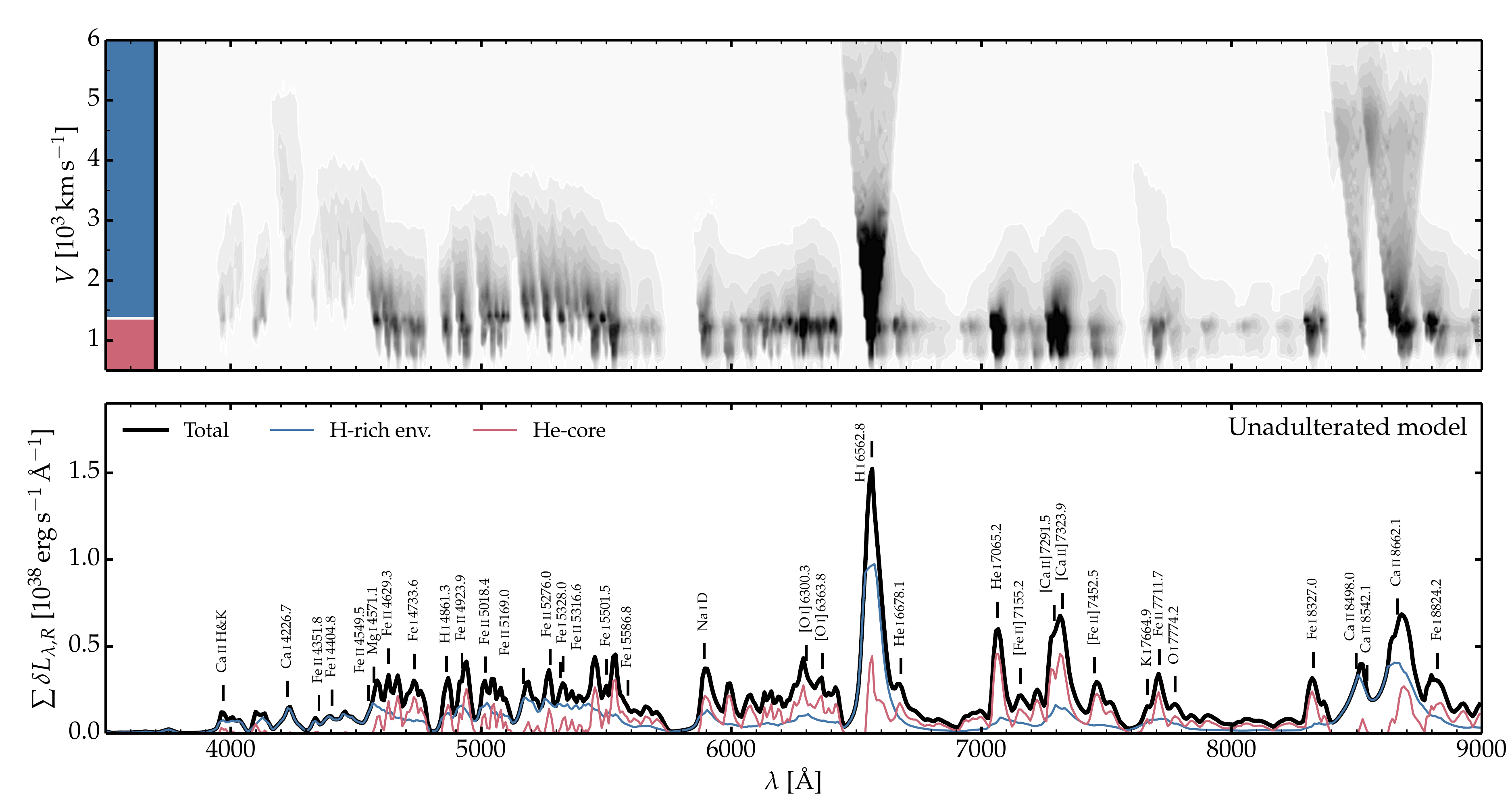, width=18cm}
\caption{Illustration of the origin of the flux in the unadulterated model. The top panel shows the observer's frame luminosity contribution $\delta L_{\lambda,R}$ (the grayscale is saturated at a third of the true maximum in order to reveal better the weaker emitting regions) versus wavelength and ejecta velocity. The two contributing shells shown at left correspond to the H-rich envelope and to the He core in the progenitor. The bottom panel shows the total luminosity as well as the fractional luminosity from each shell (some line identifications are also provided).
\label{fig_dfr}
}
\end{figure*}

We first define the 2D geometry of our hybrid model using the same approach as in \citet{D20_12aw_pol}. The angle $\beta$ is the half-opening angle of the blob.  The 2D hybrid model is composed of the properties of the unadulterated model x1p5 along all polar angles between $\beta$ and $\pi$ (or up to $\pi/2$ if we adopt mirror symmetry with respect to the equatorial plane). Since this model is taken as the reference, we refer to its electron density at $V$ as $N_{\rm e, ref}(V)$. Along the axis of asymmetry and up to $\beta$, we scale the electron density such that
\begin{equation}
N_{\rm e}(V) = N_{\rm e, ref}(V) \left( 1 + N_{\rm e, fac} \exp(-X^2) \right)  \,\, , \label{eq_nefac}
\end{equation}
where $X = (V-V_{\rm blob}) / \delta V_{\rm blob}$.  $N_{\rm e, fac}$ represents the factor by which the electron density at a particular location is enhanced relative to the spherical, unadulterated, model. This fudge conserves mass since it merely modulates the electron density. For convenience, all three parameters $N_{\rm e, fac}$, $V_{\rm blob}$, and $\delta V_{\rm blob}$  can be independently varied. With this expression, we aim to cover in a heuristic way the influence of a \nifs\ blob or an enhanced \nifs\ mixing on the ejecta ionization, which is very low in the case of weak or moderate mixing (Fig.~\ref{fig_cmfgen}; this may also occur because of a very low \nifs\ mass). When scaling the electron density, we are effectively adjusting the contribution from electron scattering to the total opacity. Other opacity and emissivity contributions remain spherically symmetric. This is not strictly correct since the adjusted electron density, and the associated change in ionization, should also affect the contributions from atoms and ions in the gas.  We thus do not investigate here the possible contribution to the line flux that could result from enhanced ionization.\footnote{A consistent modeling of the influence of a localized \nifs\ enhancement on the SN radiation and gas properties is presented by \citet{D20_12aw_pol}.} As far as polarization is concerned, only the adjustment to the electron density matters. A corollary is that the polarization we obtain in our simulations arises from a spherical emitting source (a spherical core in SN jargon) -- the only asymmetry is in the distribution of scatterers, which are primarily at large velocities of a few 1000\,\kms.

The default grid uses 19 polar angles equally spaced between zero and $\pi$ (or 10 polar angles equally spaced between zero and $\pi/2$ if two blobs are assumed, i.e., if we adopt mirror symmetry). Along polar angles up to and including $\beta$ we assign the model with the high-velocity \nifs, while beyond $\beta$ we assign the model without the high-velocity \nifs. At the junction, we interpolate between the two, so that effectively, the high-velocity \nifs-rich blob opening angle is a little greater than $\beta$ by about 5\,deg since we use the same angular resolution of 10\,deg in all simulations.

Our set of simulations includes variations in $N_{\rm e, fac}$ of 2, 5, 20, and 100 (we also include the case of a negative value corresponding to a \nifs\ ``hole", i.e., an ionization deficit; see Section~\ref{sect_hole}), in half-opening angle $\beta$ from 10 to 40\,deg, in blob velocity $V_{\rm blob}$ from 2000 to 5000\,\kms, and in blob velocity width from 200 to 2000\,\kms. Table~\ref{tab_set} summarizes the various blob configurations treated as well as the associated radial electron-scattering optical depth of the blob. Since the interface between the metal-rich inner ejecta and the H-rich outer ejecta is located at $\sim$\,1400\,\kms\ (Fig.~\ref{fig_dfr}), these enhancements are mostly located in the H-rich layers of the SN. Not all permutations are considered. Instead, we explore from a default model characterized by $N_{\rm e, fac}=$\,20, $\beta=$\,10\,deg, $V_{\rm blob}=$\,4000\,\kms\ (with two values of $\delta V_{\rm blob}$). We also do not consider smaller half-opening angles for the blob.  First, this would require a higher angular resolution (hence numerically more costly) while not changing the qualitative results. Quantitatively, a smaller blob opening angle will yield a smaller polarization that may be unobservable ($P$ should scale with $\sin^2\beta \approx \beta^2$ for small $\beta$; see Eq.~\ref{eq_pol_BM77}).

Figure~\ref{fig_blob_properties}  illustrates some properties of the various blob configurations adopted (i.e., the properties of the 2D model along the pole, together with the unadulterated model for comparison). Since the mass density is unchanged in our toy model, the mean number of electron per nucleon is proportional to the free-electron density. Hence, the $\gamma_e$ scales with $N_{\rm e, fac}$ in the models with modified free-electron density. As shown in the bottom panel of Fig.~\ref{fig_cmfgen} for the unadulterated model, all regions beyond about 2500\,\kms\ have $\gamma_e \lesssim 0.01$. Hence, the electron density in this region would increase by a factor $\gtrsim$\,100 if it were fully ionized.
 So, in Fig.~\ref{fig_blob_properties}, a model with $N_{\rm e, fac}$ of about 10 still corresponds to a partial ionization of H and He. For $N_{\rm e, fac}$ of 100, H would be ionized and He would be once ionized. The former is within the predictions of the 1D \cmfgen\ model with a high-velocity \nifs-rich shell presented in Section~\ref{sect_cmfgen}. The latter might be difficult to achieve with a \nifs\ blob alone. In that case, a concomitant rise in mass density (as for example produced in a more energetic explosion) would be needed to bring the electron density to the corresponding level.  Figure~\ref{fig_2d_ed} illustrates what the 2D distribution of the electron density might resemble in an ejecta with a high velocity \nifs\ blob. It shows a meridional cut through a 2D model with a blob located at 4000\,\kms.

Figure~\ref{fig_dfr} provides some important information on the regions that contribute to the total emergent flux from the SN ejecta in the unadulterated model. When a ``blob" is introduced, its impact on the ejecta radiation depends strongly on its location relative to the spectrum formation region. To scatter the full spectrum in this model, a blob would need to be located at a velocity greater than about 6000\,\kms. A slower moving blob would overlap with the emitting region of strong lines like H$\alpha$ and the Ca\two\ NIR triplet. In the following sections, it will be useful to go back to Fig.~\ref{fig_dfr} to compare the blob spatial properties with the spectrum formation regions.

\begin{figure}
\vspace{-0.3cm}
\epsfig{file=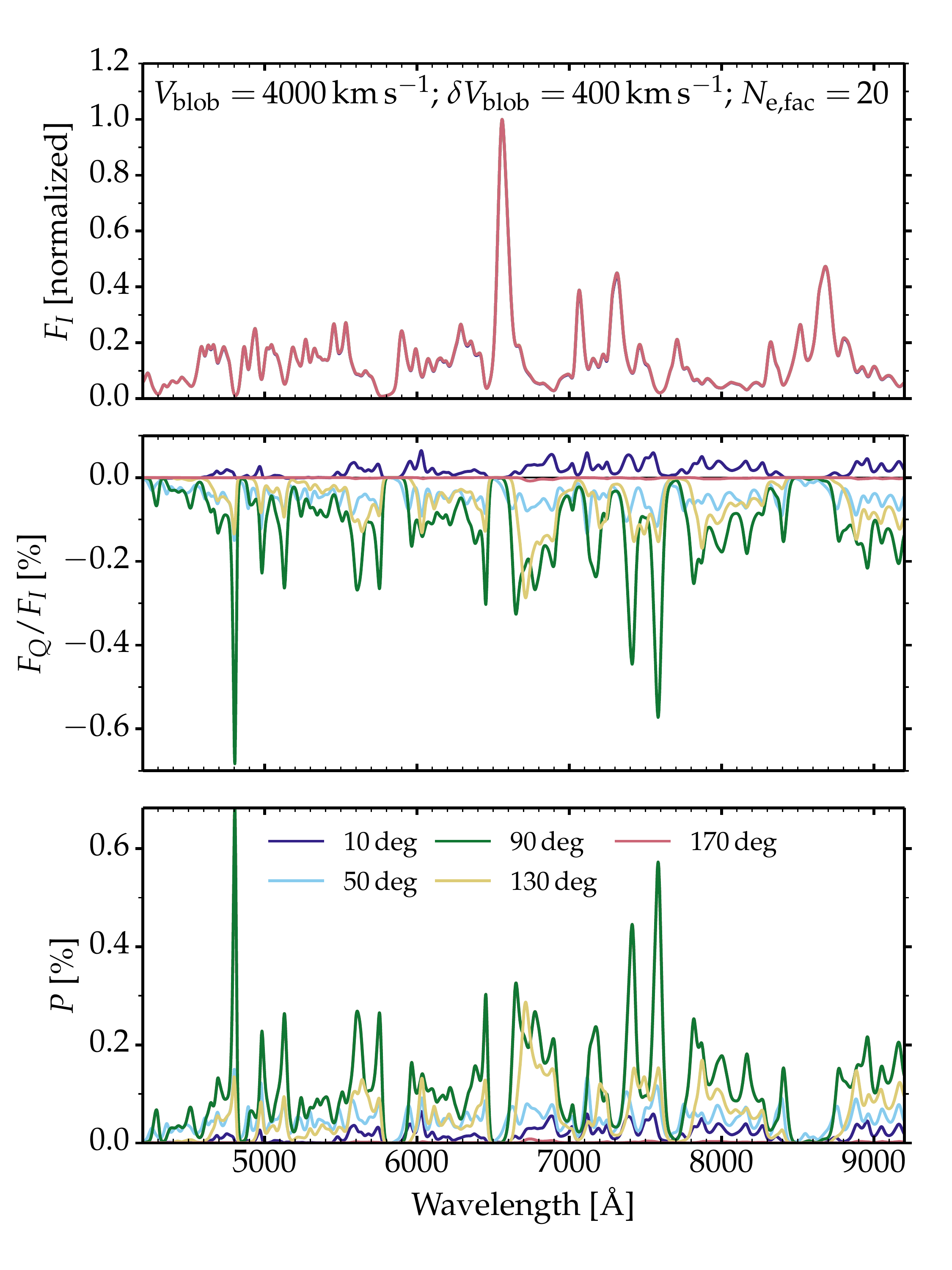, width=9.2cm}
\caption{Illustration of the normalized total flux $F_I$ (top), the percentage $F_Q/F_I$, and the polarization $P$ for inclinations of 10, 50, 90, 130, and 170\,deg. The electron-density enhancement is characterized by $V_{\rm blob}$=\,4000\,\kms, $\delta V_{\rm blob}=$\,400\,\kms, and $N_{\rm e,fac}$=\,20; the blob radial electron-scattering optical depth is 0.098 and the blob opening angle is 20\,deg. To reduce the strength of the narrow spikes in the polarization and to mimic the approximate resolution of typical spectropolarimetry data, the model has been smoothed with a gaussian kernel (FWHM of 23.5\,\AA).
\label{fig_ex_model_fi_fq_p_nbeta}
}
\end{figure}

\begin{figure}
\epsfig{file=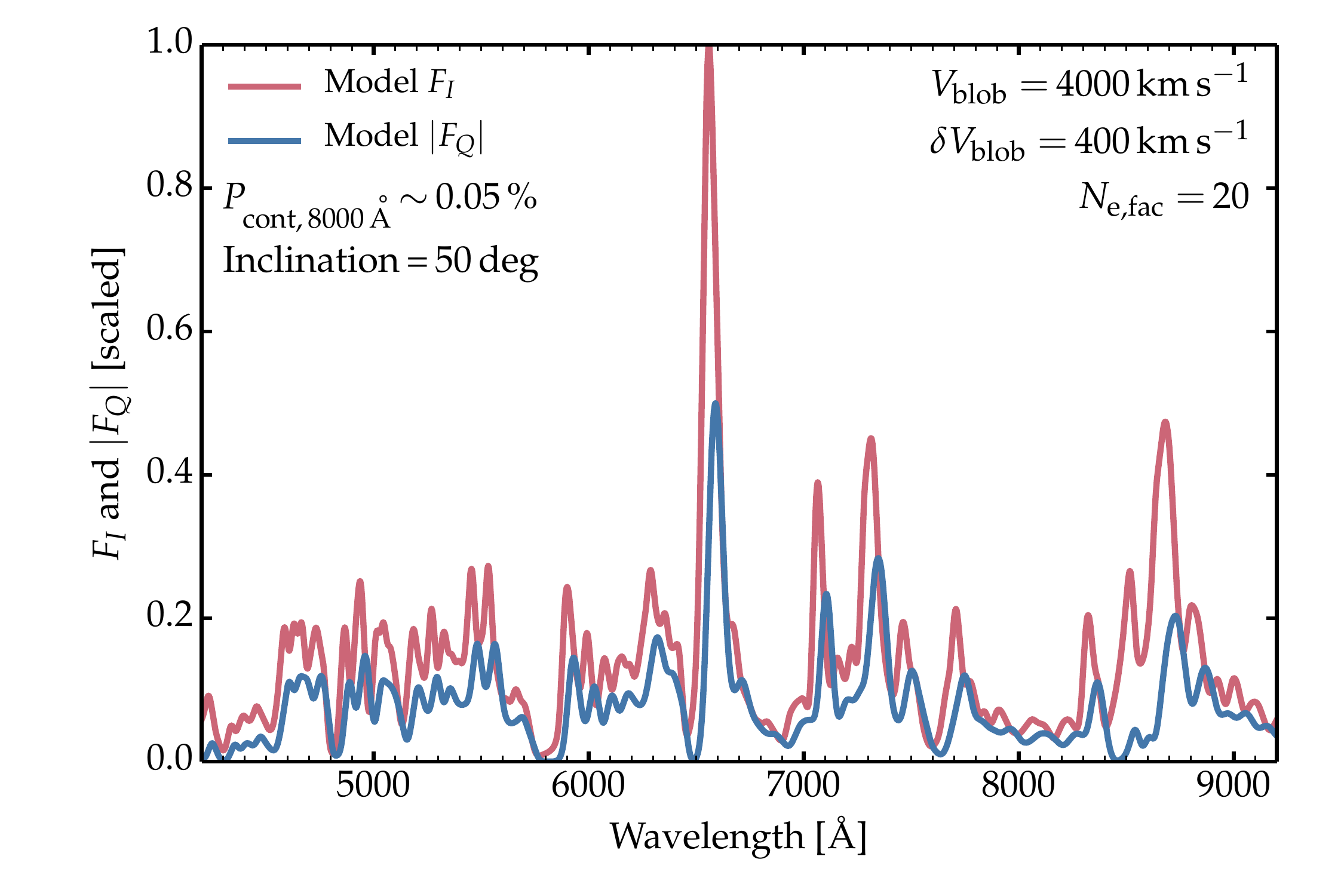, width=9.2cm}
\epsfig{file=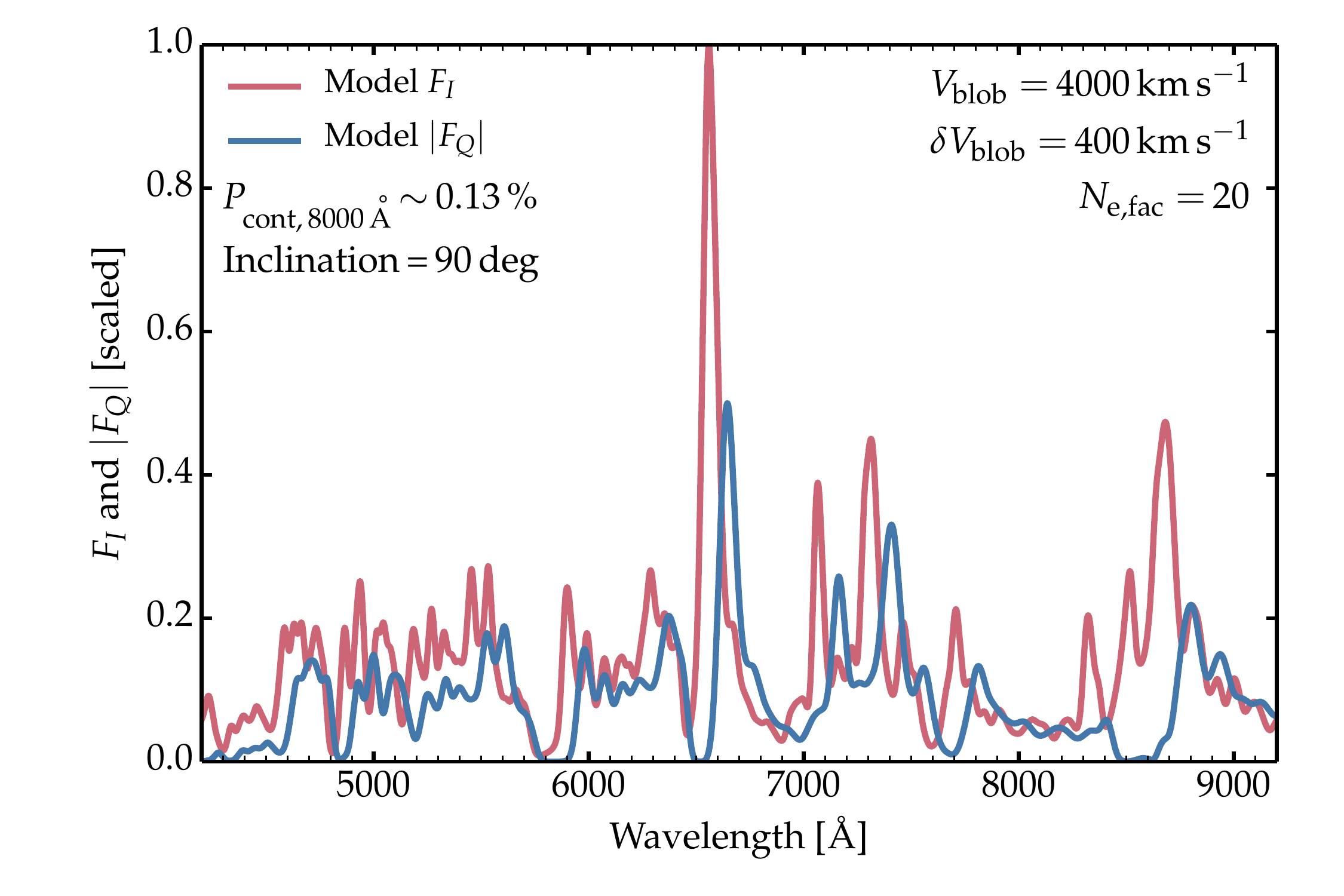, width=9.2cm}
\epsfig{file=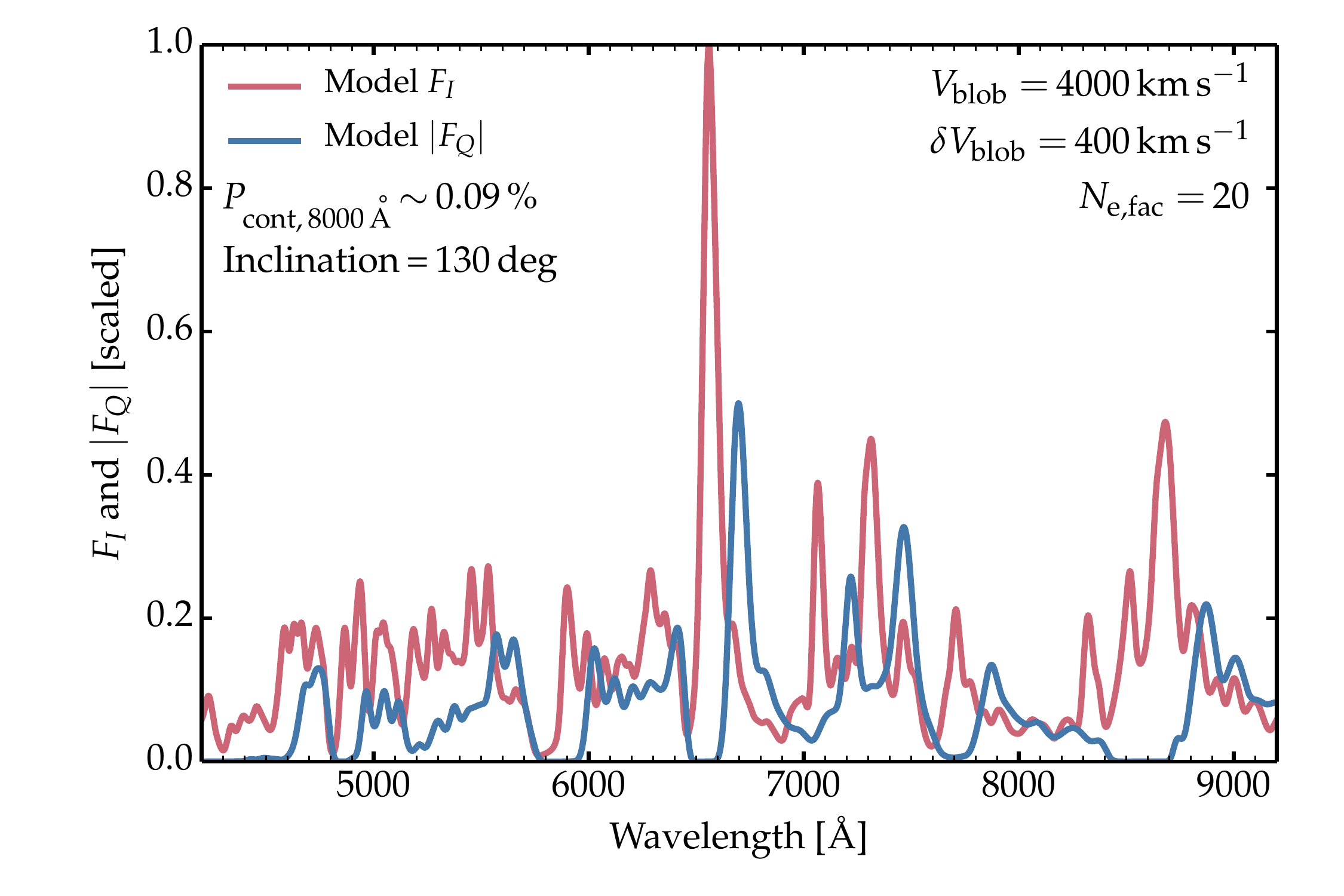, width=9.2cm}
\caption{Same model as in Fig.~\ref{fig_ex_model_fi_fq_p_nbeta}, but now showing the scaled total flux $F_I$ (the scaling places the max value at one) and polarized flux $|F_Q|$ (i.e., its absolute value, with a scaling such that it peaks at a value of 0.5) for an inclination of 50, 90, and 130\,deg to the symmetry axis. The polarized flux corresponds closely to the total flux with an offset in magnitude (by a factor $1/P$, hence about 10$^2$ to 10$^3$ smaller -- see Fig.~\ref{fig_ex_model_fi_fq_p_nbeta} for the exact offset)  and a redshift whose magnitude increases with inclination. The clean, distinct redshift seen in the scattered flux, corresponding to $V_{\rm blob} (1 - \cos \alpha_{\rm los})$, is a fundamental prediction of this work for a high-velocity scatterer (see also Appendix~\ref{sect_dop}).  To aid in the direct comparison between features seen in the polarized (i.e., scattered) flux compared with the total flux, in several future figures we shall artificially apply the appropriate blueshift to the predicted polarized flux spectra, so that it overlaps the total flux spectrum.
\label{fig_ex_model_fi_fq_p}
}
\end{figure}

 In all simulations, the SN age is 200\,d. This choice was motivated by some yet unpublished nebular-phase spectro-polarimetric observations (Leonard et al., in prep.) but the results described here would apply generically to any Type II SN at nebular times.

A detailed presentation of the polarized radiation transfer technique and conventions is provided in \citet{hillier_94,hillier_96}, \citet{DH11_pol}, and \citet{D20_12aw_pol}. Because of the axial symmetry, and following our angle convention, all the model polarization is contained in the Stokes flux $F_Q$, while the Stokes flux $F_U$ is zero. The polarization angle is therefore identically zero.

\section{Detailed description for one configuration}
\label{sect_ref}

In this section, we explore the properties for a representative model. The blob characteristics are $N_{\rm e, fac}=$\,20, $\beta=$\,10\,deg, $V_{\rm blob}=$\,4000\,\kms\ and $\delta V_{\rm blob}=$\,400\,\kms. The blob optical depth is 0.098 (hence optically thin), which is about ten times smaller than the total radial electron-scattering optical depth of the ejecta (which is 0.84; see also the bottom row panels in Fig.~\ref{fig_blob_properties}).

\begin{figure*}
\begin{center}
\epsfig{file=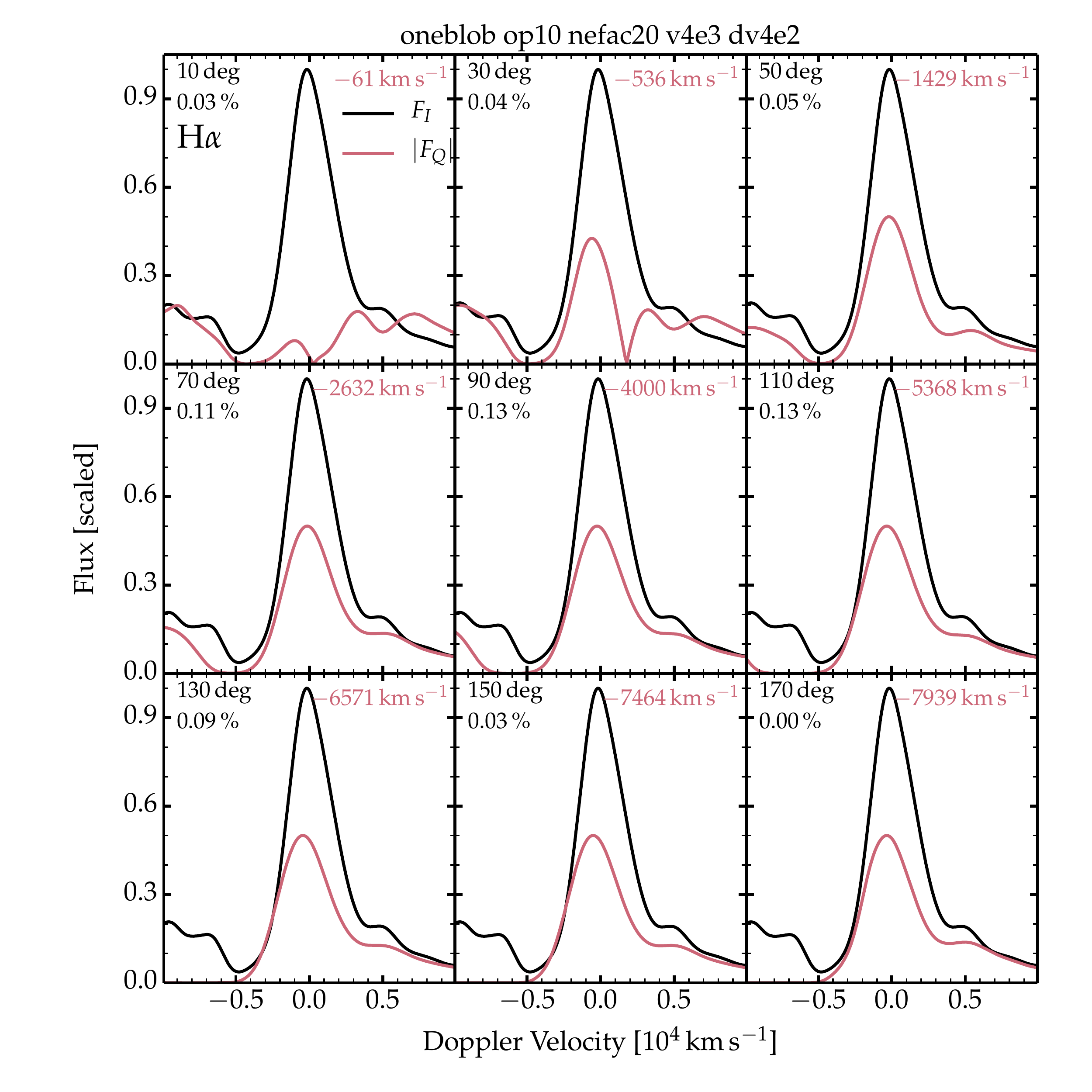, width=15cm}
\caption{Illustration for the same model as in Figs.~\ref{fig_ex_model_fi_fq_p_nbeta}\,$-$\,\ref{fig_ex_model_fi_fq_p}, but now showing the evolution of the scaled total flux $F_I$ (black) and the polarized flux $|F_Q|$ (red; scaled and blue shifted) versus Doppler velocity (zero velocity corresponds to the rest wavelength of H$\alpha$). Nine different inclinations, increasing from left to right and from top to bottom, are shown. The blueshift was applied to the polarized spectra to compensate for the redshift induced by scattering in a  blob moving radially at 4000\,\kms, and its numerical value is indicated in the top right corner of each panel. It corresponds to $-4000\,(1 - \cos \alpha_{\rm los})$\,\kms, where  the inclination angle $\alpha_{\rm los}$ is indicated in the top left corner of each panel. The percentage value in the upper left corner of each panel corresponds to the polarization in the relatively line free region around 8000\,\AA.
\label{fig_ex_model_fi_fq_halpha}
}
\end{center}
\end{figure*}

Figure~\ref{fig_ex_model_fi_fq_p_nbeta} shows the total flux $F_I$ (top), the normalized flux $F_Q/F_I$ (middle) and the polarization $P$ for five inclinations between 10 and 170\,deg. As the inclination is varied, the blob influence occurs at a different Doppler shift. In our simulations, the blob influence on the total flux is negligible (i.e., the total flux appears nearly identical for all inclinations),  which arises from the fact that in our approach  only the electron scattering emissivity is aspherical (and computed explicitly).\footnote{This property depends on the blob properties. If the blob optical depth is about one and the sight line strikes through the blob the observer would see a flux change. For other viewing angles, the observed flux does not change much due to a combination of low optical depth and the small solid angle of the blob. Some changes would be expected if the blob optical depth or its angular extent were increased.} Compared to photospheric-phase conditions, the spectrum does not exhibit true continuum regions free of line contamination. Instead, we see the dominance of lines, which appear strongly in emission in the total flux as well as through hills, spikes, and valleys in the polarized spectrum. Numerous regions show a strong deficit in both total and polarized flux, such as the central parts of the absorption troughs of Na\one\,D, H$\alpha$, O\,\one\,7774\,\AA\ or the Ca\two\, NIR triplet (this property is reminiscent of what happens during the photospheric phase). In regions around 6900 and 8000\,\AA, we find (not shown) that the continuum flux is about a third of the total model flux but lines can still be identified there through the variations in $F_Q$. The polarization flips in sign as the inclination is increased. For a small inclination, the polarization is positive everywhere, thus aligned with the axis of symmetry, which is the opposite of what should happen in the optically thin, point source, limit, whereby the polarization should be perpendicular to the axis of symmetry and thus negative. The latter holds for inclinations larger than 20\,deg (see also discussion in Appendix~\ref{appendix_BM77}).

Apart from the sign reversal for small inclinations, there is a trend of increasing polarization (throughout the optical range except in the core of the troughs of strong lines) for inclinations closer to 90\,deg. When considering $P$, the polarization exhibits spikes at the blue edge of the trough in the strongest P-Cygni profiles (these spikes appear even sharper without the smoothing applied in Fig.~\ref{fig_ex_model_fi_fq_p_nbeta}).

In the literature, more attention is generally paid to the quantity $P$ and in particular its value in line-free regions. However, at nebular times the continuum flux and the continuum optical depth are small and the bulk of the flux emerges in lines so studying the continuum polarization at that time is somewhat difficult and complicated by the poor signal. The bulk of the flux being in lines, it is much easier to study the polarized radiation associated with line photons. These photons are initially emitted with no preferred polarization but they may scatter with free electrons (probably once at such a late time) and therefore they can yield a net polarization. So, instead of inspecting $P$, one can extract information from the polarized flux $F_Q$.

Figure~\ref{fig_ex_model_fi_fq_p} compares the morphology of the total flux $F_I$ and of the polarized flux $|F_Q|$ for three different inclinations of 50, 90, and 130\,deg. The scaling applied to each flux in each panel is chosen to improve visibility (the absolute offset between $F_I$ and $F_Q$ as well as the sign of $F_Q$ can be inferred from Fig.~\ref{fig_ex_model_fi_fq_p_nbeta}). The striking property revealed by Fig.~\ref{fig_ex_model_fi_fq_p} is that the polarized flux appears as a replica of the total flux with only a shift to longer wavelengths. Physically, the blob at 4000\,\kms\ scatters the incoming radiation from the inner ejecta and in the process imparts a systematic redshift equal to  $V_{\rm blob} (1 - \cos \alpha_{\rm los})$ (see also Section~\ref{sect_mod}). The clean, distinct redshift seen in the scattered flux is a fundamental prediction of this work for a high-velocity scatterer  (see also Appendix~\ref{sect_dop}). This redshift is best revealed by inspecting the total and polarized fluxes in the spectral region around a strong emission line.

\begin{figure*}
\begin{center}
\epsfig{file=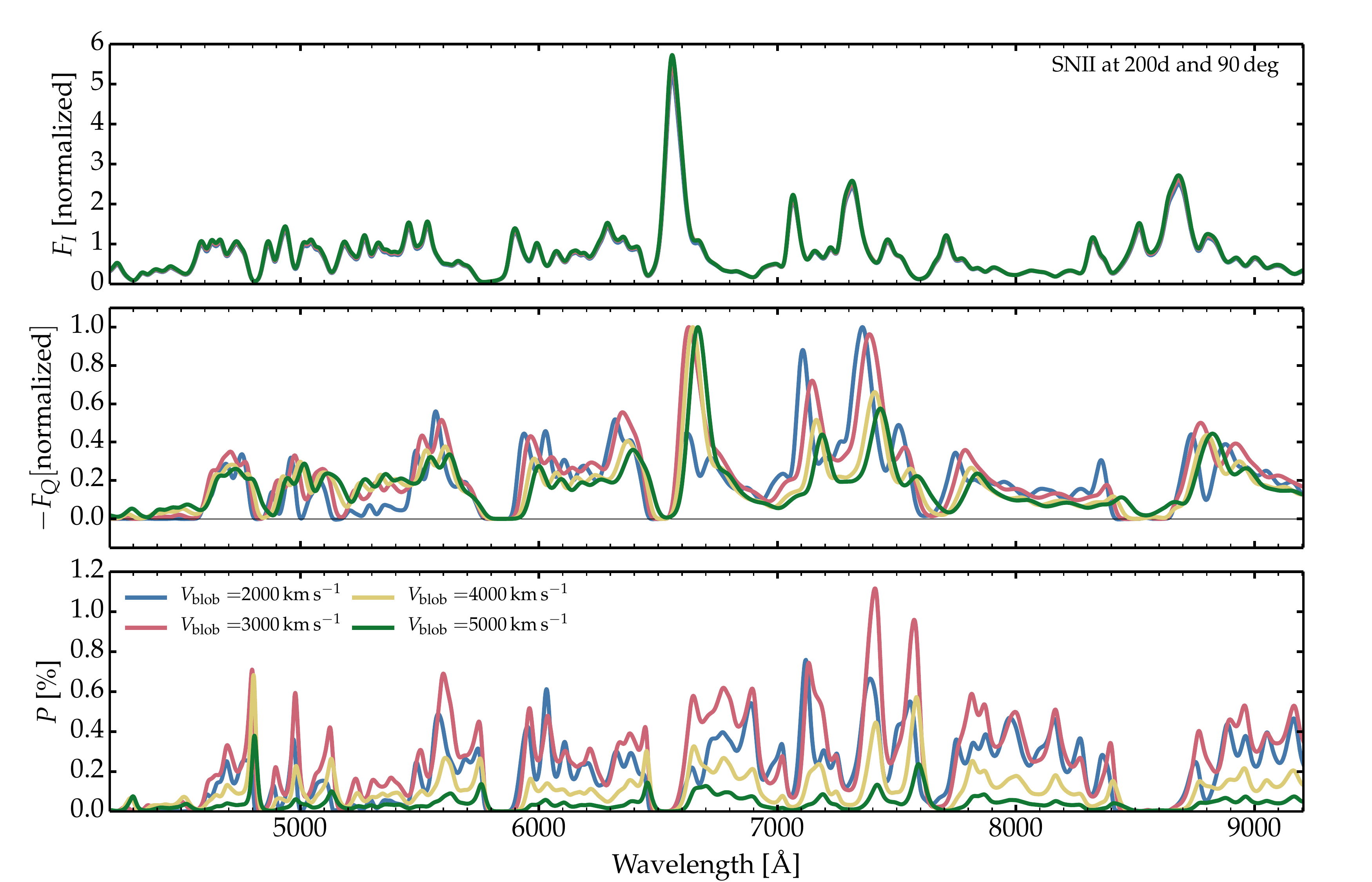, width=19cm}
\end{center}
\caption{Comparison of the normalized total flux $F_I$ (top; in this and in all similar figures the normalization is done at 7100\,\AA), the normalized polarized flux $F_Q$ (we show the quantity $-F_Q$/{\rm max}($|F_Q|$) to facilitate the comparison with $F_I$), and the polarization $P$ along a 90-deg inclination for the set of models differing in the value of \vblob. The radial optical depth of the blob decreases from 0.933 to 0.383, 0.098, and 0.029 as \vblob\ is increased from 2000, to 3000, 4000, and 5000\,\kms\ (the fractional blob velocity width is kept fixed). As discussed in the text, the polarization does not vary linearly with radial optical depth. Also apparent from the figure is the systematic increase in redshift of the polarized spectra with the increasing velocity of the blob. The blob opening angle is 20\,deg in all four cases. The models have been smoothed with a gaussian kernel (FWHM of 23.5\,\AA).
\label{fig_vblob}}
\end{figure*}

Figure~\ref{fig_ex_model_fi_fq_halpha} shows the total flux $F_I$ and polarized flux $|F_Q|$ in the H$\alpha$ region and in velocity space, with the latter blueshifted by $-V_{\rm blob} (1 - \cos \alpha_{\rm los})$. In practice, line photons scatter off free electrons (causing the red-wing excess seen in strong lines; see also \citealt{hillier_91}), but only the asymmetric distribution of these free electrons shows up in polarized flux. In Fig.~\ref{fig_ex_model_fi_fq_halpha}, the velocity offset of the $|F_Q|$ profile with respect to $F_I$ is a little less than adopted because the scattering of line photons is more efficient on the inner side of the blob (i.e. at lower velocity), where the electron density is greater (see Fig.~\ref{fig_blob_properties}) than on the outer side of the blob (i.e. at larger velocity) -- this effect is exacerbated when the blob velocity width is increased (section~\ref{sect_dvblob}).

In the top panel of Fig.~\ref{fig_ex_model_fi_fq_p}, the inclination is 50\,deg and the redshift is about 1400\,\kms. In that case, the replication of the total flux appears nearly exact throughout the optical, with the exception of the Ca\two\ NIR triplet region and the obvious redshift of line features. The code predicts low polarization there, probably because the lines are still too optically thick and thus preserve a strong depolarization power, even at the large velocity where the blob is located.\footnote{In addition, for multiplet lines, the photons emitted in the bluer lines can be absorbed by the longer wavelength members as a result of ejecta expansion and associated Doppler shifts.} Quantitatively, the replication is not perfect  since the flux ratios seen in $F_Q$ do not match those seen in $F_I$. One can see a large discrepancy for H$\alpha$ and O\one\,6300\,\AA. One explanation for this is that these lines form in very different regions of the ejecta, with O\one\,6300\,\AA\ forming deep in the ejecta (and thus being well reflected by the distant blob) whereas H$\alpha$ forms over a large volume overlapping in part with the blob location -- the blob is not exterior to the emitting region for a significant fraction of H$\alpha$ photons.

\begin{figure}
\epsfig{file=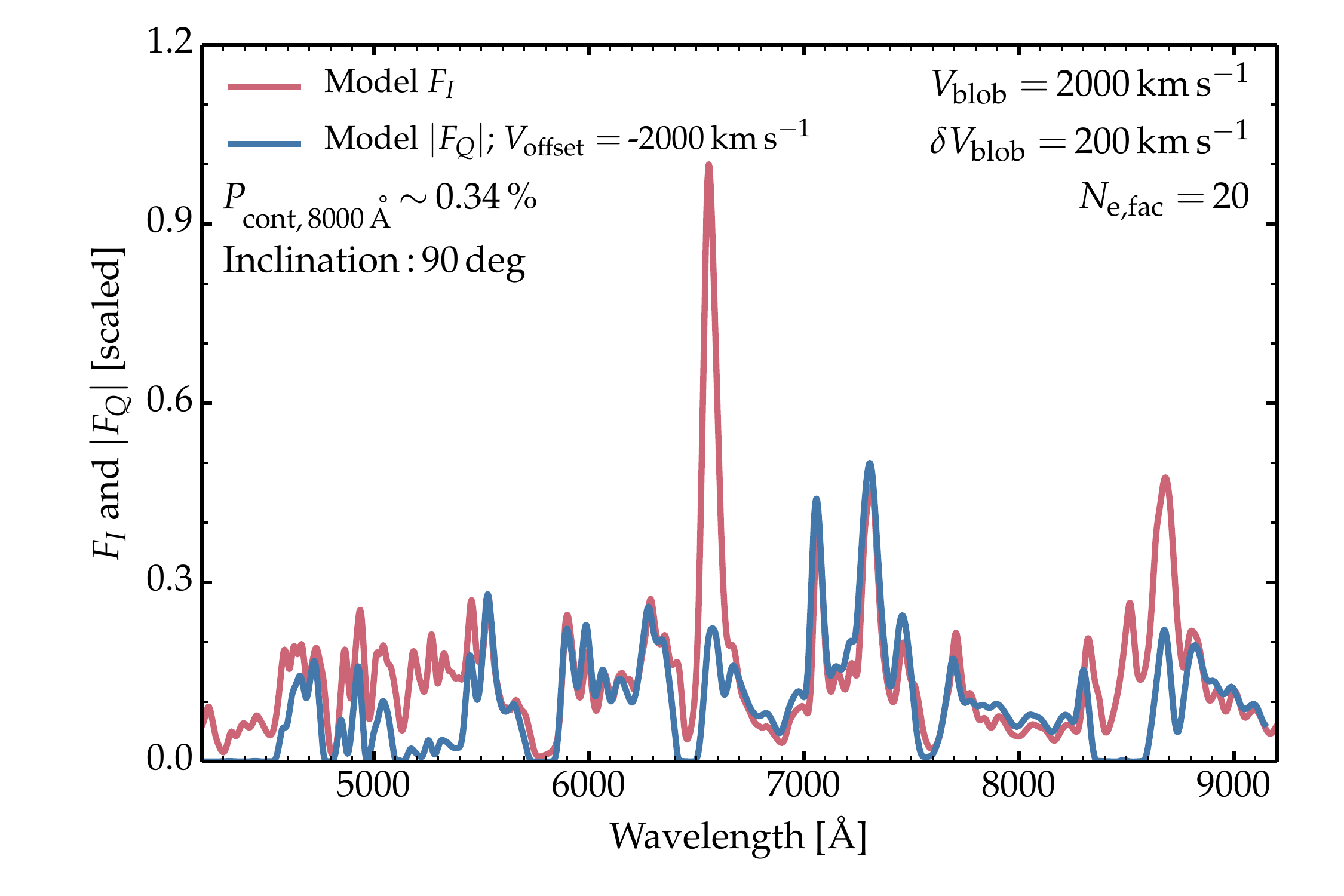, width=9.2cm}
\epsfig{file=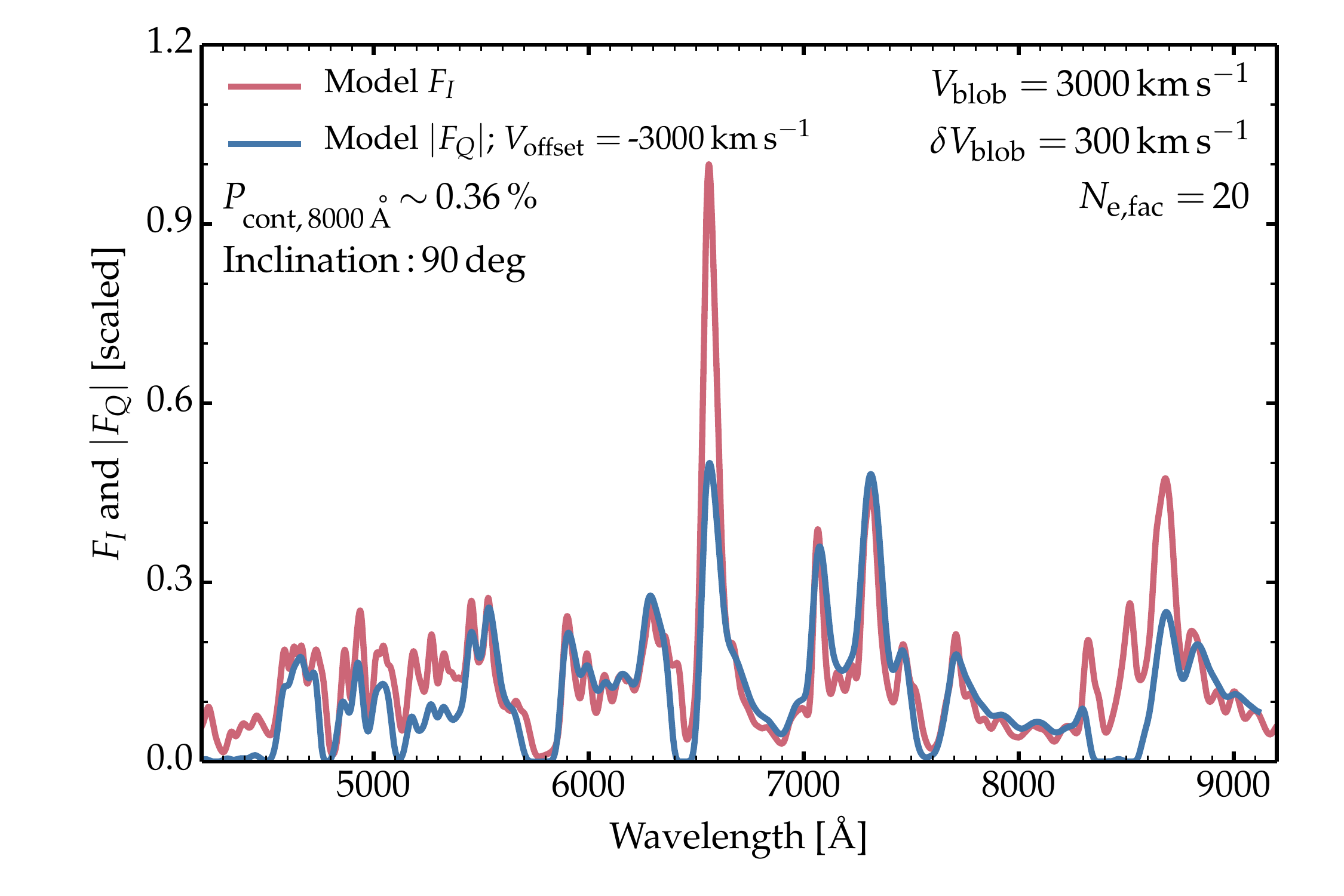, width=9.2cm}
\epsfig{file=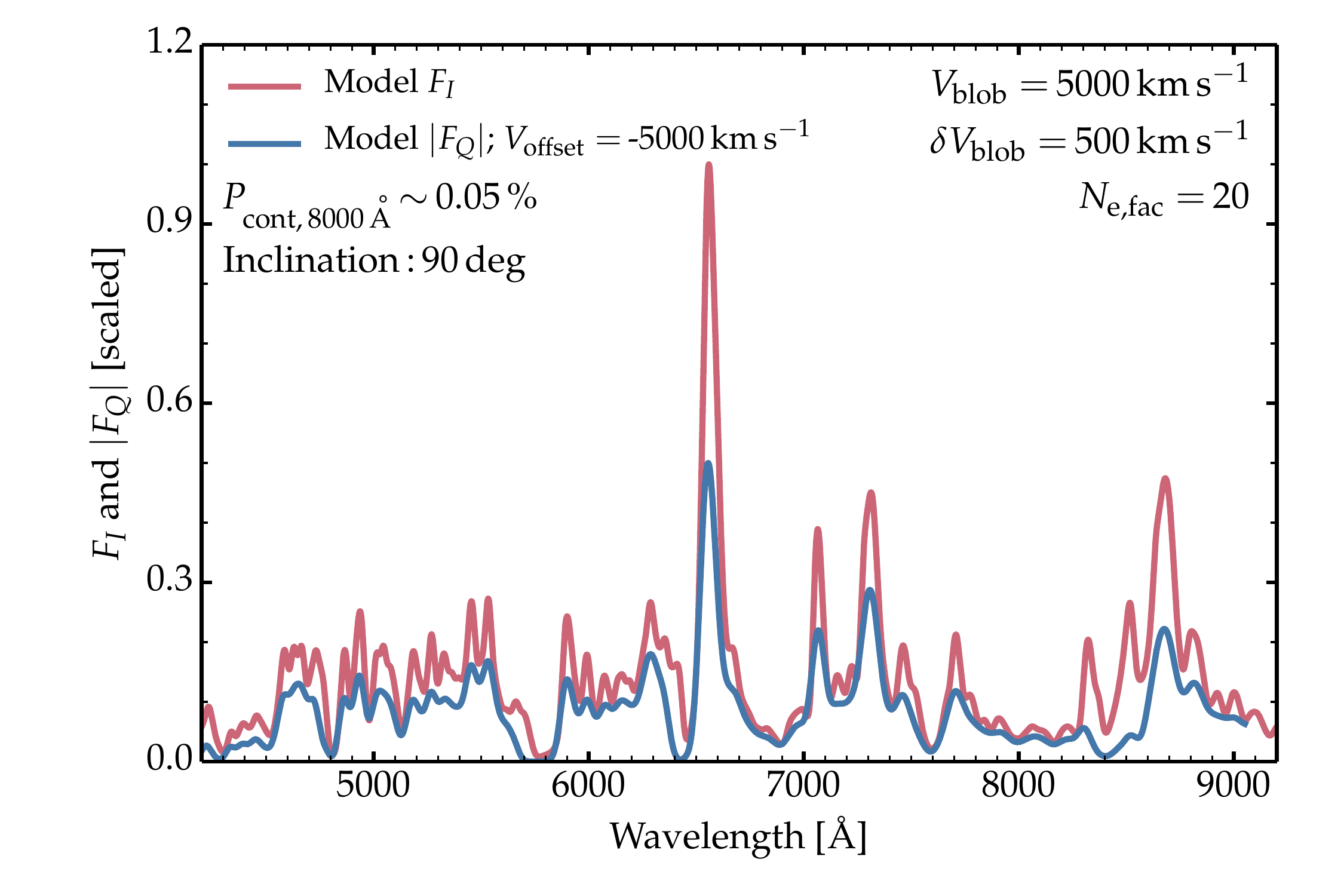, width=9.2cm}
\caption{Same models as in Fig.~\ref{fig_vblob}, but now showing the evolution of the total flux and of the blueshifted polarized flux for three different values of \vblob\ (the model with \vblob\ of 4000\,\kms\ is omitted since already shown in the middle panel of Fig.~\ref{fig_ex_model_fi_fq_p}). The adopted inclination is 90\,deg in all three panels and the velocity offset applied to $F_Q$ is  thus $V_{\rm offset} = -V_{\rm blob}$.
\label{fig_vblob_lam}
}
\end{figure}

\section{Influence of the blob velocity}
\label{sect_vblob}

   In this and subsequent sections, we vary the parameters characterizing the blob and discuss the sensitivity of the polarization properties that result. This helps to illustrate the principles that were discussed above for one reference case.

   In this section, we start by varying the blob velocity \vblob\ from 2000 to 3000, 4000, and 5000\,\kms. For simplicity, we keep the fractional blob velocity width fixed to 0.1\,\vblob\ and the blob opening angle is 20\,deg in all four cases. Figure~\ref{fig_vblob} is a counterpart of Fig.~\ref{fig_ex_model_fi_fq_p_nbeta}, but now for this new set of models and for an inclination of 90\,deg only (for which the polarization is maximum). We see that the total flux is essentially independent of blob location. However, the polarized flux varies both qualitatively and quantitatively. First, the level of polarization varies non-monotonically with \vblob. It is maximum nearly everywhere for a \vblob\ of 3000\kms, and then decreases for \vblob\ of 2000\,\kms, then 4000 and 5000\,\kms. The decrease is not necessarily uniform across the optical. For example, for the models with \vblob\ of 2000 and 3000\,\kms, the polarization is nearly identical redward of 8000\,\AA\ but significantly different below 5500\,\AA. The decrease in polarization for a large \vblob\ is easily explained. It arises from the decrease in blob optical depth  (see section~\ref{sect_BM77}).

\begin{figure}
\epsfig{file=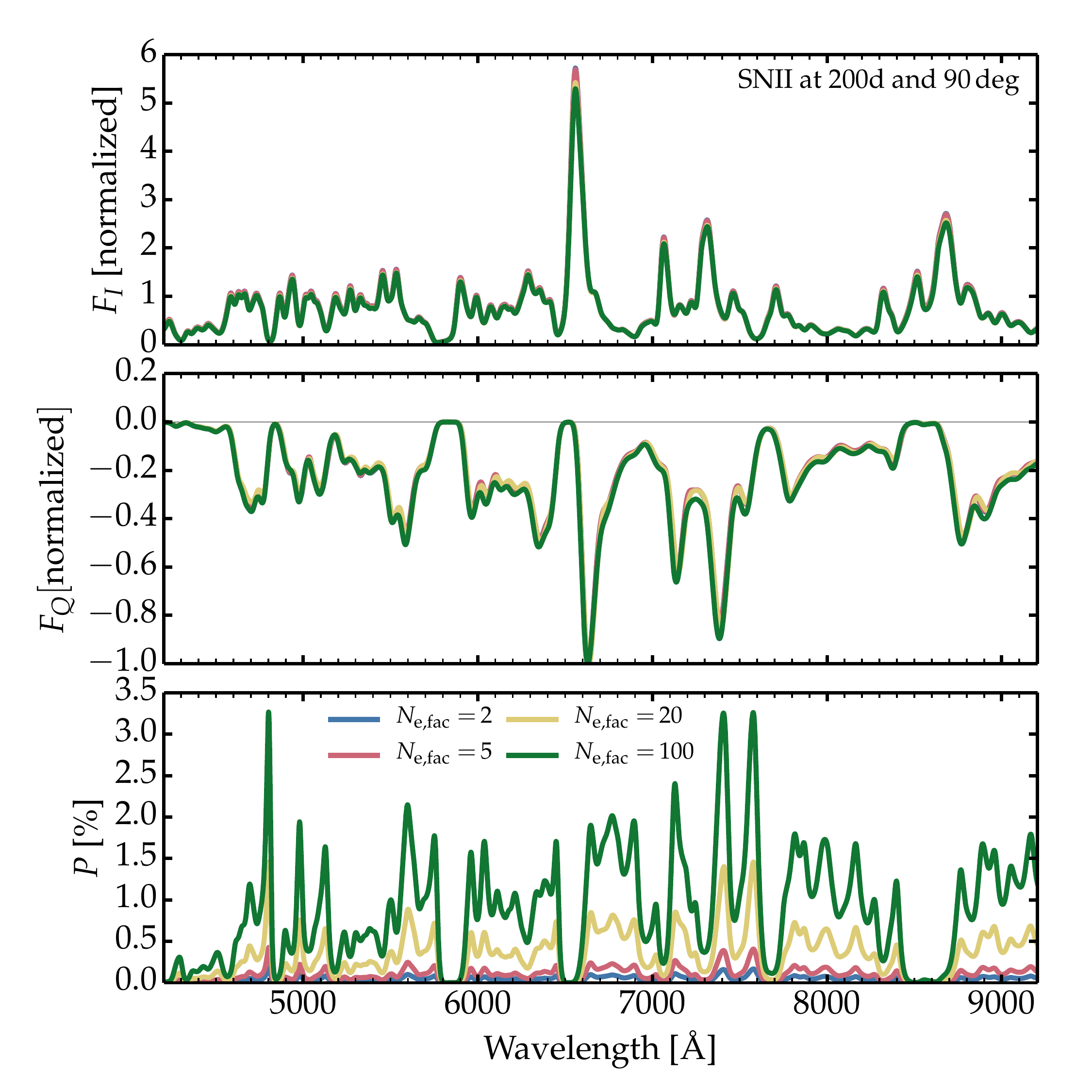, width=9.2cm}
\caption{Comparison of the normalized total flux $F_I$ (top), the normalized polarized flux $F_Q$ (the normalization is done at the location where $|F_Q|$ is maximum), and the polarization $P$ along a 90-deg inclination for the set of models differing in the value of $N_{\rm e,fac}$, which increases from 2 to 5, 20, and 100 (the blob radial optical depth increases from 0.044, to 0.109, 0.436, and 2.178). The electron-density enhancement is otherwise characterized by $V_{\rm blob}$=\,4000\,\kms, $\delta V_{\rm blob}=$\,1000\,\kms, and an opening angle of 20\,deg. The synthetic spectra have been smoothed with a gaussian kernel (FWHM of 23.5\,\AA). Note that the normalized $F_Q$ is independent of $N_{\rm e,fac}$ but that $P$ strongly increases with $N_{\rm e,fac}$. Varying $N_{\rm e,fac}$ impacts the level of polarization but not its qualitative behavior. We are in a linear regime whereby $P_{\rm cont} \propto \tau_{\rm blob} \propto N_{\rm e,fac}$.
\label{fig_nefac}}
\end{figure}

   The polarization is not the largest for the largest $\tau_{\rm blob}$ (model with \vblob\ of 2000\,\kms). In that case, the blob overlaps strongly with the emission region rather than being external to it. This tends to increase the scattering from within the emission region, enhancing the isotropy of the scattered flux (moving the blob even deeper or all the way to the ejecta center would eventually yield zero polarization). So one sees from this that there is an optimal blob distance or velocity for maximum polarization, so that the blob is sufficiently exterior relative to the emission region but not too far to be too optically thin and cause insufficient scattering. For a given blob location, the correspondence between $F_I$ and $|F_Q|$  varies with wavelength because the extent of the spectrum formation region varies with wavelength (see Fig.~\ref{fig_dfr}).

   This is further demonstrated by comparing the total flux $F_I$ and the polarized flux $|F_Q|$ for an inclination of 90\,deg and over the full optical range (Fig.~\ref{fig_vblob_lam}). Here, we blueshift the polarized flux by the velocity $-$\vblob\ and obtain an alignment with features seen in the total flux $F_I$. As the blob is migrated to a larger ejecta velocity, the polarized flux resembles more and more the total flux, while the magnitude of the polarization drops. The polarized flux remains \say{sharp} since the blob extent remains small both in velocity and in latitude (opening angle of 20\,deg). The strongest change with \vblob\ is seen for H$\alpha$  because of its large optical depth and its large associated emission volume. Even for \vblob\ of 5000\,\kms\ the polarized flux associated with H$\alpha$ is relatively weaker compared to the adjacent polarized ``continuum'' flux.

\section{Influence of the blob free electron-density enhancement}
\label{sect_nefac}

Figure~\ref{fig_nefac} shows the influence of the electron density enhancement associated with the blob on the normalized total flux $F_I$, the normalized polarized  flux $F_Q$ and the polarization $P$ over the optical range. Four values of $N_{\rm e, fac}$ are used and equal to 2, 5, 20, and 100 (see also Fig.~\ref{fig_blob_properties}). As discussed in Section~\ref{sect_motivation} and \ref{sect_mod}, an $N_{\rm e, fac}$ value of 100 might be difficult to produce with a high-velocity \nifs-rich blob alone. In that case, the model would probably require both a more energetic explosion and a high-velocity \nifs\ enrichment along the blob direction.

As was the case for other blob configurations, the impact on the total flux is negligible. Interestingly, the normalized polarized flux $F_Q$ is also identical for all four cases, despite the very different level of polarization, which varies linearly with increasing $\tau_{\rm blob}$ for $N_{\rm e, fac}$ of 2, 5, and 20. For the largest $N_{\rm e, fac}$, the polarization falls below a linear scaling, most likely because the corresponding blob is no longer optically thin ($\tau_{\rm blob}$ is 2.2 in that case, hence more than twice larger than the total ejecta optical depth in the unadulterated model with no blob; see Fig~\ref{fig_blob_properties}). The fact that the normalized $F_Q$ is identical for all four cases implies that the blob influence is qualitatively the same, and only the magnitude of the associated scattered flux varies, growing with increasing $\tau_{\rm blob}$ in the optically thin regime and eventually saturating as the blob optical depth exceeds one.

We saw in section~\ref{sect_intro} that the presence of \nifs\ at large velocity in the ejecta can yield an increase in electron density by a factor of 10 (without any change in the ejecta mass density), so intermediate with the $N_{\rm e, fac}$ values of 5 and 20. For these two cases, the polarization $P$ in the regions between strong lines (and away from saturated absorptions in $F_I$) can reach about 0.5\,\% (Fig.~\ref{fig_nefac}), thus well in line with what has been observed in Type II SNe at similar nebular epochs \citep{leonard_04dj_06,chornock_pol_10,leonard_08bk_12,leonard_iauga_15}.

\begin{figure}
\epsfig{file=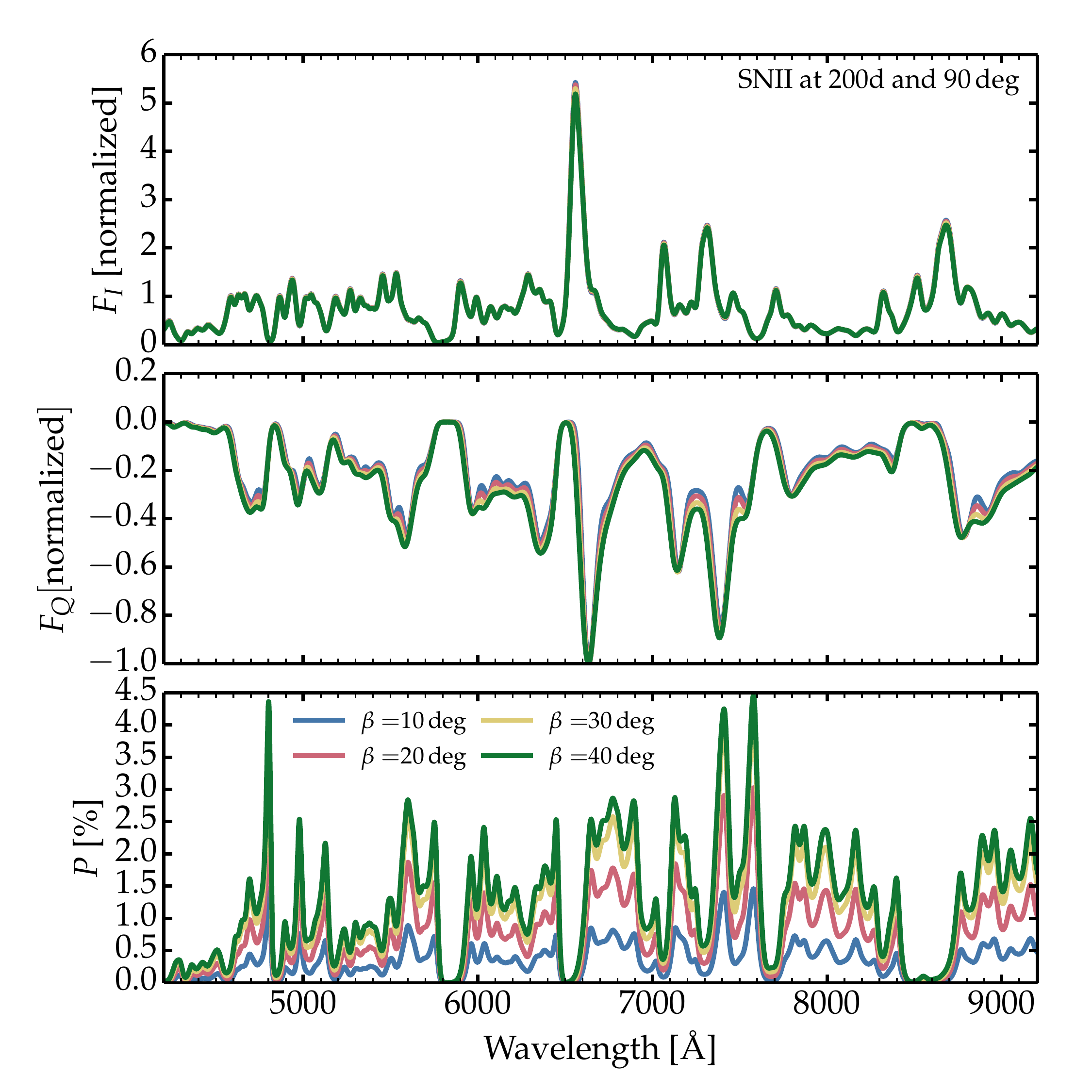, width=9.2cm}
\epsfig{file=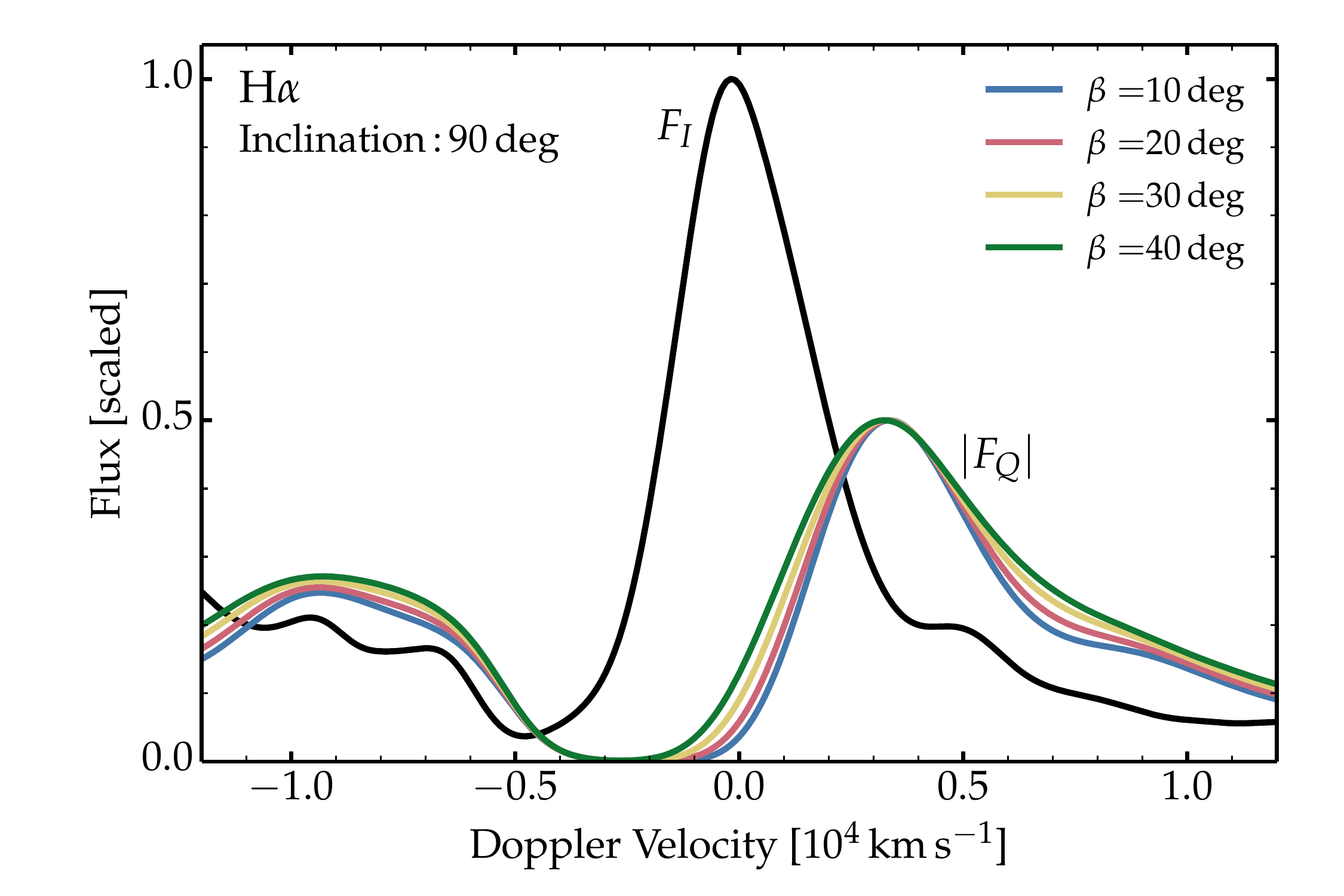, width=9.2cm}
\caption{Top: Comparison of the normalized total flux $F_I$ (upper panel), the normalized polarized flux $F_Q$ (middle panel), and the polarization $P$ (bottom panel) along a 90-deg inclination for the set of models differing in blob opening angle. The models have been smoothed with a gaussian kernel (FWHM of 23.5\,\AA). In all cases, the electron-density enhancement is characterized by $V_{\rm blob}$=\,4000\,\kms, $\delta V_{\rm blob}=$\,1000\,\kms, and $N_{\rm e,fac}$=\,20. The radial blob optical depth is 0.44. Bottom: Comparison between the total flux $F_I$ (black; only one curve is shown since $F_I$ is essentially independent of the choice of opening angle -- see top panel) and the scaled polarized flux $|F_Q|$ across the H$\alpha$ profile for the same model set as shown at top.
\label{fig_opening_angle}
}
\end{figure}

\section{Influence of blob opening angle}
\label{sect_opening_angle}

Figure~\ref{fig_opening_angle} illustrates the impact on the total and polarized flux when the blob half-opening angle is increased from 10 to 20, 30 and 40\,deg. In all four cases, the blob is characterized by \vblob\ of 4000\,\kms, a \dvblob\ of 1000\,\kms\ and a $N_{\rm e, fac}$ of 20, and thus has the same radial electron scattering optical depth of 0.44 (recall that this is the electron scattering optical depth difference for ejecta directions crossing the blob and those that do not; see also Fig.~\ref{fig_blob_properties}). The adopted inclination is 90\,deg. Physically, by varying the blob opening angle, we vary the angular extent of the enhancement in free-electron density. The reader should keep in mind that in nature, this could happen for a small but more massive \nifs\ blob, which would heat and ionize a greater volume of the H-rich material.

With increasing opening angle, the polarization level away from strong lines increases from about 0.5 to 2.5\,\%. This occurs as a result of the increasing subtended angle of the blob, which increases the probability that the incoming radiation from the inner ejecta gets scattered by free electrons within the blob. The mean redshift of the polarized flux is the same and about 3500\,\kms, thus close to the blob velocity. However, for increasing opening angle, $|F_Q|$ does broaden,  although perhaps not as much as would be naively expected given the factor of four in opening angle (the geometric extent should broaden the range of Doppler shifts by an amount $2 \sin \beta$ for a 90-deg inclination). Hence, $|F_Q|$ remains quite sharp, in contrast with the intuitive idea that a broader blob should increase the smearing of the polarized flux; nonetheless, an observed {\it lack} of broadening of lines in the polarized flux can be used to set some rough limits on the geometric extent of the scatterers.

The relative lack of broadening in the polarized flux arises from the dependence of the polarization with inclination. Polarization is maximum for regions in the plane containing the ejecta center and perpendicular to the line of sight. In other words, asymmetries away from that plane contribute much less polarization. In our setup, the blob radial optical depth is the same in all ejecta-centered directions so indeed regions in the mid-plane are favored. This emphasizes one limitation of polarization, in the sense that we are biased towards the detection of asymmetries in the mid-plane, while asymmetries closer to the line of sight yield a lower polarization that may be undetectable (the term $\sin^2 i$ in Eq~\ref{eq_pol_BM77} implies a four times lower polarization for a given blob at 30\,deg compared to the same blob seen at 90\,deg to the line of sight, all else being the same).

\begin{figure}
\epsfig{file=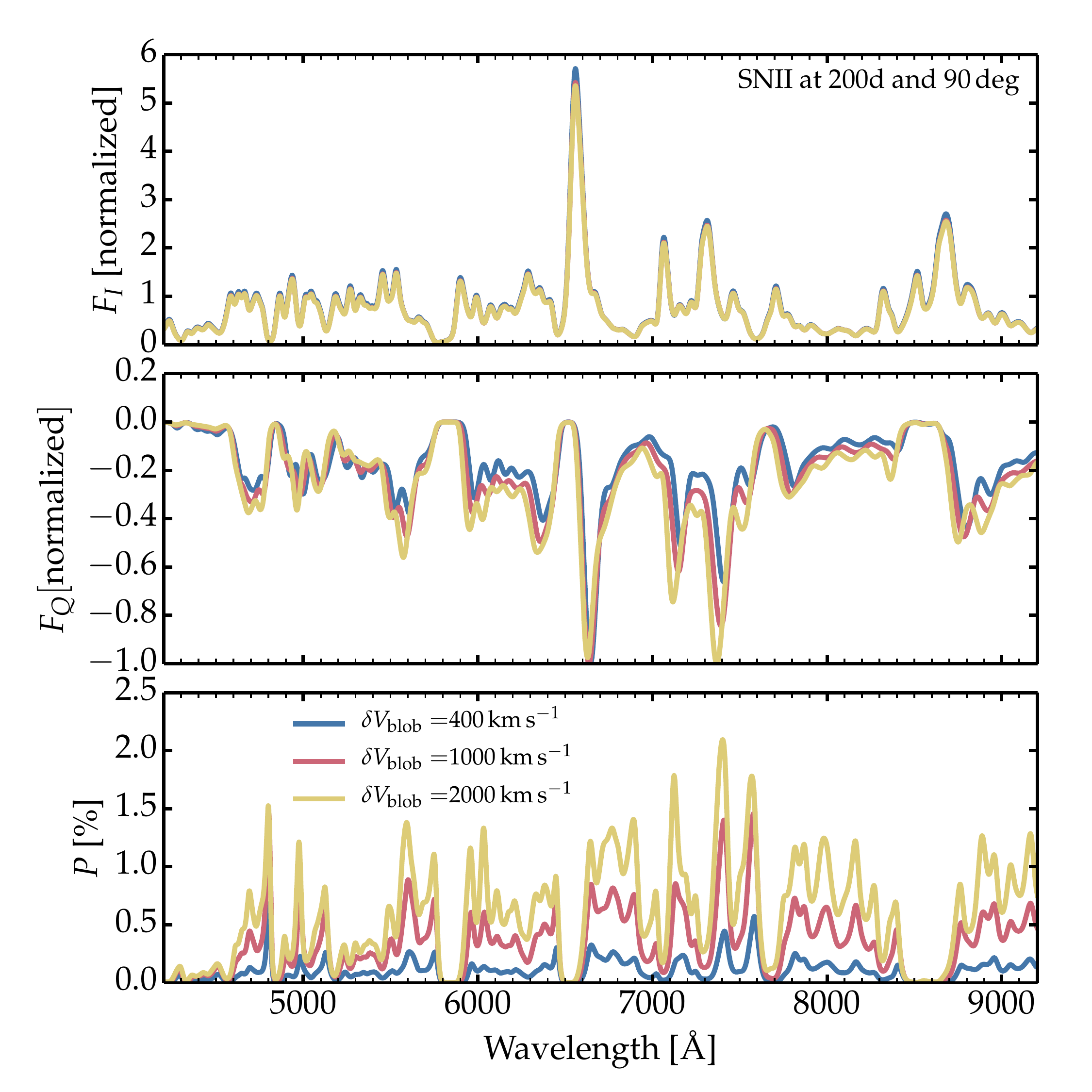, width=9.2cm}
\epsfig{file=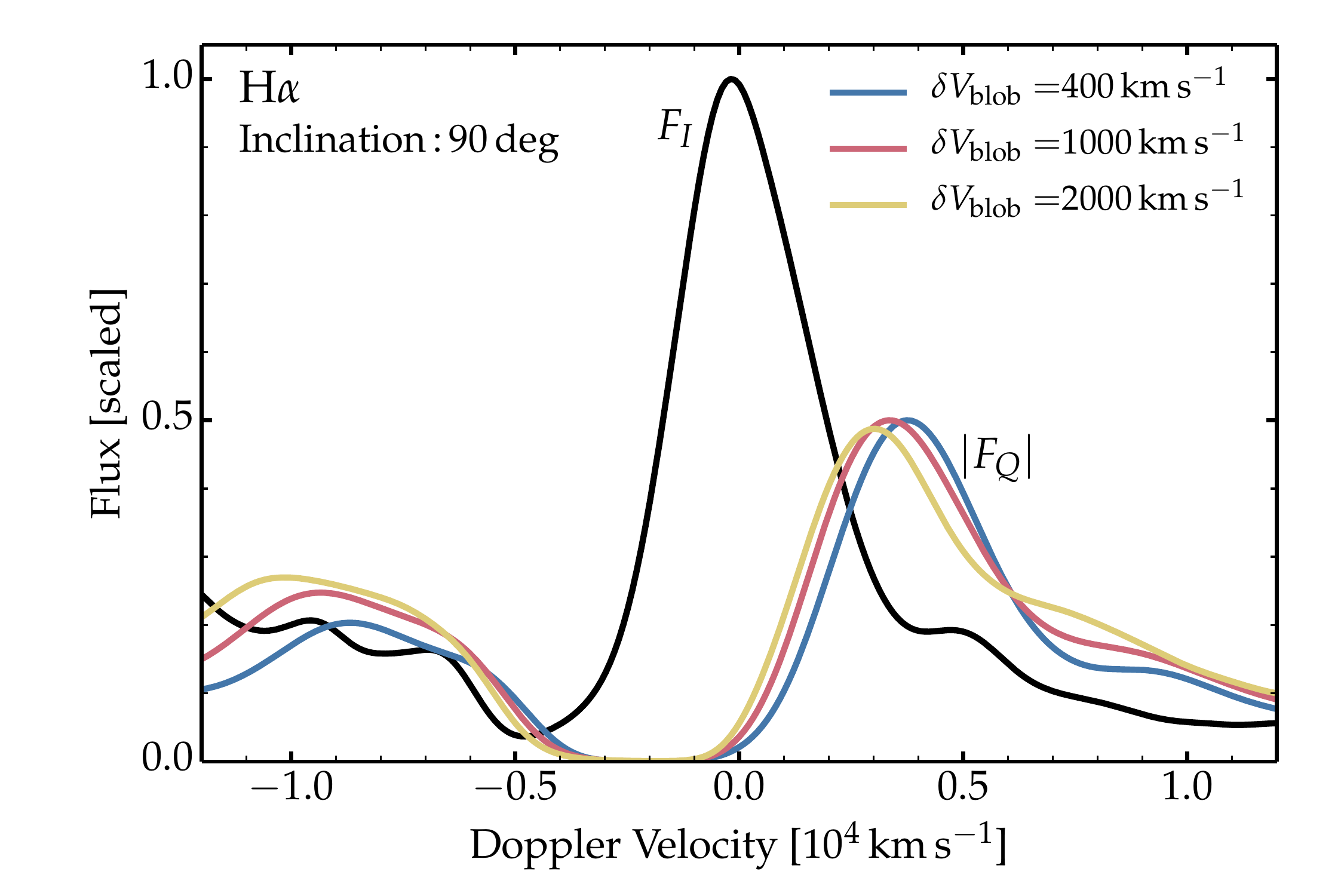, width=9.2cm}
\caption{Same as Fig.~\ref{fig_opening_angle}, but now for cases in which the blob velocity width is increased from 400 to 1000, and 2000\,\kms\ (the radial blob optical depth increases from 0.098, to 0.436 and 2.934). The blob is characterized in all cases by $V_{\rm blob}=$\,4000\,\kms, $\beta=$\,10\,deg, and $N_{\rm e,fac}=$\,20\,deg. The adopted inclination is 90\,deg.
\label{fig_dvblob}}
\end{figure}

\section{Influence of blob width}
\label{sect_dvblob}

Figure~\ref{fig_dvblob} illustrates the properties of the total flux and the polarized flux when the blob velocity width is increased from 400 to 1000 and 2000\,\kms. The largest blob width may correspond to a configuration in which the explosion energy was higher along the blob direction since this raises the mass density, and consequently the electron density (for a more consistent treatment of this configuration, see Section~6.3 and Fig.~15 of \citealt{D20_12aw_pol}). The results for a varying blob width are similar to those described in previous sections and are not repeated. One interesting feature is that the polarized flux $|F_Q|$ associated with H$\alpha$ is not broader for increasing $\delta V_{\rm blob}$ (though it does have a higher velocity tail). We see instead a lower redshift for increasing  $\delta V_{\rm blob}$. Indeed, as $\delta V_{\rm blob}$ is increased, a growing fraction of the polarization arises from the denser regions at lower velocity, while the regions at larger velocities make a negligible contribution because of their lower density (see right column of Fig.~\ref{fig_blob_properties}).

Constraining the geometry of the asymmetric scatterers at the origin of the polarization is thus complicated because of the existing bias in favor of denser regions located at 90-deg inclination to the observer. For example, a very dense blob at large velocity and low inclination may yield the same polarization (magnitude and redshift) as a lower density blob at small velocity in the mid-plane. And of course, many such blobs may be present along multiple ejecta-centered directions.

\begin{figure*}
\epsfig{file=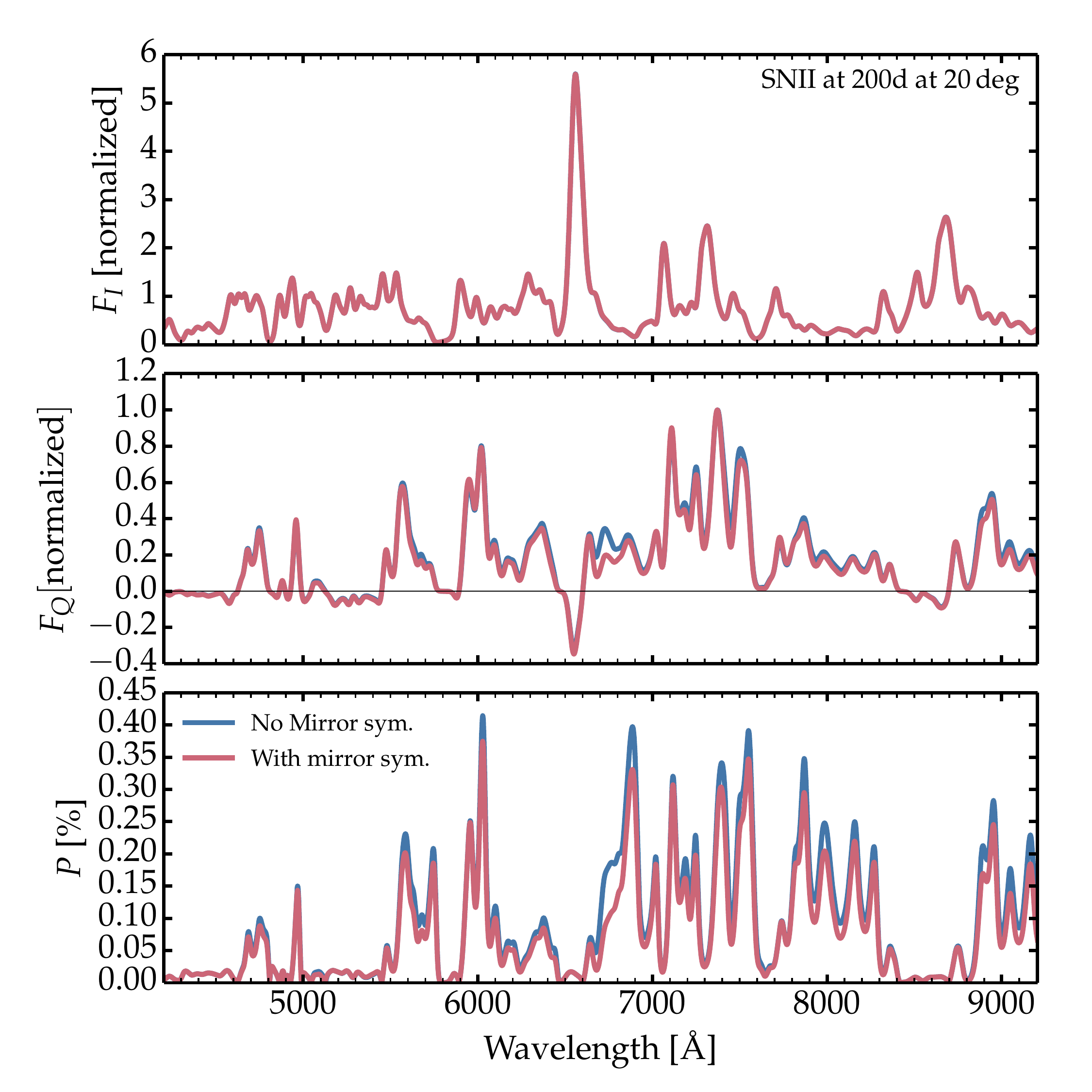, width=9.2cm}
\epsfig{file=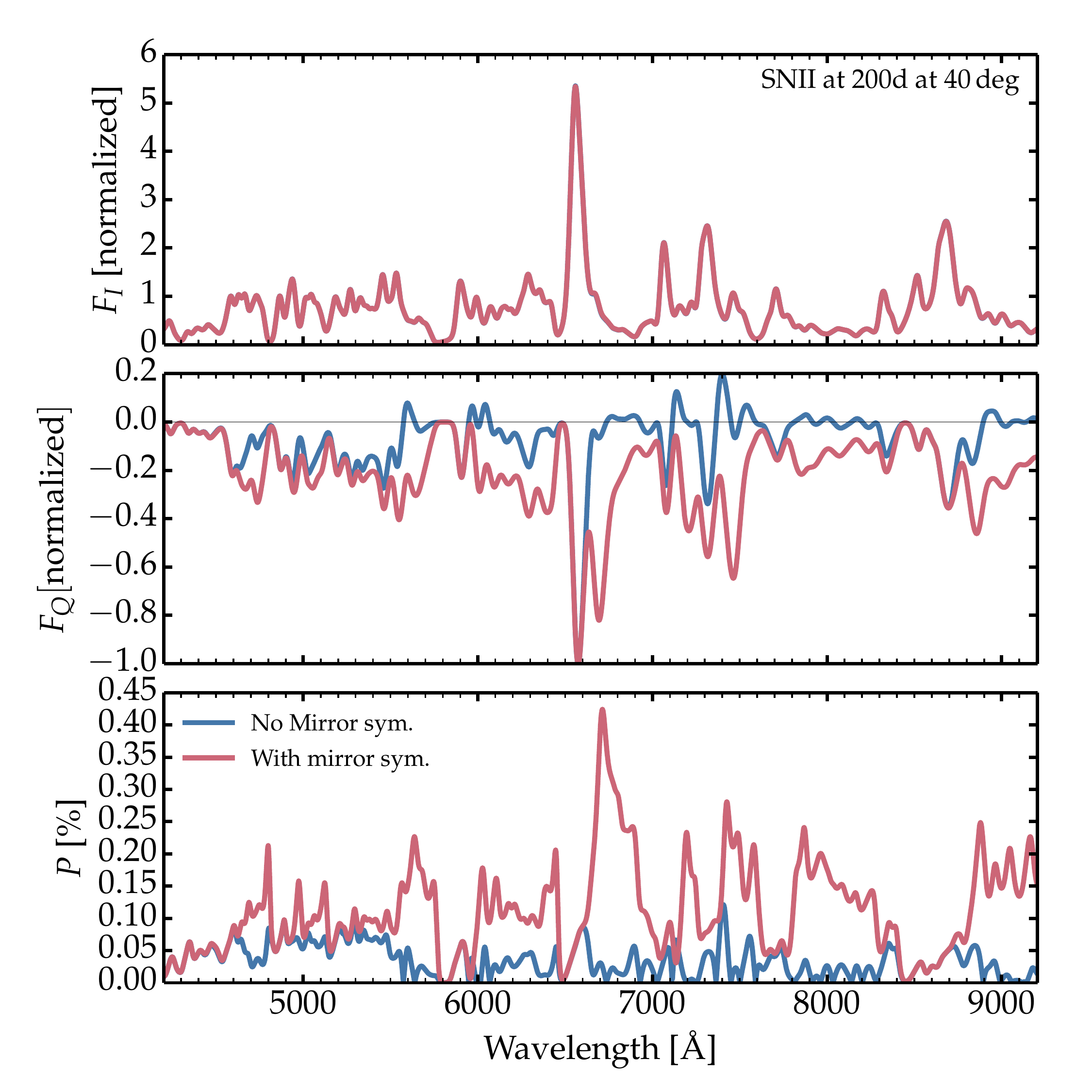, width=9.2cm}
\epsfig{file=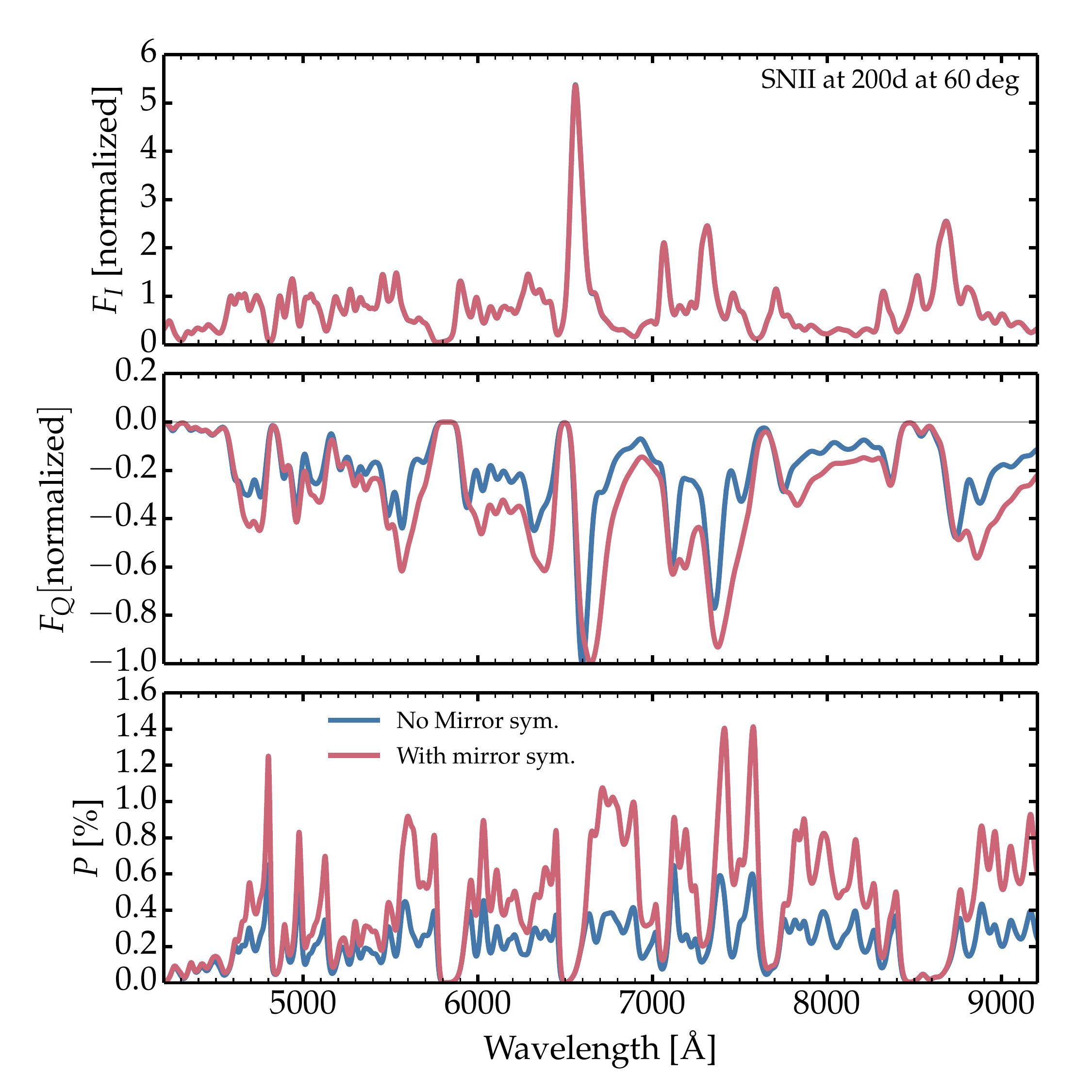, width=9.2cm}
\epsfig{file=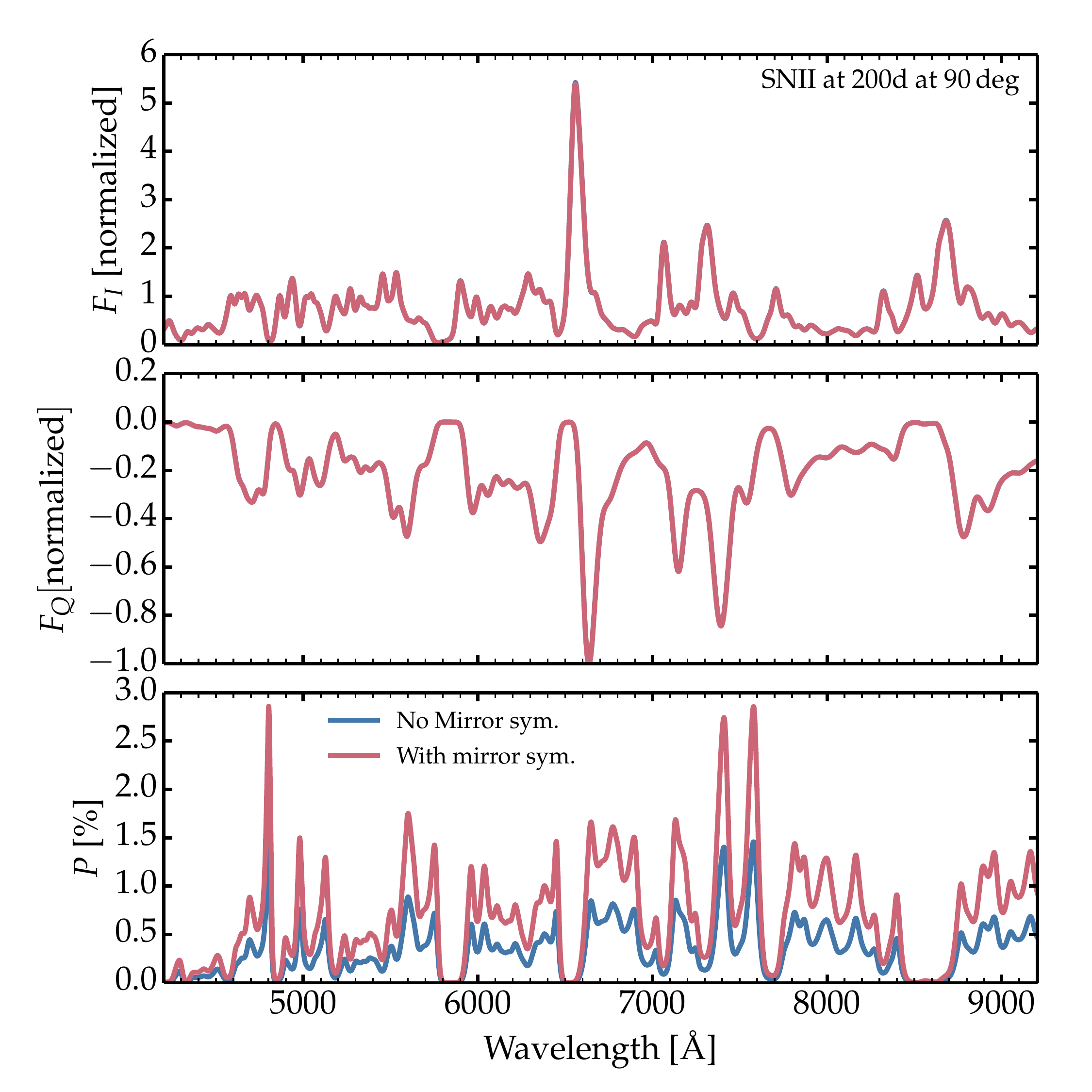, width=9.2cm}
\caption{Comparison of the normalized total flux $F_I$ (top), the normalized polarized flux $F_Q$, and the polarization $P$ seen along an inclination of 20 (top left), 40 (top right), 60 (bottom left) and 90\,deg (bottom right) for the model with mirror symmetry (aka two blobs) and without (aka one blob). The blob(s) is (are) characterized in all cases by $V_{\rm blob}=$\,4000\,\kms, $\delta V_{\rm blob}=$\,1000\,\kms, $\beta=$\,10\,deg, and $N_{\rm e,fac}=$\,20, corresponding to a radial blob optical depth of 0.436.
\label{fig_mirror_sym}}
\end{figure*}

\begin{figure*}
\begin{center}
\epsfig{file=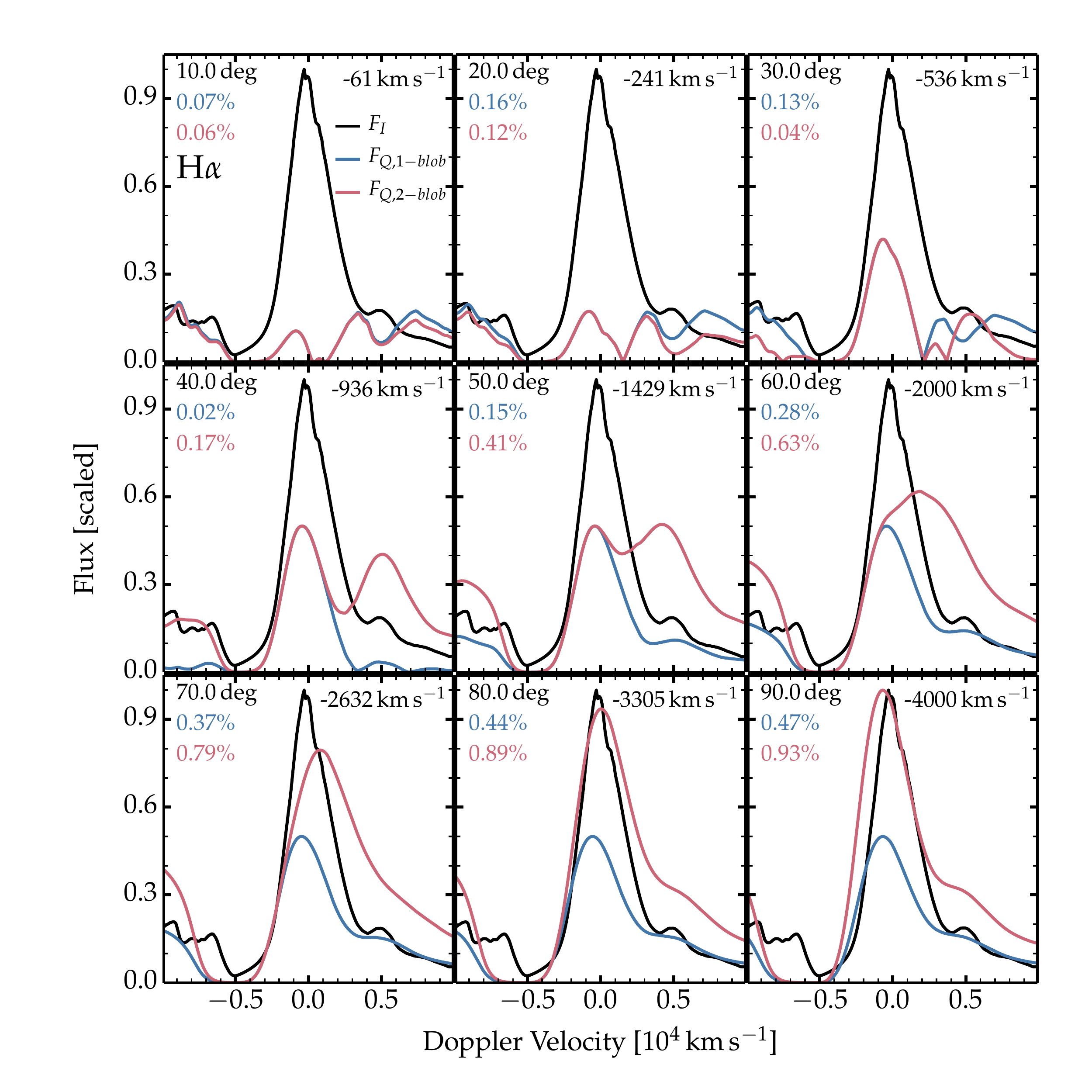, width=15cm}
\end{center}
\vspace{-0.5cm}
\caption{Same models as in Fig.~\ref{fig_mirror_sym}, but now showing the evolution of the normalized total flux $F_I$ and scaled, blueshifted polarized flux $|F_Q|$ in the H$\alpha$ region (zero Doppler velocity corresponds to the line rest wavelength) as a function of the inclination (progressing clockwise from 10 to 90\,deg; the value of the inclination is given at the top left of each panel). The flux $F_{Q,{\rm 1-blob}}$ corresponds to the single-blob case in which a blob is placed along the symmetry axis and the computation is done over 180\,deg in latitudes (no mirror symmetry). In contrast, the flux $F_{Q,{\rm 2-blob}}$ corresponds to the case in which a blob is placed along the symmetry axis but the simulation is carried over from 0 to 90\,deg in latitude and assuming mirror symmetry -- this is then equivalent to having two blobs 180\,deg apart. The value of the blueshift is the same for each polarized flux and set to  $-V_{\rm blob} (1 - \cos \alpha_{\rm los})$, where $\alpha_{\rm los}$ is the inclination of the single blob in the 1-blob case, and of the approaching blob in the 2-blob case. We use the same scaling of $|F_Q|$ for both models, so that the difference in polarization level between the two models is preserved (the continuum polarization in the 8000\,\AA\ region is given at the top left of each panel, below the value of the inclination).  The enhancement is characterized in all cases by $V_{\rm blob}=$\,4000\,\kms, $\delta V_{\rm blob}=$\,1000\,\kms, $\beta=$\,10\,deg, and $N_{\rm e,fac}=$\,20, corresponding to a radial blob optical depth of 0.436.
\label{fig_mirror_sym_ha}
}
\end{figure*}

\section{Influence of mirror symmetry: unipolar versus bipolar explosion}
\label{sect_mirror}

All simulations presented so far in this study have assumed a single blob. This ideal configuration simplifies the interpretation since there is only one isolated source for the residual polarization. Multiple blobs along the axis but in one hemisphere would be analogous to a configuration in which we have just one blob but with a large velocity width (this case was studied in the preceding section).  In Nature, the 3D distribution of the \nifs\ may, however, be very complex (see Section~\ref{sect_motivation}). So, to make a first attempt towards treating this complexity, we consider the configurations in which two distinct blobs are present at the same ejecta velocity \vblob\ but in opposite hemispheres (one blob along polar angles of 0 and 180\,deg). In practice, this corresponds to simulating for the presence of one blob at \vblob\ and adopting mirror symmetry with respect to the equatorial plane. The blob(s) is (are) characterized in all cases by $V_{\rm blob}=$\,4000\,\kms, $\delta V_{\rm blob}=$\,1000\,\kms, $\beta=$\,10\,deg, and $N_{\rm e,fac}=$\,20.

Figure~\ref{fig_mirror_sym} illustrates the influence of the assumption of mirror symmetry on the total flux and polarized flux. Simplistically, the configuration with two blobs is equivalent to combining the results for one blob for an inclination $i$ with those for an inclination $\pi - i$ (see results for different inclinations in section~\ref{sect_ref} and Fig.~\ref{fig_ex_model_fi_fq_p_nbeta}).
For an inclination of 20\,deg, there is moderate polarization, and it is dominated by the blob in the near hemisphere (compare the results for inclinations 10 and 170\,deg in Fig.~\ref{fig_ex_model_fi_fq_p_nbeta}), so that the results are very similar for one and two blobs. For an inclination of 40\,deg, it is the reverse with the overall polarization level arising from the near blob being much weaker than that of the far blob (compare the results for inclinations 50 and 130\,deg in Fig.~\ref{fig_ex_model_fi_fq_p_nbeta}). For an inclination of 60 and 90\,deg, the overall polarization level is qualitatively similar in both, but nearly twice as large in the two-blob case. For an inclination of 90\,deg, the polarization is exactly twice as large (and otherwise identical) for the two-blob case -- the contribution of each blob is identical and simply adds up in $F_Q$ and $P$.

Inspecting more closely the morphology of $|F_Q|$ for the two-blob case in the middle panels of Fig.~\ref{fig_mirror_sym}, and in particular the region around 6500\,\AA, one sees the presence of a single narrow peak with small redshift (inclination of 20\,deg), a double peak (two distinct redshifts; inclination of 40\,deg), a broad single peak (inclination of 60\,deg), and back to a narrower single peak (inclination of 90\,deg). To illustrate this more clearly, we show the variation of $F_I$ and $|F_Q|$ in the H$\alpha$ region for the two models with a single blob and two blobs 180\,deg apart in Fig.~\ref{fig_mirror_sym_ha}. In all panels, the polarized flux $|F_Q|$ has been blueshifted by the amount $-V_{\rm blob} (1 - \cos \alpha_{\rm los})$, where $\alpha_{\rm los}$ is the inclination with respect to the blob along the polar angle of zero. This blueshift thus corrects for the redshift of the scattered (polarized) flux associated with the blob along the polar angle of zero degree. There is an additional redshift for the other blob (if present) by $V_{\rm blob} \cos \alpha_{\rm los}$. Thus, irrespective of the applied blueshift, each blob contribution to $|F_Q|$ is separated by  $2 V_{\rm blob} \cos \alpha_{\rm los}$. For near-zero inclination, the separation is $\sim$\,2\vblob\ but the polarization is very small. For an edge-one view, the separation is zero and the polarization level is twice as large as for the single-blob case. For intermediate inclinations, the two contributions are separated and we see two distinct peaks, in particular for inclinations around 40\,deg (for an inclination of 40\,deg, the separation is equal to 6130\,\kms\ in the two-blob model).

This result indicates that the polarized flux at nebular times carries valuable information on the kinematics of the high velocity scatterers and may be one direct way to constrain the highest velocities reached by \nifs\ blobs in core collapse SNe  -- the fastest moving \nifs\ material should by construction be asymmetrically distributed and therefore conducive to a residual polarization. The continuum polarization yields no information of this nature at late times, although the first appearance of continuum polarization during the photospheric phase is indicative of the maximum ejecta velocity at which asphericity exists. The scattering of line photons is not a process limited to nebular times. During the photospheric phase, line photons are also scattered by free electrons and yield a jump in $|F_Q|$ in the red-wing of strong lines (the so-called electron scattering wings; see \citealt{hillier_91}). But at these earlier times, the continuum flux is strong and makes this feature less conspicuous. At nebular times, the continuum flux is weak and the bulk of polarized photons are associated with lines.

\section{Conclusions}
\label{sect_conc}

We have presented numerical simulations of the influence of a \nifs\ blob on the emerging total flux and polarized flux from a 2D axially symmetric ejecta. The ansatz of our study, supported by 1D non-LTE radiative transfer simulations with \cmfgen, is that the primary influence of a localized enhancement in \nifs\ is to boost the free electron density. In 1D, this boost corresponds to a shell, with no associated polarization. It can however strongly impact the strength of some lines such as H$\alpha$ \citep{D20_12aw_pol}. If the \nifs\ enhancement is limited to a confined blob, a sizable polarized flux may result from the asymmetry introduced in the distribution of scatterers, but with little influence on the total flux. The exact level of influence on the flux will depend on the location of the \nifs\ blob and its impact on the emissivity. The deeper it is, the greater the impact on the emissivity and the total flux, but the weaker the residual polarization. We limit the study to 200\,d after explosion. By design, the mass density, the decay power, or the emissivity of the ejecta retain a spherical symmetry, and in that sense the ``core'' of the ejecta is symmetric. The polarization here results exclusively from the distribution of scatterers in the H-rich envelope, which typically reside beyond 2000\,\kms. The ejecta \nifs\ is not limited to the \nifs\ blob, but the blob is what remains of the \nifs\ distribution once the spherical part has been subtracted off (only the aspherical part can contribute polarization).

Our simulations were  designed to mimic the presence of a \nifs\ blob in SN ejecta. To cover a wide parameter space we varied the location of the blobs in the ejecta, their widths and opening angles, and their free electron density. However, in our heuristic study not all blob configurations may exactly match what the influence a \nifs-rich blob would have in a 3D explosion, and some of the electron-density enhancements used may not be realistic. Nevertheless the study provides crucial insights into expected polarization signatures. Further, the models may provide useful representations of asymmetric but axisymmetric explosions.  For example, a very large blob width might better represent a \nifs\ finger rather than a blob. A large and strong electron-density enhancement may be more typical of a higher energy explosion with strong \nifs\ mixing. In that sense, the association we make between \nifs-rich blob and electron-density enhancement is loose and our results should be interpreted at a more qualitative level. A quantitative study would require a physically consistent 3D radiative transfer model and a 3D explosion model, which is beyond the scope of this paper, and is not the philosophy of the present work.

We find that the fundamental polarization signatures of a \nifs\ blob (or equivalently a free electron-density enhancement) are largely independent of its properties. Qualitatively, the blob acts as an asymmetric scatterer, yielding a polarized flux that appears like a replica of the total flux, but scaled down by a factor 100 to 1000 depending on blob properties, and also redshifted by an amount that scales as \vblob\ $(1 - \cos \alpha_{\rm los})$. Once blueshifted by this amount and scaled by $1/P$, the polarized and total fluxes match closely for a high-velocity blob. For a blob located deeper in the ejecta, the overlap with some line emission regions (as is the case, for example, for H$\alpha$) leads to a decrease of the polarization in the corresponding spectral region. For a blob located far out in the ejecta, the replication is nearly perfect. The redshift of the polarized line features as well as the width and strength of these redshifted polarized line photons give some contraints on the maximum extent or velocity of the asymmetric distribution of the \nifs\ nucleosynthesised during the explosion.

Overall, this study suggests there is much to learn from the polarized flux at nebular times, in particular from the inspection of the polarization associated with line photons. This is in stark contrast with the literature that focuses nearly exclusively on the discussion of continuum polarization and total flux spectra, and also conflicts with the generally held belief that line photons are unpolarized. They may well be at the time of emission but like other photons, they may interact with free electrons on their way to escape.

A second aspect of our study is that it is possible to produce significant polarization without invoking any asymmetry in the core. Here, the asymmetry is limited to that of the blob, which was consistently placed at a velocity greater than 2000\,\kms, thus within the H-rich layers of the ejecta. This confirms the more physically consistent model presented for SN\,2012aw \citep{D20_12aw_pol} in which strong \nifs\ mixing was invoked along one direction (thus less contrived than adopting a localized blob).

\begin{figure}
\epsfig{file=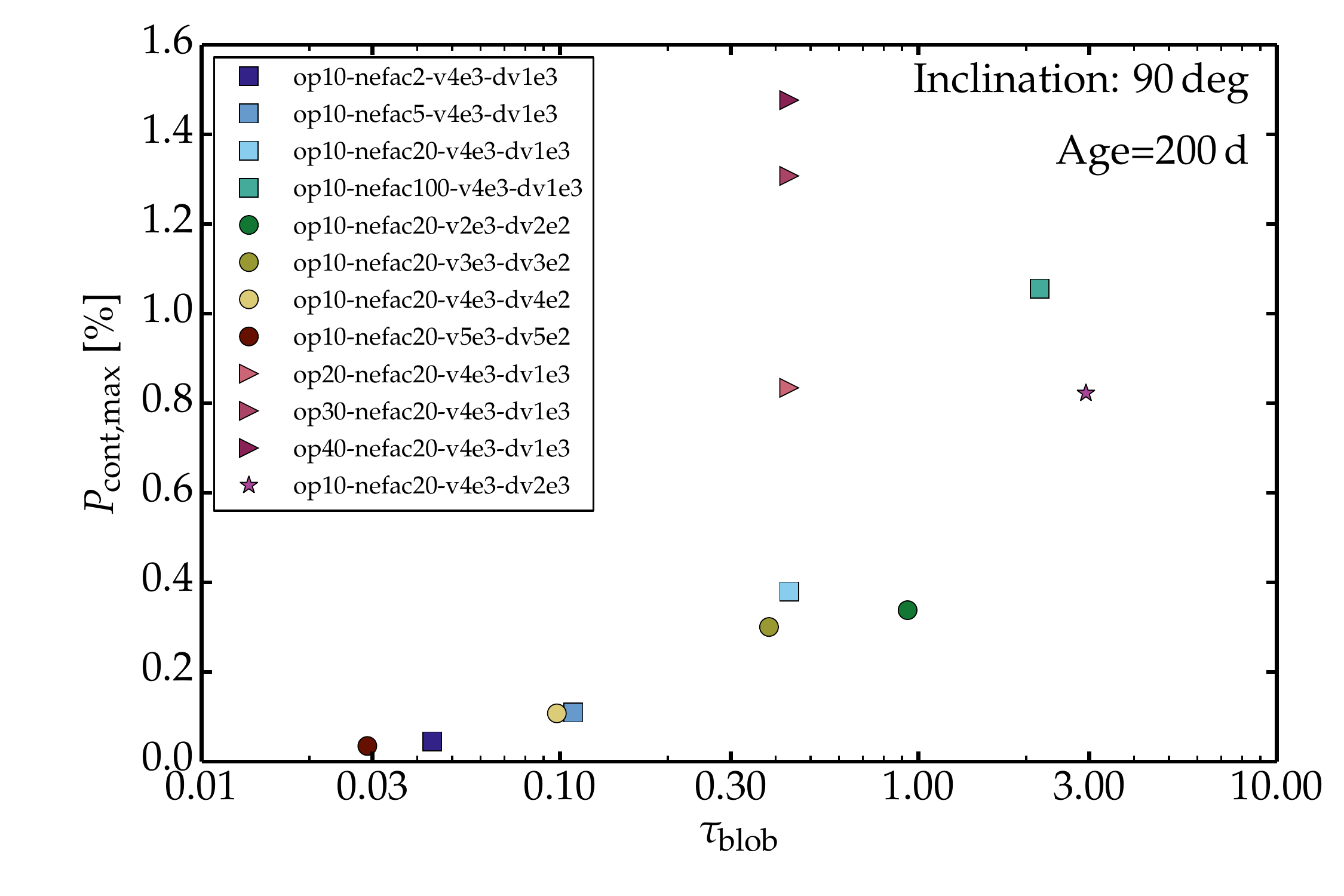, width=9.2cm}
\caption{Maximum continuum polarization (i.e., for a 90-deg inclination) versus blob radial electron scattering optical depth obtained for our set of models. Our simulations suggest that core-collapse SNe may routinely produce 0.5\,\% continuum polarization at 200\,d from the presence of a high velocity \nifs\ blob.
\label{fig_cont_all}}
\end{figure}

The continuum polarization produced by a blob of \nifs\ depends on the properties of the associated boost in free electron density, its magnitude, its size, or inclination to the observer.  For a given inclination, the continuum polarization scales with the blob optical depth and angular size (see Eq.~\ref{eq_pol_BM77}). In our set of simulations, the maximum continuum polarization (for a 90-deg inclination) ranges between 0.03 up to 1.47\,\% (Fig.~\ref{fig_cont_all}). The smallest value of 0.03\,\% corresponds to a blob at 5000\,\kms, thus with a low optical depth (and a small opening angle of 20\,deg). The largest value of 1.47\,\% corresponds to a spatially extended blob at 4000\,\kms\ (velocity width of 1000\,\kms, opening angle of 80\,deg) and with an optical depth of 0.44. For less ad-hoc conditions such as those obtained with the \cmfgen\ simulation described in section~\ref{sect_motivation} (for example, $V_{\rm blob}=$\,4000\,\kms, $\delta V_{\rm blob}=$\,1000\,\kms, $\beta=$\,10\,deg, and $N_{\rm e,fac}$ between 5 and 20), a continuum polarization of 0.5\,\% is easily explained. This suggests that high velocity \nifs\ blobs can explain the late time polarization of some Type II SNe, provided these blobs are few and only partially cover the full solid angle.

This paper is concerned with the polarization signature of a confined electron density enhancement at high velocity (high $V_{\rm blob}$, small $\delta V_{\rm blob}$, small opening angle). However such a configuration can mimic the presence of dust formed in clumps in the outer parts of an ejecta that originally interacted with CSM (e.g., for example as could occur in Type II SNe like SN2013fs; \citealt{yaron_13fs_17}). In that case, the localized boost to the scattering opacity by dust arises not from the enhanced density of free electrons but from the much greater opacity introduced locally (by means of the blob) through the putative presence of dust. The polarization from a dust blob would behave somewhat differently because dust and electron scattering do not have the same phase functions, and polarization induced by dust is wavelength dependent. However, the wavelength shifts predicted in the polarized flux would be similar (since it is induced by the expansion of the SN ejecta), and the level of continuum polarization could be of a similar magnitude, provided enough dust forms in the outer ejecta.

Although all models were performed at a single epoch of 200\,d, we can anticipate the evolution with time, in particular for the continuum polarization (the line polarization is more complicated since the line emission will significantly change during the nebular phase). The ejecta ionization at nebular epochs is essentially constant so the electron-scattering optical depth drops as $1/t^2$, where $t$ is the elapsed time since explosion (see, for example, \citealt{DH11_2p}, their Fig~7). Since our adopted electron density enhancement scales with the local electron density of the unadulterated model, the blob optical depth would also drop as $1/t^2$. And since the polarization scales with the blob optical depth, it would also drop as $1/t^2$. This is consistent with the more sophisticated simulations presented in  \citet[][their Fig 12, which covers only the early nebular phase]{D20_12aw_pol} for SN\,2012aw. This late-time behavior has also been studied in a similar context in \citet[][see for example their Fig.~22]{DH11_pol}. SN\,2004dj is probably the best observed Type II SN to exhibit a continuum polarization with such a $1/t^2$ dependence at nebular times \citep{leonard_04dj_06}.

Future studies should explore the influence of \nifs\ blobs in a wider range of progenitors and explosion properties. In low energy explosions, the ejecta are denser and thus tends to be more recombined. The lower \nifs\ mass at low energy may also inhibit the influence of a \nifs\ blob, although this may not apply strictly since the polarization is sensitive to the relative offset in ionization between different ejecta locations. Similarly, for higher explosion energies with more \nifs, the offset in ionization caused by a \nifs\ blob may not be so large because the ``background'' ionization is higher. These considerations need to be checked with realistic progenitor, explosion, and radiative transfer models, as recently done for SN\,2012aw \citep{D20_12aw_pol}.

\begin{acknowledgements}

D.J.H. thanks NASA for partial support through the astrophysical theory grant 80NSSC20K0524.  D.C.L. acknowledges support from NSF grants AST-1009571, AST-1210311, and AST-2010001, under which part of this research was carried out. This work was granted access to the HPC resources of  CINES under the allocation  2018 -- A0050410554, 2019 -- A0070410554, and 2020 -- A0090410554 made by GENCI, France. This research has made use of NASA's Astrophysics Data System Bibliographic Services.

\end{acknowledgements}

% \bibliographystyle{aa}
% \bibliography{/Users/ldessart/Bibliography/new_sn_library_luc}

s

\appendix

\section{Doppler shift of the scattered flux}
\label{sect_dop}

\subsection{General considerations}

By scattering through the fast expanding ejecta, a photon is systematically red-shifted relative to its original wavelength. For single scattering, the observed redshift depends on the projected velocity of the scatterer. In the ideal configuration of a central emitting source (at rest) and a scatterer at velocity $V$ along direction $\vec{r}$, the final redshift of the emerging photon is $V (1 - \cos \alpha_{\rm los})$, where $\alpha_{\rm los}$ is the angle between $\vec{r}$ and the direction to the observer. If  $\alpha_{\rm los}=$\,0, the scatterer is moving towards the observer and the redshift is zero. For a scatterer moving away from us, the redshift is maximum and equal to $2V$.

In the context of a single blob moving at $V$ along the symmetry axis of a 2D ejecta, the polarization is zero if the inclination is zero or 180\,deg. Maximum polarization occurs at a 90-deg inclination, and the photons that are scattered within the blob should reach the observer with a redshift of $V$.  So, while the singly-scattered photons will be redshifted from zero to $2V$, the polarized flux (which originates from scattering and is enhanced for asymmetries in the plane perpendicular to the line of sight) will tend to be redshifted by $V$. In other words, for a distribution of blobs at different line-of-sight angles, the residual polarization they produce will tend to be dominated by the contribution of those blobs nearer the mid-plane, for which the total redshift is close to $V$. This Doppler shift can help reduce cancellation effects since scattered line photons are no longer coincident with the line rest wavelength. It may also bring additional polarization to regions that in a static case would be made exclusively of continuum photons (as in the spectral regions redwards of strong lines).

In a SN, the emitting source is not a localized region at rest at the center of the ejecta. The source of emission is both extended and moving. If the emitting source is, however, roughly spherical, there will be a broad distribution of Doppler shifts, although the mean Doppler shift received by photons on their way out may be the same as if the central source was confined and at rest. This broadening of the distribution will reflect the range of velocities of the emitting sites. In other words, an extended blob and a point source may yield the same redistribution of photons in velocity space as an extended source and a confined blob.

\subsection{Example for an idealized case}
\label{sect_ex_redshift}

\begin{figure*}
 \centering
\epsfig{file=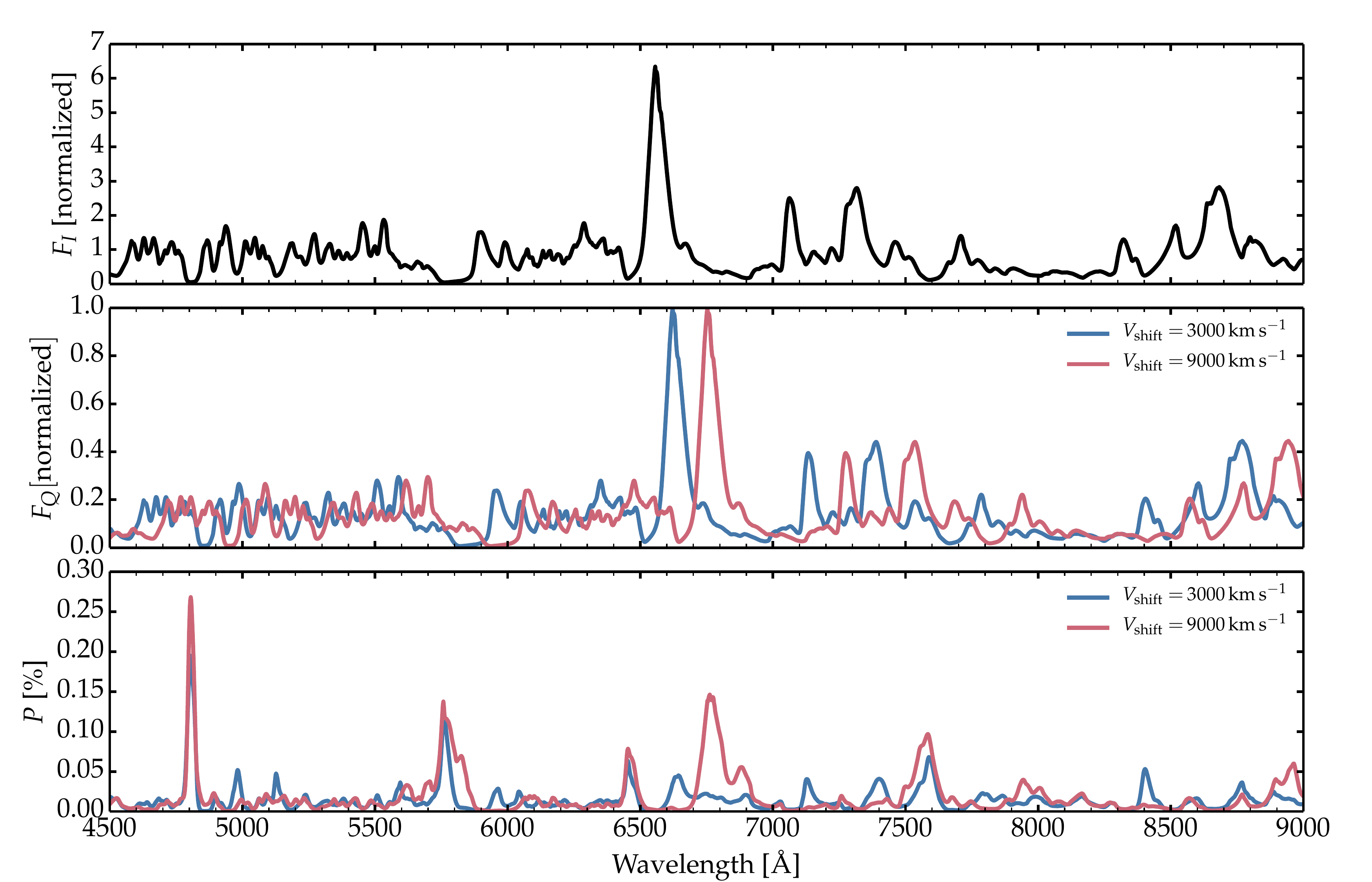, width=17cm}
\caption{Same as Fig.~\ref{fig_vblob}, but now showing the results for the idealized case of a high-velocity scatterer and a point source discussed in Section~\ref{sect_ex_redshift}. The polarized spectrum is identical to the flux spectrum except that it is redshifted. If we increase the velocity of the scatterer from 3000 (blue) to 9000\,\kms\ (red), $F_Q$ remains the same except for the increased redshift. The similarity between the intrinsic and scattered spectra is easily seen in the middle panel. However, the polarization $P$ (bottom panel) is much more complicated because of the velocity shifts. Line photons can be scattered into wavelengths where there is little flux, and hence produce a strong enhancement in the polarization, and even spikes in polarization, at those wavelengths. Retrieving information on the scatterer by analyzing $P$ is more difficult than by inspecting $F_Q$.
\label{fig_redshift}}
\end{figure*}

Let us now consider a point source and a (optically thin) scatterer moving away from the source at a velocity $V$. The polarized spectrum will show identical characteristics to the SN spectrum, except it will be redshifted\footnote{In this discussion we ignore the redistribution caused by the thermal motions of the electron whose random velocities are much smaller than the velocities of the ejecta.}. If we double the velocity $V$ we will observe exactly the same polarized spectrum, except it will have double the redshift. Because of this property,  $F_Q$  can provide more fundamental insights than the percentage polarization $P$. This is illustrated in Fig.~\ref{fig_redshift} which shows $F_I$, $F_Q$, and $P$ for two idealized cases -- the first with the polarized spectrum corresponding to $F_I$ redshifted by 3000\,\kms, the second redshifted by 9000\,\kms. As readily apparent, there are dramatic changes in the percentage polarization spectrum. Such changes are easy to explain;  the percentage polarization at a given wavelength does not simply reflect the scattering at that wavelength -- it is strongly influenced by scattering from shorter wavelengths. This will have a large effect on nebular-phase polarization spectra where there are strong emission lines adjacent to regions with little flux. A more informative polarization spectrum may be produced by shifting $F_Q$ in velocity before dividing by $F_I$.

Unfortunately, SNe cannot be approximated by a point source, and the scattering blob will occupy a range of velocities, and a range of scattering angles. As a consequence, the polarized spectrum will not be a simple copy of the flux spectrum. However it will still exhibit a strong correlation. The differences will, for example, provide information on the extent and location of regions giving rise to individual emission lines.

\begin{figure}
\epsfig{file=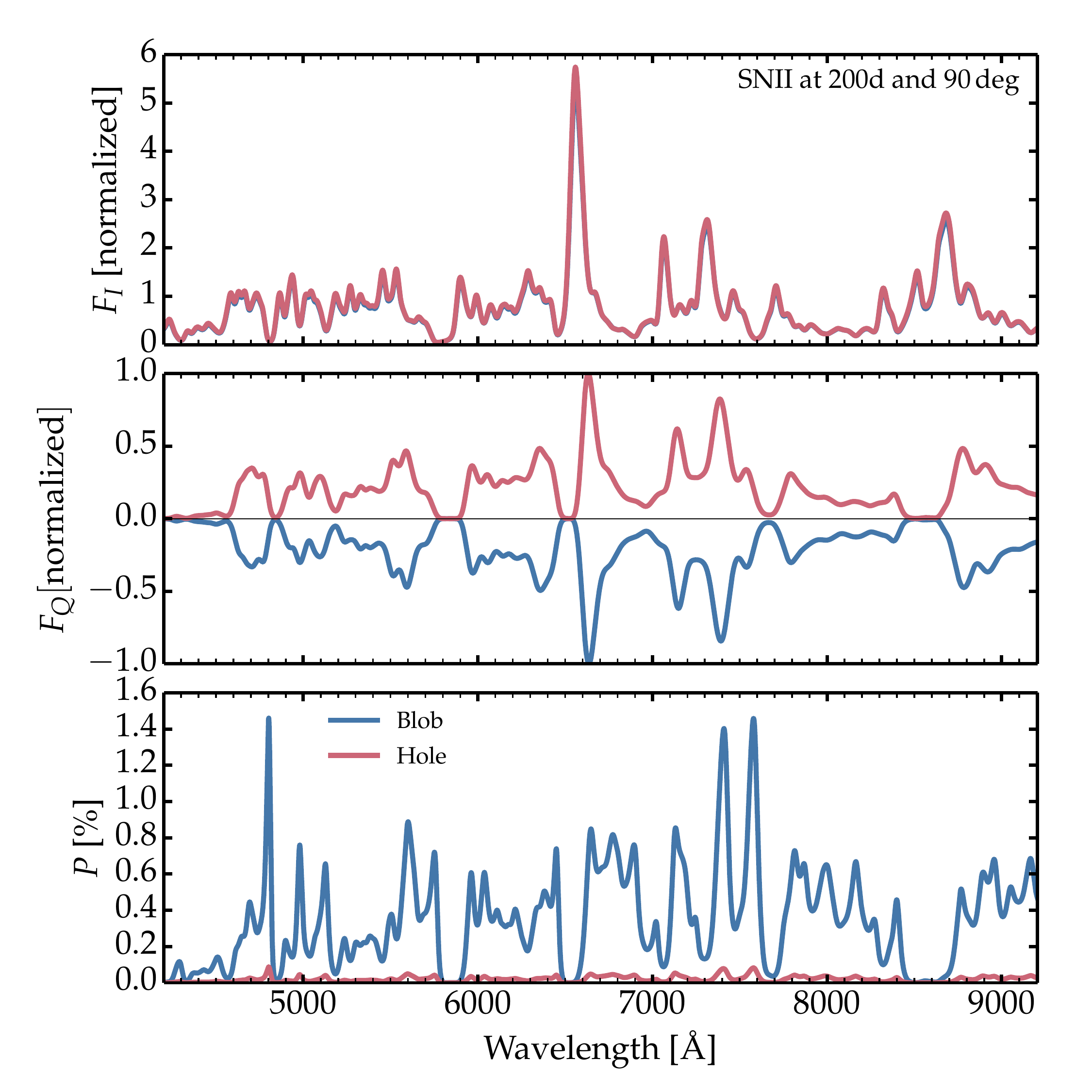, width=9.2cm}
\caption{Same as Fig.~\ref{fig_mirror_sym}, but now comparing a 2D model with a \nifs\ blob ($N_{\rm e,fac}=$\,20, corresponding to a radial blob optical depth of 0.436) and a \nifs\ hole ($N_{\rm e,fac}=-0.95$; reduction of the electron density by a factor of 20 corresponding to an optical depth deficit of $-$\,0.021). The blob or hole are located at 4000\,\kms, have a width of 1000\,\kms, and an half-opening angle of 10\,deg.
\label{fig_hole}}
\end{figure}

\section{Signatures of a \nifs\ hole}
\label{sect_hole}

Here, we consider the complementary configuration to a \nifs\ blob surrounded by a large volume with a lower free electron density. This complementary configuration corresponds to a \nifs\ hole, in other words, a region of reduced free electron density surrounded  by a large volume with a higher free electron density. Such a localized deficit in ionization could for example arise from localized density enhancements (i.e., clumps whose greater density would enhance the recombination relative to the surroundings), or chemical inhomogeneities (for example, He-rich or O-rich clumps, which would yield a lower free-electron density than H-rich clumps at the same temperature and ionization because of their higher atomic mass). Such asymmetries are likely occurring concomitantly to asymmetries in the \nifs\ distribution since they are the result of similar fluid instabilities. We show the results for the total flux and the polarized flux in Fig.~\ref{fig_hole}.

This hole breaks the ejecta symmetry but because of the reduced density of scatterers, it causes a very modest polarization relative to the case with a blob of enhanced free electron density. The remarkable feature is that the normalized $F_Q$ for the \nifs\ blob and \nifs\ hole are the exact mirror of each other, with merely a flip in sign. Switching from blob to hole, one merely switches from a prolate to an oblate configuration and a change of the sign of the shape factor term $(1-3\gamma)$ in Eq~\ref{eq_pol_BM77}.

\section{Analytic predictions for a single blob in optically thin regime}
\label{sect_BM77}

For a point source, the continuum polarization of radiation emerging from an optically-thin scattering envelope (scatterers being free electrons) with axial symmetry is given by \citet{Brown_McLean_77} and reads as

\begin{equation}
P= {3\over 8} \bar \tau (1 - 3 \gamma) \sin^2 i   \,\, ,
\end{equation}
where $\bar \tau$ is an angle-average optical depth defined by
 \begin{equation}
 \bar \tau = {1\over 2} \int \int_{-1}^1 \sigma_{\rm Th} N_e \,d\mu dr  \,\, ,
 \end{equation}
and $\gamma$ is a shape factor defined as
 \begin{equation}
 \gamma = { { \int \int_{-1}^1 N_e\, \mu^2 d\mu dr} \over { \int \int_{-1}^1 N_e \,  d\mu dr}} \,\, .
 \end{equation}
In these expressions, $r$ is the radius and $\mu$ is the cosine of the angle relative to the axis of symmetry.

\noindent
  The inclination angle $i$ corresponds to our line-of-sight angle $\alpha_{\rm LOS}$. Now consider a single bob of radial extent $r_{\rm min}$ to $r_{\rm max}$ with fixed angular size, located on the  symmetry axis,  and extending an angle $\beta$ from the axis  (i.e., the opening angle of the blob is $2\beta$). Then
 \begin{equation}
 \bar \tau = {1 \over 2} \tau_{\rm blob} [1-\cos \beta]  \,\, ,
 \end{equation}
where  $\tau_{\rm blob}$ is the radial optical depth of the blob and
 \begin{eqnarray}
 \gamma &=&  { {1 \over 3} (1-\cos^3 \beta) \over  (1-\cos \beta) }  \\
              &=& {1 \over 3}( 1+ \cos \beta + \cos^2 \beta)   \,\, .
 \end{eqnarray}
 Hence, after a little algebra, we obtain

\begin{eqnarray}
 P  &=& -  {3\tau_{\rm blob} \over 16} \cos \beta \sin^2 \beta \sin^2 i  \,\, .  \label{eq_pol_BM77}
\end{eqnarray}

For an edge-on view, the maximum  polarization occurs when $\cos^2 \beta= \frac{1}{3}$ (i.e., $\beta= 54.7$\,deg). The minus sign simply indicates that the electric vector is perpendicular to the symmetry axis. Expressions for the corresponding \nifs\ blob mass are given in appendix~\ref{sect_blob_mass}.

This polarization estimate ignores lines, which dominate the flux in a nebular phase spectrum. It also ignores optical depth effects, which can remain sizable because the ejecta radial electron-scattering optical depth stays close to 1 until several hundred days (assuming constant ionization from the onset of the nebular phase, the optical depth will drop by a factor of 4 from 150 to 300\,d.). The single-scattering limit postulated in the analysis of \citet{Brown_McLean_77} holds for optical depths below about 0.1 (see \citealt{hillier_94}). Finally, the approach assumes that the emission arises from a point source whereas in Type II SN ejecta in the nebular phase, the spectrum forms over an extended region spanning from the inner ejecta to several thousands of \kms\ (i.e. from the metal rich inner regions to the H-rich layers above). All these effects would tend to reduce the resulting polarization. A comparison of our simulations with the predictions of \citet{Brown_McLean_77}, and in particular Eq.~\ref{eq_pol_BM77}, is presented in the Appendix~\ref{appendix_BM77}.

\begin{table}
\caption{Variation of the continuum polarization normalized by the radial electron-scattering optical depth of the blob for various blob half-opening angles (calculation based on Eq.\ref{eq_pol_BM77}).}
\begin{center}
\begin{tabular}{lll}
\hline
$\beta$ & {$-P / \tau_{\rm blob}$}(\%)  \\
\hline
10 & 0.557 \\
20 & 2.06 \\
30& 4.06 \\
40& 5.93 \\
50&  7.07 \\
54.7& 7.22 \\
60& 7.03 \\
\hline
\end{tabular}
\end{center}
\end{table}

\section{Expression for the blob mass}
\label{sect_blob_mass}

Following on from the polarization analysis in section~\ref{sect_BM77}, the mass of the blob (i.e., the mass of the region in which the free-electron density is boosted by the localized enhancement in \nifs) is given by
\begin{eqnarray*}
M_{\rm blob} = 2\pi \int \int^1_{\cos \beta} r^2 \rho \, d\mu dr
                          = 2\pi (1 -\cos \beta) \int  r^2 \rho \, dr
\end{eqnarray*}

Assume that in the blob density $\rho =\rho_i (r_i /r)^n$ where $r_i$ is the inner radius of the blob (and $r_m$ is the  outer radius).
Thus
\begin{eqnarray}
M_{\rm blob} &=& 2\pi (1 -\cos \beta)\int^{r_m}_{r_i} r^2 \rho_i (r_i /r)^n   dr  \nonumber \\
         &=& 2\pi \rho_i (1 -\cos \beta) r_i^n \int^{r_m}_{r_i} (1/r)^{n-2} dr  \nonumber \\
         &=& {2\pi \rho_i (1 -\cos \beta) r_i^3  \over n-3 } \left[ 1 - \left( {  r_i \over r_m } \right)^{(n-3)} \right]
 \end{eqnarray}
Similarly the optical depth of the blob is
\begin{eqnarray*}
\tau_{\rm blob} &=& \int^{r_m}_{r_i} \sigma_{\rm Th}  N_{ei} (r_i /r)^n dr\\
            &=& {N_{ei}r_i \over (n-1) } \left[ 1 - \left( {  r_i \over r_m } \right)^{(n-1)} \right]
 \end{eqnarray*}
 Writing $N_{ei}= {\rho_i \gamma_e \over \bar{A} m_{\rm H}} $ gives
 \begin{equation}
 \tau_{\rm blob} = {\rho_i \gamma_e \sigma_{\rm Th}  r_i \over \bar{A} m_{\rm H} (n-1) }  \left[ 1 - \left( {  r_i \over r_m } \right)^{(n-1)} \right] \,\,.
\end{equation}
where $\bar{A}$ is the mean nucleon mass in amu, and $\gamma_e$ is the mean number of electrons per nucleon.
Solving for $\rho_i$ we have
\begin{equation}
\rho_i =  { \tau_{\rm blob} \bar{A} m_{\rm H} (n-1) \over r_i  \sigma_{\rm Th} \gamma_e}  \left[ 1 - \left( {  r_i \over r_m } \right)^{(n-1)} \right]^{-1}
\end{equation}
and hence
\begin{equation}
M_{\rm blob} = 2\pi (1 -\cos \beta) r_i^2 \tau_{\rm blob} \left({ \bar{A} m_{\rm H}(n-1) \over (n-3) \sigma_{\rm Th}  \gamma_e}\right)
 {  \left[ 1 - \left( {  r_i \over r_m } \right)^{(n-3)} \right] \over  \left[ 1 - \left( {  r_i \over r_m } \right)^{(n-1)} \right] }
\end{equation}
\noindent
Writing $r=Vt$ gives

\begin{equation}
\begin{aligned}
M_{\rm blob} = {} &  0.147 \,M_\odot (1 -\cos \beta) \tau_{\rm blob} \left({ \bar{A} (n-1) \over (n-3)  \gamma_e}\right) \\
         & \times  \left({ V \over 5000\,{\rm km\,s}^{-1}}\right)^2 \left({ t \over 100\,{\rm d}}\right)^2 \\
         & \times {  \left[ 1 - \left( {  V_i \over V_m } \right)^{(n-3)} \right]   \left[ 1 - \left( {  V_i \over V_m } \right)^{(n-1)} \right]^{-1} }
\end{aligned}
\end{equation}

\section{Comparison between our polarization results and the analysis of \citet{Brown_McLean_77}}
\label{appendix_BM77}

\begin{figure}
\epsfig{file=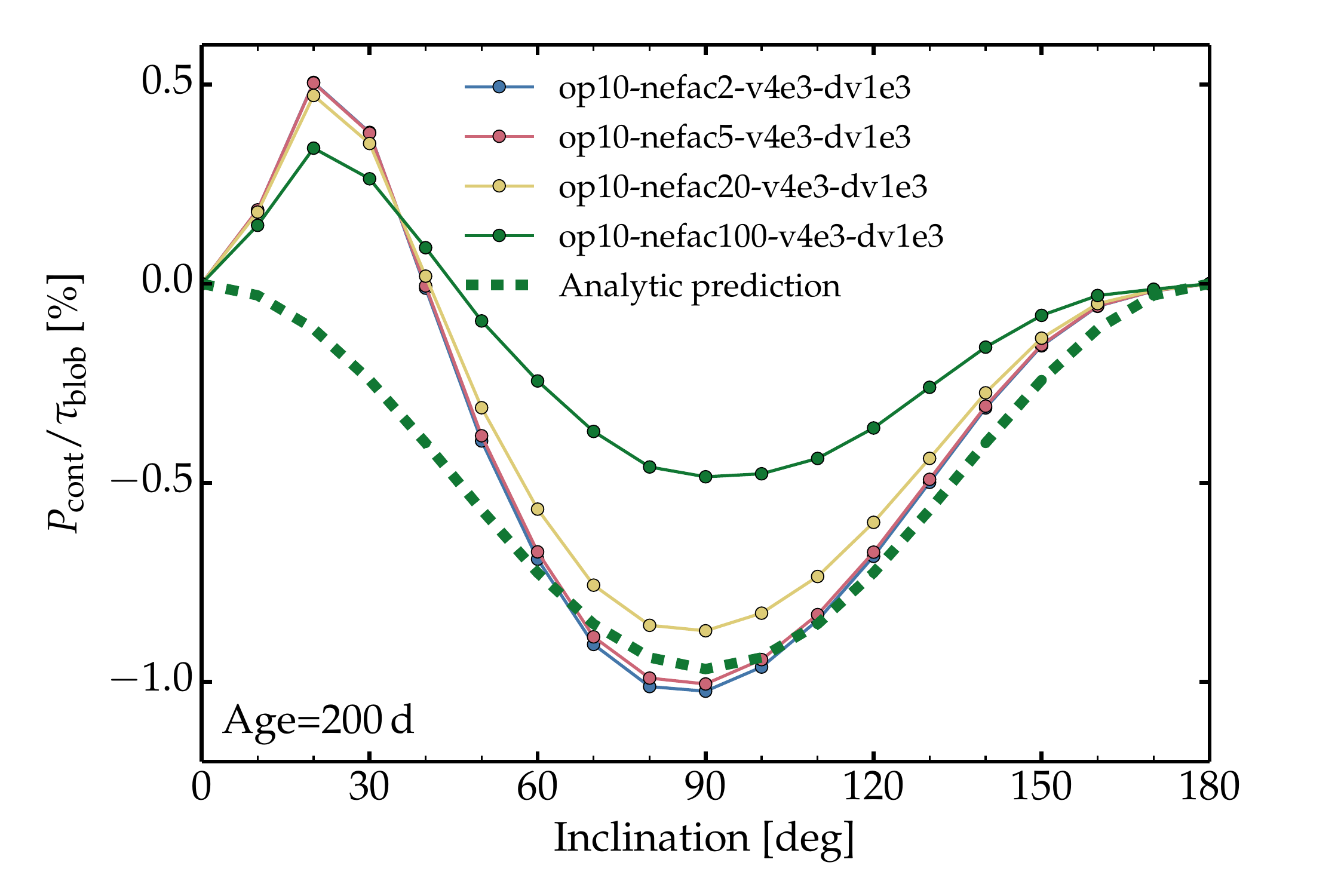, width=9.2cm}
\caption{Illustration of the continuum polarization at 5000\,\AA\ normalized to the blob optical depth and shown for inclinations between zero and 180\,deg, with the model aged to 200\,d. The simulations are characterized by various values of the electron density enhancement associated with the blob (see label; the corresponding blob radial optical depth increases from 0.044, to 0.109, 0.436, and 2.178 for increasing $N_{\rm e,fac}$). The dashed curve shows the prediction of  \citet{Brown_McLean_77}, scaled by a factor 0.75, using the characteristics of model op10-nefac100-v4e3-dv1e3 (green curve).
\label{fig_200d_pol_BM77}}
\end{figure}

\begin{figure}
\epsfig{file=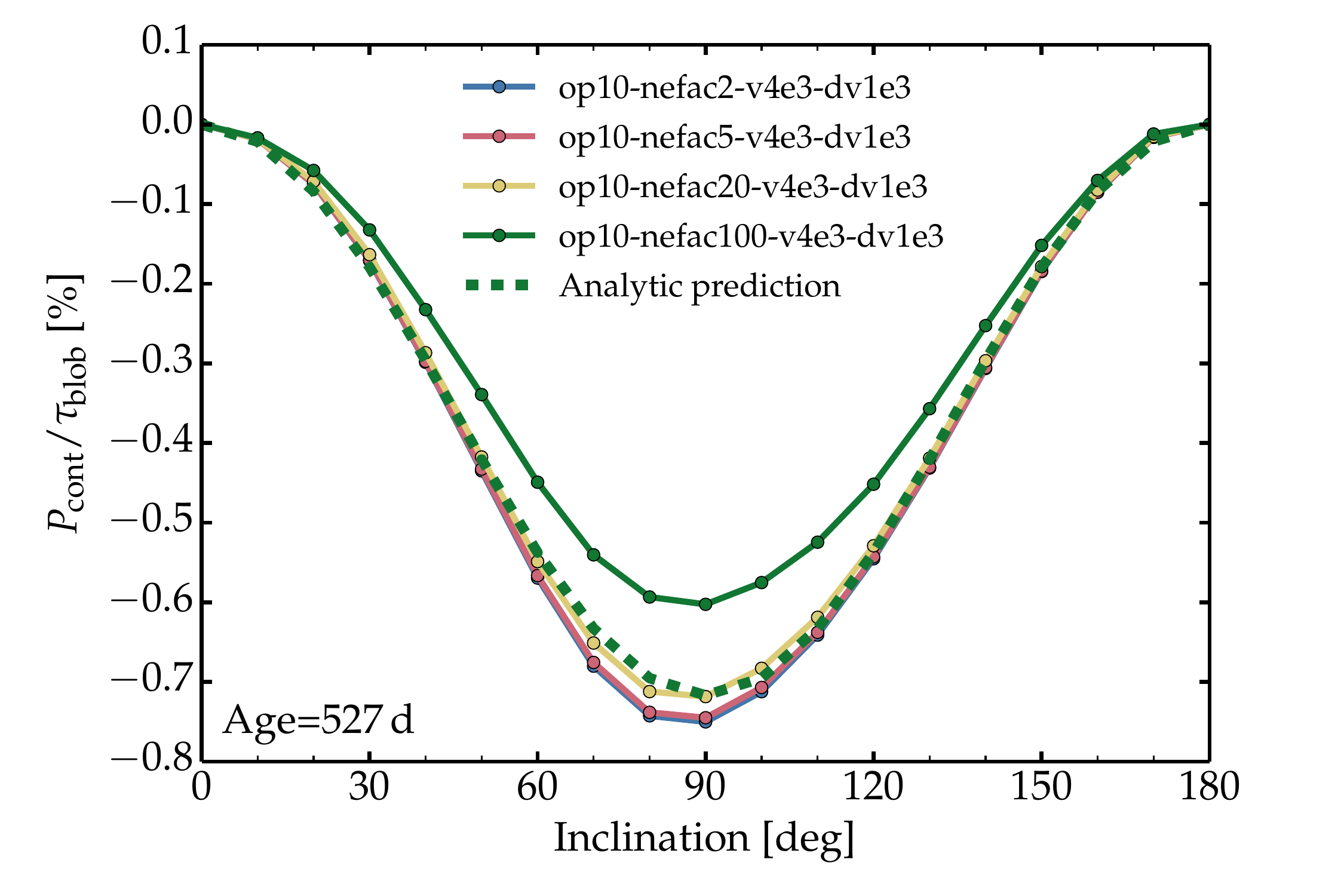, width=9.2cm}
\epsfig{file=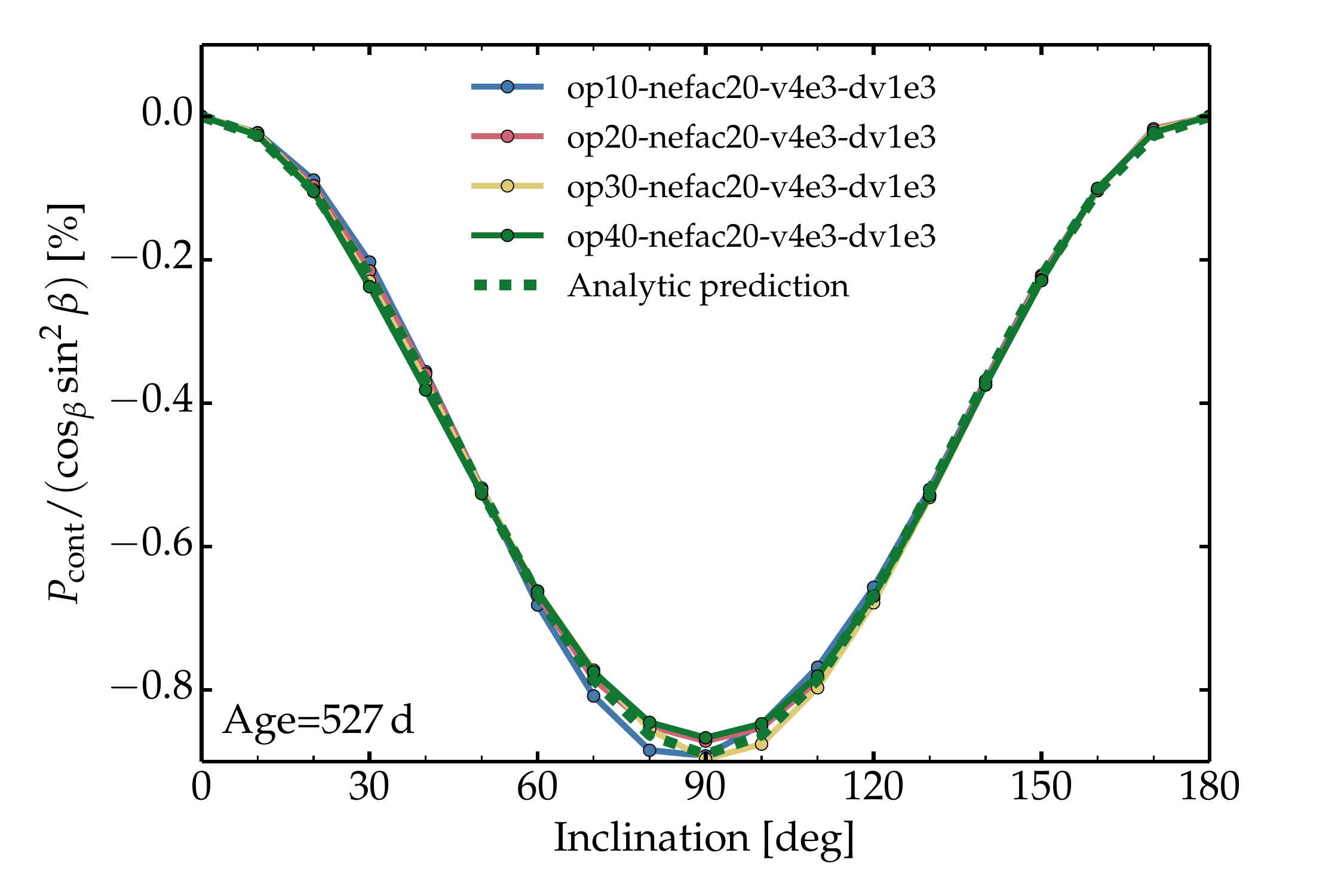, width=9.2cm}
\caption{Same as Fig.~\ref{fig_200d_pol_BM77}, but all calculations are now done at a much later time in the nebular phase, namely 527\,d after explosion.  The top panel shows the equivalent of Fig.~\ref{fig_200d_pol_BM77} at that older time. The bottom panel shows $P_{\rm cont}/(\cos_\beta \sin^2\beta)$ versus inclination. In both panels, the dashed line gives the analytic prediction.\label{fig_527d_pol_BM77}}
\end{figure}

In this section, we compare some of the results obtained in this study with the analytical predictions of \citet{Brown_McLean_77}. Equation~\ref{eq_pol_BM77} gives the expected magnitude and sign of the polarization for a blob (i.e., an electron density enhancement) of radial optical $\tau_{\rm blob}$, half-opening angle $\beta$, and seen at an inclination $i$ with respect to the axis of symmetry. This formula also gives scalings with these quantities.

Figure~\ref{fig_200d_pol_BM77} shows the continuum polarization (normalized by $\tau_{\rm blob}$) for the models characterized by different electron-density enhancements in the blob. The blob opening angle is the same in all four cases so the model predictions should all overlap when plotting $P/\tau_{\rm blob}$. The two models with the lowest enhancements do overlap. The third one (with ``nefac20'') si slightly off, but the last one (with ``nefac100'') is way off. Another feature present in all models and incompatible with Eq.~\ref{eq_pol_BM77} is the sign reversal at low inclinations. In the optically-thin limit and for a point source, such sign reversals are not expected.

The offsets discussed above are likely caused by an optical depth effect and because we do not have a point source of emission (i.e. if a blob of the same optical depth was placed at larger radii, the agreement would be better). Although the nebular phase starts as soon as a Type II falls off the plateau and lands on the nebular tail (with the bolometric luminosity equal to the total decay power absorbed), the mean optical depth (whether associated with the Rosseland-mean or the electron-scattering opacity) is only $\sim$\,1 at that time. In the model shown here, the ejecta without a blob has an electron-scattering optical depth of 0.84 at 200\,d so not small. The blob itself turns optically-thick for the largest enhancements used. For model ``nefac100'', the blob optical depth is about 2.2, so  Eq.~\ref{eq_pol_BM77} is no longer adequate and the prerequisites in the model of \citet{Brown_McLean_77} are not met.

To confirm that optical depth is the likely cause of the offsets, we repeat the same exploration but this time using a Type II SN model at 527\,d (this standard Type II SN model was computed with \cmfgen\ as part of the study by \citealt{HD19}). In that model at a later time, the total electron-scattering optical depth is now only 0.044, so much more optically thin. We recreate the same hybrid model configurations by introducing a blob of various electron-density enhancements or opening angle. The left panel of Fig.~\ref{fig_527d_pol_BM77} is the counterpart of Fig.\ref{fig_200d_pol_BM77} but now for this ejecta at 527\,d. A scaling by a factor of 0.6 has been applied to the curve corresponding to Eq.~\ref{eq_pol_BM77}, indicating that the model polarization is lower than predicted. This likely arises because at such late times, the $\gamma$-ray mean free path is much larger than at 200\,d, so we are even further away from the configuration of a point source (i.e., the emitting source is extended). The large extent of the emitting source necessarily reduces the polarization. Apart from this persisting (but understandable offset), the sign reversal is now gone and all curves follow qualitatively the predictions of \citet{Brown_McLean_77}. The bottom panel of Fig.~\ref{fig_527d_pol_BM77} shows the results for models characterized by the same electron-density enhancement but a range of blob opening angle. Hence, $\tau_{\rm blob}$ is the same for all four cases so the polarization normalized by the opening-angle term in Eq.~\ref{eq_pol_BM77} should follow a $\sin^2i$ curve. They do with good fidelity.

Besides demonstrating that the code behaves as expected, this exploration emphasizes that true optically-thin conditions are met only at very late times in Type II SNe. But then the point-source assumption no longer holds. In the context of SNe, this regime is not really suitable for discussing the properties of the continuum radiation since nebular phase spectra are instead notorious for their strong emission lines. In our model, the continuum flux at 200\,d represents only 9\,\% of the total flux and the polarization of continuum photons is typically less than 1\,\%.

\end{document}